\documentclass{article}[11pt]
\usepackage{jheppub}
\usepackage[toc,page]{appendix}
\pdfoutput=1
\usepackage{graphicx}
\usepackage{amsmath,amssymb,mathrsfs}
\usepackage{bbm}
\usepackage{color}
\usepackage{xcolor}
\usepackage{multicol}
\usepackage[thinlines]{easytable}

\usepackage{dsfont}
\usepackage{cancel}
\usepackage{dsfont}
\usepackage{epstopdf}
\usepackage{epsfig}
\usepackage{bm}
\usepackage{dcolumn}
\usepackage{hyperref}
\usepackage{enumitem}
\usepackage{multirow}
\usepackage{lineno}
\usepackage{mathtools,slashed}
\usepackage{url}
\usepackage{lineno}
\usepackage{wrapfig}
\usepackage{float} 
\usepackage{arydshln} 
\usepackage{xspace}
\usepackage{makecell}

\usepackage[load-configurations=abbreviations, redefine-symbols=true]{siunitx}

\usepackage{lineno}

\newcommand{\ben}{\begin{enumerate}}
\newcommand{\een}{\end{enumerate}}
\newcommand{\bit}{\begin{itemize}}
\newcommand{\eit}{\end{itemize}}

\newcommand{\beqa}{\begin{eqnarray}}
\newcommand{\eeqa}{\end{eqnarray}}
\newcommand{\beq}{\begin{equation}}
\newcommand{\eeq}{\end{equation}}
\newcommand{\bay}{\begin{array}}
\newcommand{\eay}{\end{array}}

\def\ifmath#1{\relax\ifmmode #1\else $#1$\fi}

\def\gsim{\ \rlap{\raise 3pt \hbox{$>$}}{\lower 3pt \hbox{$\sim$}}\ }
\def\lsim{\ \rlap{\raise 3pt \hbox{$<$}}{\lower 3pt \hbox{$\sim$}}\ }

\def\ls#1{\ifmath{_{\lower1.5pt\hbox{$\scriptstyle #1$}}}}
\def\lsup#1{^{\lower 6pt\hbox{$\scriptstyle#1$}}}

\def\bracket#1#2 {\mathinner{\langle{#1}|{#2}\rangle}}

\def\bracket#1#2 {\mathinner{\langle{#1}|{#2}\rangle}}

\newcommand{\be}{\begin{equation}}
\newcommand{\ee}{\end{equation}}
\newcommand{\bea}{\begin{eqnarray}}
\newcommand{\eea}{\end{eqnarray}}

\DeclareMathOperator*{\argmax}{arg\,max}
\DeclareMathOperator*{\argmin}{arg\,min}

\newcommand{\npop}{\ensuremath{n_\text{P}}\xspace}
\newcommand{\ntest}{\ensuremath{n_\text{T}}\xspace}
\newcommand{\ninf}{\ensuremath{n_{\text{I} \wedge \text{T}}}\xspace}
\newcommand{\rinf}{\ensuremath{r_\text{I}}\xspace}
\newcommand{\nfatal}{\ensuremath{n_\text{F}}\xspace}
\newcommand{\rfatal}{\ensuremath{r_\text{F}}\xspace}

\graphicspath{{figs/}}

\begin{document}



\title{Statistical techniques to estimate the SARS-CoV-2 infection fatality rate}

\abstract{The determination of the infection fatality rate (IFR) for the novel SARS-CoV-2 coronavirus is a key aim for many of the field studies that are currently being undertaken in response to the pandemic. The IFR together with the basic reproduction number $R_0$, are the main epidemic parameters describing severity and transmissibility of the virus, respectively. The IFR can be also used as a basis for estimating and monitoring the number of infected individuals in a population, which may be subsequently used to inform policy decisions relating to public health interventions and lockdown strategies. The interpretation of IFR measurements requires the calculation of confidence intervals. We present a number of statistical methods that are relevant in this context and develop an inverse problem formulation to determine correction factors to mitigate time-dependent effects that can lead to biased IFR estimates. We also review a number of methods to combine IFR estimates from multiple independent studies, provide example calculations throughout this note and conclude with a summary and ``best practice'' recommendations. The developed code is available online. }

\author[1]{M.~Mieskolainen,}
\author[1]{R.~Bainbridge,}
\author[1]{O.~Buchmueller,}
\author[1]{L.~Lyons,}
\author[1]{N.~Wardle}
\affiliation[1]{Department of Physics, Blackett Laboratory, Imperial College, Prince Consort Road, London, SW7 2AZ, UK}

\emailAdd{m.mieskolainen@imperial.ac.uk}

\maketitle

\begin{flushleft}
\end{flushleft}
\medskip
\noindent

\newpage
\section{Introduction} 

A critical task in the context of the SARS-CoV-2 pandemic is to
determine the infection fatality rate (IFR), defined as the proportion 
of deaths among all infected  individuals. The IFR is one of several 
characteristic measures that form the basis for epidemiological models such as \cite{imperial_report_9}, 
which are subsequently used to shape and justify government policy on 
public health interventions. The basic mean reproduction number $R_0$ and its full distribution extensions characterize the multiplicative process rate on the production side of the epidemic and the IFR is defined on the decay side of the individual infections. In this paper we concentrate our efforts on the IFR and do not consider the estimation of $R_0$, even if the total risk and harm caused by the virus is driven by both factors.

A more widely reported metric is the case fatality rate (CFR),
defined as the ratio of the number of deaths attributable to
SARS-CoV-2 and the number of documented infections. The reported CFR
can vary significantly between regions and countries, due in a large
part to the varying ability of local authorities to comprehensively
screen all suspected cases of infection. Further, there is significant
evidence for a substantial cohort of asymptomatic carriers of
SARS-CoV-2, an important subpopulation that is only partially (if at
all) represented by the CFR. Studies to determine the IFR are
typically supported by cross-sectional seroprevalence surveys in
population samples to also account for asymptomatic (and mildly or
atypically symptomatic) infections. Hence, the IFR is considered to be
a more reliable variable than the CFR for the purposes of
epidemiological modelling.

There are numerous serological surveys underway, or recently
concluded, that aim to estimate the IFR for SARS-CoV-2. It is crucial
that these studies consider all relevant sources of uncertainty, of
both statistical and systematic origin, to establish the confidence
intervals on individual estimates of the IFR. This in turn allows for
meaningful comparisons between (and potential combinations of)
independent results. The reported confidence interval in Ref.~\cite{Streeck2020.05.04.20090076}
appears to neglect the dominant uncertainty in the study, namely the
variance in the underlying statistical distribution used to model the
number of fatalities. This omission may have implications for policy
decisions made on the basis of estimates from these types of
study. An accurate and unbiased estimate of IFR can be also difficult to obtain
during an evolving epidemic due to various time delays that must be
correctly accounted for: from exposure to the virus to the onset of
symptoms following the incubation period, to the reporting of a
positive case following a PCR test (polymerase chain reaction), to the development of sufficient antibodies to be identified by a test (seroconversion), to the
recording of a fatality.

The body of research on SARS-CoV-2, and the resulting COVID-19
disease, grows at an astonishing rate. The number of studies from
which an estimate of the IFR can be drawn is now plentiful and
meta-analyses are now being performed that aggregate information from
several sources. For example,
Ref.~\cite{Meyerowitz-Katz2020.05.03.20089854} considers 25 estimates
of the IFR that are derived from studies of various types, including
serological surveys and epidemiological modelling. It reports a
point-estimate of 0.68 [0.53, 0.82]\% for the IFR, with the interval
stated at a 95\% confidence level.\footnote{All subsequent intervals
  reported in this note are also provided at a 95\% CL, unless stated
  otherwise.} However, the study acknowledges a high degree of
heterogeneity in the individual estimates. Indeed, there are many
factors that may influence the results of the individual
studies. Ref.~\cite{Meyerowitz-Katz2020.05.03.20089854} comments on
the variability in the quality and rigour of the surveys, and the lack
of peer review for some studies. Also noted is the challenge of
determining the IFR from serological surveys for which there is an
absence of an associated, reliable fatality count. There are many
local factors, such as population density, demographics, and health,
and the ability of healthcare services and government policy to
protect the local population. Perhaps one of the most important
factors is the age demographic of a population, as the IFR appears to
be highly dependent on age. Accurate estimates of IFR stratified by
age are highly desirable in the near future.

The meta-analysis reported in
Ref.~\cite{Meyerowitz-Katz2020.05.03.20089854} relies on a common
`method of moments' method by DerSimonian and
Lard~\cite{dersimonian1986meta} to provide a point-estimate of the
IFR. We review several approaches to combine results from individual
studies of the IFR into a single estimate. These include the method 
of moments and a normal likelihood based classic meta-analysis, a joint likelihood combination, and methods to combine full probability densities such as optimal transport and the product or mean of Bayesian posteriors.

In Section~\ref{sec:Heinsberg}, we introduce the Gangelt field study~\cite{Streeck2020.05.04.20090076}, which we use as an example in the first part of the paper. Section~\ref{sec:methods} reviews several methods that can be used to determine a confidence interval for the IFR, based on a single binomial proportion, a ratio of binomial proportions, a profile likelihood ratio, Monte Carlo simulation, and Bayesian constructs. Section~\ref{sec:comparisons} compares the confidence intervals from the various methods. Section~\ref{sec:time_evolution} presents a time-dependent deconvolution calculus that accounts for intrinsic time delays through the determination and application of a correction factor (and associated uncertainty) to the IFR estimate. Section~\ref{sec:combination} provides first a general perspective on how multiple data points can be combined, before a set of seroprevalence studies from around the globe are introduced, which are then used as concrete examples for the aforementioned combination techniques. Finally, Section~\ref{sec:summary} summarises this work and concludes by providing ``best practice'' recommendations in the context of confidence interval calculations for the IFR of the SARS-CoV-2 coronavirus.

\section{The Gangelt field study}
\label{sec:Heinsberg} 

A sero-epidemiological study of a small German community exposed to a
super-spreading event was undertaken to determine the
IFR~\cite{Streeck2020.05.04.20090076}. The municipality of Gangelt is
located in the district of Heinsberg in the German state of North
Rhine-Westphalia. The municipality reported unusually high levels of
SARS-CoV-2 infections following local Carnival festivities held on
15$^{\text{th}}$ February 2020. Strict social distancing measures,
which included an advisory curfew, were introduced on 26$^{\text{th}}$
February to suppress further infections. The Carnival festivities held
in Gangelt are attended predominantly by people living locally and the
community is relatively closed, with low levels of tourism and travel.

An estimate of the IFR is obtained from the double ratio
\begin{equation}
  \label{eq:main_formula}
  \widehat{\text{IFR}} = \frac{\rfatal}{\rinf} = \frac{\nfatal/\npop}{\ninf/\ntest} =
  \frac{\nfatal}{\npop\rinf},
\end{equation}
where the raw fatality rate \rfatal is the ratio of the number of
fatalities \nfatal and individuals \npop in a given population, and
the raw infection rate \rinf is the ratio of the number of infections
\ninf identified in a representative cross-sectional sample of tested
individuals \ntest. Assuming the test sample of \ntest individuals is
representative of the population under study, in terms of both
demographic and seroprevalence measures, the product $\npop\rinf$
provides an estimate of the total number of infected individuals in
the population \npop. The authors of
Ref.~\cite{Streeck2020.05.04.20090076} identified the Gangelt
municipality as a unique opportunity to accurately estimate the IFR,
because of its stable, relatively isolated population and an
appreciable number of fatalities arising from the high level of
infections present in its community.

The Gangelt municipality has a population of $\npop =
12597$. Inhabitants were randomly polled to participate in testing for
both SARS-CoV-2 virus RNA (PCR) and antibodies to identify the number
of both present and past infections, respectively. The number of
inhabitants that were tested and satisfied all survey requirements is
$\ntest = 919$, and the number of identified infections (\ninf) in
this sample is $\ninf = 138$. Within the duration of the study, the
number of deaths recorded in the Gangelt municipality that were
associated with a SARS-CoV-2 virus infection is $\nfatal = 7$. Hence
the study yields a raw infection rate of 15.0\% and an IFR of 0.37\%.

Following the application of corrections for various identifiable
biases and the associated statistical and systematic uncertainties,
Ref.~\cite{Streeck2020.05.04.20090076} reports $\rinf = 15.5$
$[12.3, 19.0]$\%. Hence, the total number of infected individuals in the
Gangelt municipality is estimated to be 1950 $[1550, 2390]$. The
reported interval for the IFR of 0.37\% is $[0.29, 0.45]$\%, which can be
used to estimate the number of infected individuals in other places
with population characteristics similar to that used in the Gangelt
study. Several other key findings are reported in
Ref.~\cite{Streeck2020.05.04.20090076}, which are beyond the scope of
this note. We restrict the discussion in this note to the reported
confidence interval for the IFR estimate.

A total of 6575 deaths associated with SARS-CoV-2 were recorded in
Germany by 2$^{\text{nd}}$ May 2020. Assuming the Gangelt sample
population is representative of the German population as a whole, the
IFR can be used to estimate the total number of infections -- at some
point earlier in the pandemic -- that led to the reported death toll
on 2$^{\text{nd}}$ May 2020. The study yields an estimate of 1.8
$[1.5, 2.3]$ million infected individuals in the German
population. Using the methods described below, we determine a confidence interval (CI95) of $[0.9, 3.7]$ million on the estimate of 1.8
million infected individuals, based on a simple method (the Wilson
score) applicable to a single binomial proportion.

\section{Methods}
\label{sec:methods}

Estimation of confidence intervals of parameters with statistical tests goes as follows. The test statistic for the null hypothesis $H_0$ is known to asymptotically follow an analytic distribution, which is taken as an approximate proxy to the problem. The exact solution is available only in special cases. The analytic approximation can be relaxed with more modern numerical bootstrap and Monte Carlo techniques which provide the most appropriate tools when data needs to be also re-weighted, propagated through a chain of analysis algorithms or manipulated in more complex ways.

\paragraph{Notation} ~ The likelihood function is $L(\theta) \equiv L(\theta; x_1,x_2,\dots,x_n) = f(x_1,x_2,\dots,x_n;\theta) = \prod_j f(x_j;\theta)$, where $f$ is the underlying sampling probability density (pdf) and the last equality holds for $n$ independent and identically distributed (iid) observations in the sample $\{x_j\}$. Algebraically, the likelihood and the density have the same origin, the difference being only if the parameter $\theta$ \textit{or} the observable $x$ is treated as the variable of the function. The density unit normalization holds only over $x$. This difference should make the mathematical meaning unambiguous, even if the term likelihood is used often in a relaxed way. The null hypothesis $H_0$ parameter $\theta$ values, which are not fixed, are denoted with $\theta_0$ and the maximum likelihood estimates (MLE) with $\hat{\theta} = \argmax \, L(\theta)$. 

\subsection{Single binomial confidence intervals}

Single binomial uncertainty is the most dominating statistical uncertainty on the IFR estimate, because the fractional uncertainty on number of fatalities is typically much larger than the uncertainty in the number of infections.

\paragraph{Wald test (normal)} ~ The most common estimator for the binomial success rate confidence interval is the so-called normal approximation interval based. The interval can be derived by inverting the Wald test
\begin{equation}
\label{eq:Wald_test}
W =  \frac{(\hat{\theta} - \theta_0)^2}{\text{var}(\hat{\theta})}, \;\; Z =  \frac{\hat{\theta} - \theta_0}{\text{se}(\hat{\theta})},
\end{equation}
where the parameter $\theta \equiv p$. On the left side, the statistic for $H_0$ follows asymptotically the $\chi^2$-distribution and on the right side, the asymptotic $Z$-distribution (standard normal). The number of degrees of freedom of the $\chi^2$-distribution is $d = \dim[\theta]$, with $d=1$ here. The standard error of the binomial parameter is $\text{se}(\hat{p}) = [\hat{p}(1-\hat{p})/n]^{1/2}$. Then writing down $-z_{\alpha/2} \leq Z \leq z_{\alpha/2}$, substituting Eq.~\ref{eq:Wald_test} and re-arranging gives
\begin{equation}
\label{eq:normal_approx}
CI_S = \hat{p} \pm z_{\alpha/2} [\hat{p}(1-\hat{p})/n]^{1/2},
\end{equation}
where $\hat{p} = k/n$ is the maximum likelihood estimate of the central success rate given $k$ successes and $n$ trials. The standard normal inverse cumulative distribution quantile is $\Phi^{-1}(1-\alpha/2) = -\Phi^{-1}(\alpha/2) = z_{\alpha/2}$ for a confidence level $(1-\alpha) \times 100 $ \%. Numerically, these are $z=1$ for 68.27 \% and $z=1.96$ for 95 \% confidence levels (intervals), respectively. The construction here assumes $\sqrt{n}(\theta - \hat{\theta})$ to follow a Gaussian $N(0,\sigma^2)$ by Central Limit Theorem and the true variance $I(\theta)^{-1}$ is estimated with the plug-in estimate $I(\hat{\theta})^{-1}$, using the Fisher information $I(\theta)$ given in Eq.~\ref{eq:binomial_fisher_information}. Both assumptions are valid only under $n \rightarrow \infty$. Finite $n$ coverage of this interval estimator is weak as emphasized in~\cite{brown2001interval}, and also shown in our simulations in Appendix~\ref{appendix:sec:coverage}, and its use cannot be recommended. Because the Wald test is not scale-invariant, one may try to improve its behavior with normalizing transformations such as the log-odds transform $\phi = \ln p/(1-p) \in (-\infty, \infty)$ or a pure log-transform $\phi=\ln p$. The interval endpoints are then calculated in the transformed space and inverse transformed.

\paragraph{Wilson score} ~ Wilson derived an estimator~\cite{wilson1927probable} for the binomial proportion parameter confidence intervals using more advanced reasoning on the probabilities than the standard Wald test-based approximation, and this leads crucially to a different evaluation point. Using here a more modern construction, the score test statistic is
\begin{equation}
\label{eq:score}
S = \frac{U(\theta_0)^2}{I(\theta_0)},
\end{equation}
where the gradient of the log-likelihood (score) and the Fisher information are
\begin{align}
&U(\theta) = \partial \ln \ell(\theta)/\partial \theta = (k-np)/(p-p^2) \\
\label{eq:binomial_fisher_information}
&I(\theta)=\mathbb{E}[-\partial^2 \ln \ell(\theta) / \partial \theta | \theta] = \hat{\text{var}}[\theta]^{-1} = n/(p-p^2).
\end{align}
As originally shown by Rao~\cite{rao1948large}, this test follows $\chi^2$-distribution asymptotics like the Wald test. Also it can be shown that the score test formulation is actually equivalent with a Lagrange multipliers-based constrained optimization~\cite{silvey1959lagrangian}, used often in economics, physics and engineering.

Score intervals typically require numerical solutions. However, by setting Eq.~\ref{eq:score} equal to $z^2$ which is allowed because $\chi^2$ with one dof is equal to the standard normal squared, a quadratic closed form solution is obtained
\begin{equation}
CI_W = \frac{\hat{p} + \frac{z^2}{2n}}{1 + z^2/n} \pm z \frac{\sqrt{\hat{p}(1-\hat{p})/n + \frac{z^2}{4n^2}}}{1 + z^2/n}.
\end{equation}
The central estimate of the rate is not given by $k/n$, but $(k + z^2/2) / (n + z^2)$, which makes a significant difference with small event counts. We return to this feature of intervals with the Bayesian estimates. Typical extensions to the Wilson score interval add continuity corrections to the standard formula. Wilson score can be recommended as the de facto choice to replace the weakly performing pure Wald test based one.

\paragraph{Likelihood ratio} ~ The log-likelihood ratio based test statistic is
\begin{equation}
LLR(\theta_0) = -2\ln \frac{L(\theta_0)}{L(\hat{\theta})} = -2[\ln L(\theta_0) - \ln L(\hat{\theta})]
\end{equation}
which unlike the Wald or the score test, is a scale-invariant test. As before, one relies here on the $\chi^2$-distribution, which gives the asymptotic null hypothesis distribution according to Wilks' theorem~\cite{wilks1938large}. Non-asymptotic inference without relying on the $\chi^2$-distribution is typically only possible via Monte Carlo simulations, unless one uses techniques such as the saddlepoint approximations \cite{barndorff1994inference}. The binomial log-likelihood ratio is
\begin{equation}
\label{eq:binom_LLR_test}
LLR(p_0) = 2\left[k \ln \hat{p} + (n-k)\ln (1-\hat{p}) - (k \ln p_0 + (n-k)\ln (1-p_0)) \right].
\end{equation}
Comparing with the $\chi_1^2$-distribution $1-\alpha$ quantile gives the confidence interval
\begin{equation}
\label{eq:asymptotic_LLR_chi2}
CI_{LLR} = [\min(p_0), \max(p_0)] \; \text{such that} \; LLR(p_0) \leq \chi_{1,1-\alpha}^2,
\end{equation}
which is found numerically. In a more general case, asymptotic $d$-parameter (vector) inference requires comparing the likelihood ratio with a $\chi^2$-distribution having $d$ degrees of freedom.
\\

\begin{table}[b!]
    \centering
    \begin{tabular}{|c|c|c|}
    \hline
    Test strategy & Based on & Scale invariant \\
    \hline
    Wald & Information curvature of likelihood at $\hat{\theta}$ & No\\
    Score (Lagrange) & Information slope and curvature at $\theta_0$ & No \\
    LR & Comparing likelihoods of $\hat{\theta}$ and $\theta_0$ & Yes \\
    `Exact' & Direct integration & - \\
    \hline
    \end{tabular}
    \caption{Different frequentist confidence interval test constructions summarized.}
    \label{tab:test_table}
\end{table}

\paragraph{Clopper-Pearson} ~ This classic~\cite{clopper1934use} binomial confidence interval estimator is also known as the `exact' interval because it is based on direct integration over the binomial distribution and thus compatible with the Neyman construction of confidence intervals, given in Appendix~\ref{appendix:sec:interval_definition}. By construction, it never undercovers. The interval is usually obtained numerically by integrating over a beta distribution, which is dual to the sum over the binomial tails
\begin{equation}
\sum_{k=X}^n \begin{pmatrix} n \\ k \end{pmatrix} p^k (1-p)^{n-k} = \int_0^p dt \, \text{Beta}(t|X,n-X+1).
\end{equation}
The quantile integrals are
\begin{align}
\label{eq:clopper_pearson_lower}
\alpha/2   &= \int_0^{L} dx \, \text{Beta}(x|k, n-k+1) \;\;\;\, \text{(lower endpoint}) \\
\label{eq:clopper_pearson_upper}
1-\alpha/2 &= \int_0^{U} dx \, \text{Beta}(x|k+1, n-k) \;\;\; \text{(upper endpoint}),
\end{align}
which are numerically inverted for $L$ and $U$. The beta distribution is
\begin{equation}
\text{Beta}(x|\alpha,\beta) = \frac{1}{B(\alpha,\beta)} x^{\alpha-1} (1-x)^{\beta-1}, \;\;\; \alpha,\beta > 0, \;\; x \in [0,1]
\end{equation}
with the normalization provided by the beta function $B(\alpha,\beta) \equiv \int_0^1 dx \, x^{\alpha-1} (1-x)^{\beta-1} = \Gamma(\alpha+\beta) / [\Gamma(\alpha) \Gamma(\beta)]$, where $\Gamma$ is the gamma function. When $k=n$ the interval is $[0, (1-\alpha/2)^{1/n}]$ and when $k=n$ the interval is $[(\alpha/2)^{1/n}, 1]$. This estimator is called \textit{conservative}, because its guaranteed interval coverage is always equal to or larger than the nominal one. Related, Blyth and Still~\cite{blyth1983binomial} have shown how to construct a confidence interval for the binomial distribution with nominally optimal but conservative coverage \textit{and} minimal length. The construction is nearly equivalent with Clopper-Pearson, but different optimization criteria are being used. Downside of the alternative constructions is that different sized intervals are not always fully contained within each other, i.e., they are not necessarily nested as one would simply expect.

\paragraph{Lancaster mid-$P$} ~  In situations like the one studied here where observations are discrete (integers), the $p$-value is traditionally defined as the probability of obtaining the actual observed number $k_{obs}$ or anything more extreme. These are used, for example, in obtaining Clopper-Pearson intervals for the binomial probability of success from the given $k_{obs}$; and this results in over-coverage. A method for mitigating this~\cite{lancaster1949combination} is to consider only half of the probability of obtaining $k_{obs}$ in calculating the $p$-value, i.e.
\begin{equation}
\text{mid-}p = \frac{1}{2} \times p(k = k_{obs}) + p(k > k_{obs})
\end{equation}
for the right-hand tail. Using this results in intervals that are shorter than those for the standard Clopper-Pearson intervals. The price to pay for the shorter intervals is that the mid-$p$ method has undercoverage for specific ranges of the parameter of interest $p$, which vary with the total number of binomial tests $N$. Since $N$ carries no useful information about $p$, suggestions have been made to average the coverage over $N$ which should be acceptable even to frequentists. This procedure results in much reduced undercoverage for the mid-$p$ approach (see, for example, ref.~\cite{cousins2010frequentist}).

\paragraph{Characteristics}~Chaotic coverage properties of classic binomial uncertainty interval estimators were studied in detail in~\cite{brown2001interval}, where the Wald test based interval estimator was shown to be completely unsuitable when it comes to its practical coverage. In general the chaotic properties are due to the underlying binomial distribution spanning a discrete lattice structure, not a continuum. Table~\ref{tab:test_table} summarizes different test construction strategies. The Wald, the score and the likelihood ratio all have the $\chi^2$-distribution as their null hypothesis $H_0$ asymptotic distribution. The frequentist coverage aspects behind these estimators have been also studied in~\cite{cousins2010frequentist}.

\subsection{Generating non-asymptotic test statistics}
\label{sec:toymc}
Confidence intervals without relying on the asymptotic $\chi^2$-statistic can be obtained using Monte Carlo. A common choice with optimal properties in this context is the likelihood ratio based construction or `ordering principle' of acceptance sets, which has been studied theoretically and computationally since the Neyman-Pearson lemma, see e.g. Refs.~\cite{lehmann1959testing, kendall1961advanced, spjotvoll1972unbiasedness, owen1990empirical, feldman1998unified}. It relies on the (exact) duality between statistical tests and confidence intervals. Now, the well known brute-force algorithm to construct the so-called (Neyman) confidence belt is as follows.
\begin{enumerate}
\item The parameter $\theta_0$ space is discretized or randomized uniformly.

\item For each point value of $\theta_0$, a sample of toy MC values $\{\widetilde{k}\}$ is generated by drawing from the corresponding sampling model density, for example $\widetilde{k} \sim$ Binom($\theta_0,n$). Drawing from the underlying density is typically strictly necessary (only) if no analytic or parametric pdf is available, or when their evaluation is difficult.

\item Ordering principle (acceptance region) in the sample space: A. Generated MC points can be used without any intermediate test to construct the exact empirical CDF quantiles, similar to Clopper-Pearson central intervals. B. Targeting optimal properties induced by Neyman-Pearson lemma, the (profile)-likelihood ratio such as Eq.~\ref{eq:binom_LLR_test} can be calculated as an intermediate step to provide an ordering, using $\hat{\theta} \leftarrow \widetilde{k}/n$ for different $\widetilde{k}$. The exact empirical CDF quantile points are then taken using the generated test statistics. Return to Step~2.

\item Finally, the parameter uncertainty region is obtained by intersecting the Neyman belt in `orthogonal direction' at the observed value of $k$, i.e. in the parameter space. This is an essential part of the construction and implements the (test) inversion, by taking a union of acceptance sets. Special care must be taken at this point while looping over the acceptance sets, because their union may yield sometimes discontinuous topologies depending on the specific ordering or acceptance principle of Step~3.
\end{enumerate}

\begin{figure}[t!]
\centering
\includegraphics[width=0.5\textwidth]{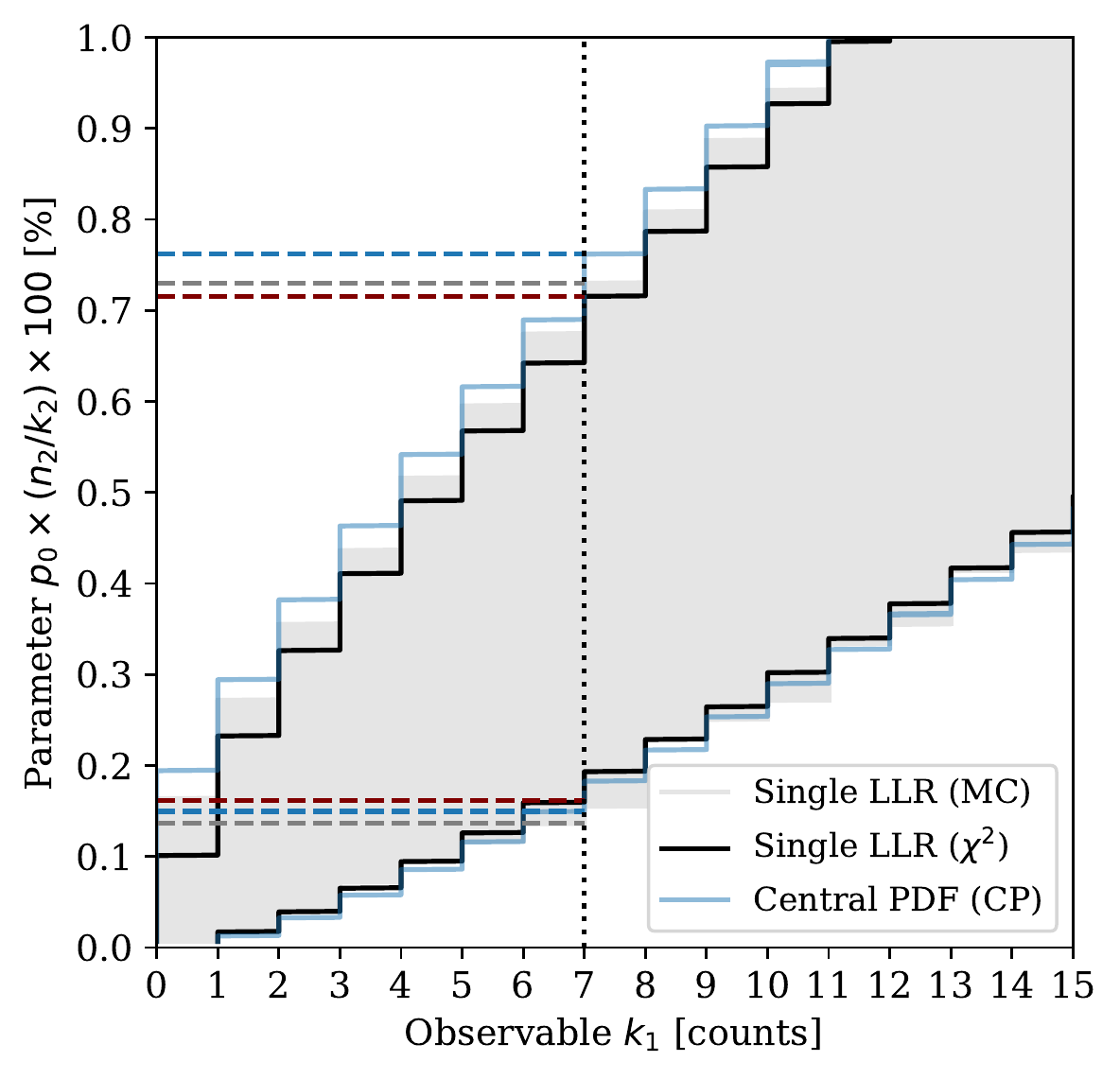}
\caption{(Gangelt setup) CI95 confidence belts using a Monte Carlo based (exact) likelihood ratio test, an asymptotic $\chi^2$-based likelihood ratio test and a central binomial pdf based construction (Clopper-Pearson). Each horizontal gray slice is computed with MC, the rest can be obtained without random numbers.}
\label{fig:neyman_belt}
\end{figure}

For more information on this, see Appendix~\ref{appendix:sec:coverage}. We illustrate this algorithm in Figure~\ref{fig:neyman_belt} for a single binomial uncertainty using the exact (MC based) log-likelihood ratio test statistic \cite{kendall1961advanced, spjotvoll1972unbiasedness, feldman1998unified} as described above, where we see that the asymptotic $\chi^2$-approximation is quite good in this case, but the relative discrepancy grows reasonably large with small numbers. This MC based construction was introduced to high energy physics by Feldman and Cousins \cite{feldman1998unified}. Clopper-Pearson procedure is also compared, which gives slightly different intervals than the exact LLR based. The basic property of the asymptotic LLR approximation is that the acceptance region threshold given by the $\chi^2$-distribution quantile, is independent of the local $\theta_0$ value, unlike the exact Monte Carlo driven LLR test statistic and the CP construction. These both are exact procedures in a sense that their coverage is always larger than or equal to $1-\alpha$. The lattice structure shows why $<$ and $\leq$ operations make a difference with discrete numbers but not with continuous parameters. When constructing Figure~\ref{fig:neyman_belt} for LLR based variants, in Step 3., we used $t \leq t_C$, where $t$ is the log-likelihood ratio test statistic and $t_C$ the cut value ($\chi^2$-quantile or Monte Carlo based). Similar care is required with the `vertical direction' in Step~4.

The approach described here is in principle generic, however, the construction in higher parameter dimensions can be technically challenging. This depends on the construction of the likelihood functions (parametric vs non-parametric) and computational complexity of the Monte Carlo procedures. With several nuisance parameters, the profile likelihood approximation is typically used in Step 3. to keep the whole approach feasible. For $K$ nuisance parameters, the profiled likelihood ratio complexity may scale (naively speaking) linearly $\mathcal{O}(K)$ using e.g. simultaneous stochastic gradient search at each likelihood evaluation point, whereas a full hyperbelt scan is exponentially hard $\mathcal{O}(N^K)$, where $N$ is the number of discretization points in each dimension. The multiple minima need to be in principle handled by the profiling procedure (an example in Section~\ref{sec:profile_likelihood}). Analogous computational challenges arise in Bayesian solutions with high dimensional integration, typically implemented with various Markov Chain MC and variational approximations.

\subsection{Two binomial sample ratio confidence intervals}

We now consider a ratio $r = p_1/p_2$ between two binomial proportions $p_1 \simeq k_1/n_1$ and $p_2 \simeq k_2/n_2$, where $k_i$ denotes the number of success and $n_i$ the total number of trials. This setup models the uncertainty in the IFR estimate given by Eq.~\ref{eq:main_formula}, the double ratio between the fatality rate and the infection rate. The full combinatorial setup of our problem is enumerated later in Table~\ref{table:MC_Bernoulli}, which is beyond the two independent binomial approximation. In this section we concentrate on the ratio between two binomial proportions -- a problem which can be described using a $2 \times 2$ table with 4 elements.

A remark for completeness; studies of $2 \times 2$ contingency tables yielding hypergeometric distributions under the table marginal constraint conditionals -- or sampling without replacement schemes -- have a long history since the Fisher's and Barnard's exact tests. The generalized contingency table analysis can be handled with various algorithms such as networks based~\cite{mehta1983network} or by algebraic statistics~\cite{diaconis1998algebraic}, but these are not used in our case.

\paragraph{Exact ratio interval} ~ Exact interval estimation of the two binomial sample ratio parameter is seemingly not theoretically fully possible within the frequentist framework~\cite{newcombe2011defence}, or the possibility to calculate generalized hypergeometric probabilities is not possible. However, it is possible within the Bayesian framework as shown in~\cite{nurminen1987exact}, which we derive in Section~\ref{sec:bayesian}.

\paragraph{Conditional ratio} ~ Nelson~\cite{nelson1970confidence} considers confidence intervals for the ratio of two unknown Poisson mean occurrence rates. He proceeds by using binomials and constructs the conditional distribution of $k_1$ given $k_1+k_2=N$, which is $\text{Bin}\left(N=k_1 + k_2, p = p_1 n_1 / (p_1 n_1 + p_2 n_2)\right)$. The maximum likelihood estimator for the ratio is $\hat{r} = (k_1/n_1) / (k_2/n_2)$, which is biased, but it can be shown that no unbiased estimator exists. To obtain the parameter $p$ confidence interval, the endpoints $L$ and $U$ are computed by inverting from Eqs.~\ref{eq:clopper_pearson_lower} and \ref{eq:clopper_pearson_upper} using Beta$(k_1,k_2+1)$ and Beta($k_1+1,k_2)$, respectively. But in principle, other interval estimators than the Clopper-Pearson can be also used. The confidence interval for the ratio $r=p_1/p_2$ is then written as
\begin{equation}
CI_N = [(n_2/n_1) L / (1 - L), (n_2/n_1) U / (1 - U) ].
\end{equation}

\paragraph{Katz et al. log} ~ This approximation~\cite{katz1978obtaining} is based on using the Wald test construction, a log-transform of the observed ratio and analytic error propagation by the standard delta method~\cite{doob1935limiting}, which combines the central limit theorem and the first-order Taylor expansion $g(\theta) + (\hat{\theta} - \theta) g'(\theta)$. In essence, the delta method is used for estimating the uncertainty on some non-linear function $g(\theta)$ of the parameter $\theta$. Based on these tools, the standard error estimate of the logarithmic ratio is
\begin{equation}
\hat{\text{se}}[\ln(\hat{r})] = \left[\frac{1}{k_1} - \frac{1}{n_1} + \frac{1}{k_2} - \frac{1}{n_2} \right]^{1/2}
\end{equation}
and the confidence interval for the ratio is
\begin{equation}
CI_K = \exp \left( \ln(\hat{r}) \pm z \hat{\text{se}}[\ln(\hat{r})] \right).
\end{equation}
This Gaussianization of the ratio in the log-space cannot be guaranteed to yield uniformly high precision results especially for small $n$, but it results in a very simple formula. Also, it is possible to combine this approach for example with a sinh$^{-1}$ transform to optimize the interval lengths as suggested in~\cite{newcombe2001logit}.

\paragraph{Bootstrap} A non-parametric Efron's bootstrap~\cite{efron1994introduction} proceeds via simulations by resampling with replacement the obtained sample and calculates the observable of interest for each bootstrap sample. Here, if we pick random numbers parametrically from two binomial distributions with parameters set to their maximum likelihood values, the results will be identical to those from non-parametric bootstrap. In general, this is not the case with more complicated distributions and sampling scenarios.

The most common first order method with a coverage correct up to terms proportional to $\mathcal{O}(n^{-1/2})$, is to obtain confidence interval estimates based on taking the quantiles (percentiles) of the generated bootstrap sample $\{\theta^*\}$, known as the percentile bootstrap (PRC). This assumes the the bootstrap distribution is a good proxy for the underlying true distribution. The confidence interval estimate is
\begin{equation}
CI_{PRC} = [\theta^*_{\alpha}, \theta^*_{1-\alpha}],
\end{equation}
obtained by ordering $B$ bootstrap sample estimates $\theta_1^* \leq \theta_2^* \leq \dots \leq \theta_B^*$. A different variant, usually known by the name `basic bootstrap', is to assume that bootstrap generates a good proxy of the error $e^* = \theta^* - \hat{\theta}$, and then obtain the confidence interval with $[2\hat{\theta} - \theta^*_{1-\alpha}, 2\hat{\theta} - \theta^*_{\alpha}]$. We do not consider the basic bootstrap further here.

Well known extensions of the percentile bootstrap are are the so-called bias corrected (BC) and bias corrected with acceleration (BCA) bootstrap~\cite{efron1987better}. Under certain assumptions and using asymptotic Edgeworth expansion techniques, it was shown by Hall that the second-order BCA bootstrap coverage is correct up to order $\mathcal{O}([n^{-1/2}]^2)$~\cite{hall1988theoretical}. The BCA confidence interval estimate is
\begin{equation}
CI_{BCA} = [\theta^*_{\alpha}, \theta^*_{1-\alpha}],
\end{equation}
where
\begin{align}
\theta_{k}^* &= \hat{G}^{-1}\left( \Phi\left(\hat{z}_0 + \frac{\hat{z}_0  + z_{k}}{1 - \hat{a} (\hat{z}_0 + z_{k}) } \right) \right), \;\;\; k \in \{ \alpha, 1-\alpha \} \\
\hat{z}_0 &= \Phi^{-1} \left( \hat{G}(\hat{\theta}) \right) \\
\hat{a} &= \frac{1}{6}  \sum_{i=1}^n d_i^3 / \left( \sum_{i=1}^n d_i^2 \right)^{3/2},
\end{align}
where $\hat{G}$ is the empirical CDF of the bootstrap sample statistics and $\Phi$ the standard normal CDF. The bias correction is $\hat{z}_0$ and the acceleration term is $\hat{a}$, which can be negative, is to account for non-uniform variance. To construct the polynomial acceleration estimate, the jackknife residuals $d_i$ are needed
\begin{align}
d_i &= \hat{\theta}_{(i)} - \hat{\theta}_{(\cdot)}, \; \text{with} \;\; \hat{\theta}_{(\cdot)} = \frac{1}{n}\sum_{i=1}^n \hat{\theta}_{(i)},
\end{align}
where $\hat{\theta}_{(i)}$ is one of the jackknife estimates obtained by dropping the $i$-th data point, and proceeding with this $n-1$ sized sample as with the original data sample. The whole construction is motivated by doing a monotone normalizing transform $m: \theta \mapsto \varphi$ with a statistics
\begin{equation}
\hat{\varphi} \sim N(\varphi-z_0 \sigma_\varphi , \sigma_\varphi^2), \; \text{with} \; \; \sigma_\varphi = 1 + a\varphi.
\end{equation}
The interval construction is done in the transformed space, and finally the endpoints are inverse mapped with $m^{-1}$. The case $\hat{z}_0 \equiv \hat{a} \equiv 0$ reduces identically to the percentile bootstrap and the case $\hat{z}_0 \neq 0, \hat{a} \equiv 0$ is the case of bias correction without acceleration.

\subsection{Profile likelihood ratio}
\label{sec:profile_likelihood}

The profile likelihood method splits the parameters into two groups: true parameters of interest $\theta$ and \textit{nuisance} parameters $\xi$, and maximizes the full likelihood over the nuisance parameters
\begin{equation}
L_{pr}(\theta) = \sup_{\xi} L(\theta,\xi),
\end{equation}
where $\sup$ denotes the supremum, the least upper bound, which is almost the same as the maximum but takes into account the possibility that the likelihood cannot be evaluated exactly at that point $\xi$. The main idea behind profiling is the dimensional reduction over the nuisance parameters, which then allows one to infer the uncertainty on $\theta$ by formally proceeding as with a usual likelihood, for example by using the score test or the likelihood ratio test which are asymptotically equivalent. Solutions based on the score test for the two binomial case have been proposed in~\cite{chan1999test, agresti2001small}. We shall now derive the likelihood ratio test based solution.

Let us parametrize $r \equiv p_1 / p_2$ and write down the joint likelihood function for two independent binomials
\begin{equation}
L(r,p_1) = \begin{pmatrix} n_1 \\ k_1 \end{pmatrix} p_1^{k_1} (1-p_1)^{n_1 - k_1} \times \begin{pmatrix} n_2 \\ k_2 \end{pmatrix} \left( \frac{p_1}{r} \right)^{k_2} \left( 1-\frac{p_1}{r} \right)^{n_2-k_2}.
\end{equation}
This re-parametrization does not involve the change of variables formula (Jacobian), because the likelihood as a sampling function and its volume normalization is over $k_1$ and $k_2$, which we left intact. We are interested in the parameter $r$ and treat the parameter $p_1$ as a nuisance parameter, which we profile out by finding a value for $p_1$ which maximizes the joint likelihood for every single value of $r$. This procedure is in principle readily generalized to arbitrary number of nuisance and true parameters of interest, which however is only a formal statement. Possible singularities depend on the exact type of nuisance parameters and models. A practical problem in higher dimensional cases is also the parameter optimization problem itself.

\begin{figure}[t!]
\centering
\includegraphics[width=0.6\textwidth]{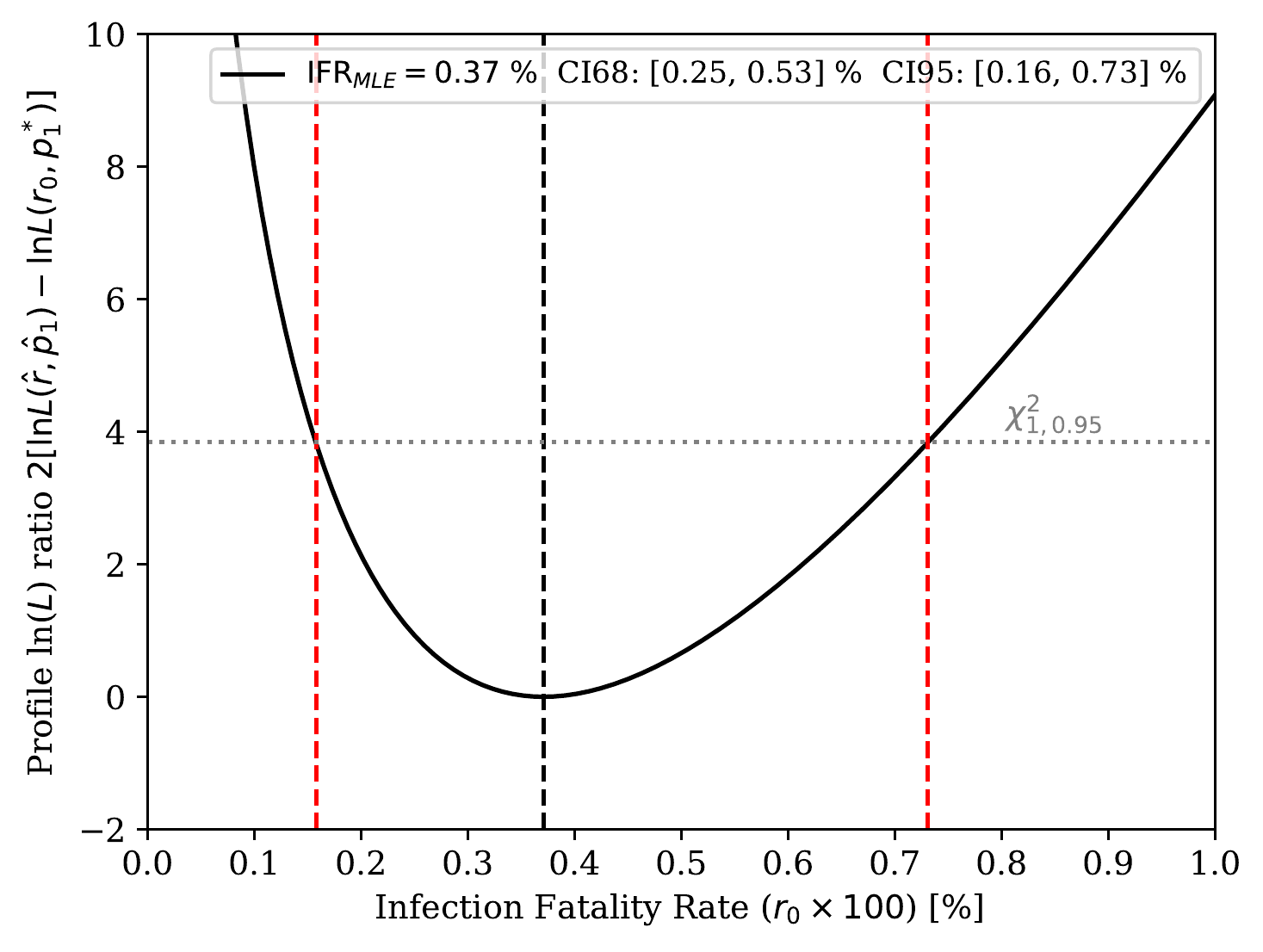}
\caption{(Gangelt setup) On the left, a profile likelihood ratio based IFR confidence interval.}
\label{fig:profile_IFR}
\end{figure}

The profile log-likelihood ratio test statistic follows here asymptotically
\begin{equation}
\label{eq:profile_likelihood}
LLR(r_0) = -2 \left[ \sup_{p_1} \ln L(r_0, p_1) - \sup_{r,p_1} \ln L(r,p_1)\right] \rightarrow \chi_1^2.
\end{equation}
The fact that $\chi^2$-asymptotics holds also for the profile likelihood inference is a non-trivial result, but propagates from Wilks' theorem under certain assumptions. After maximizing Eq.~\ref{eq:profile_likelihood}, the profile log-likelihood ratio is
\begin{align}
LLR(r_0) = -2 \left[ \ln L(r_0, p_1^*(r_0)) - \ln L(\hat{r},\hat{p}_1) \right],
\end{align}
where the maximum likelihood estimates are
\begin{align}
\hat{r} &= (k_1/n_1) / (k_2/n_2) \\
\hat{p}_1 &= (k_1/n_1)
\end{align}
and the local profile extremum solution $p_1^*$ \textit{conditioned} at point $r_0$ has two roots
\begin{equation}
p_1^*(r_0) = \frac{k_1 + n_2 + k_2r_0 + n_1 r_0 \pm [(-k_1 -n_2 -k_2r_0 -n_1r_0)^2 -4(n_1+n_2)(k_1 r_0 + k_2 r_0) ]^{1/2}}{2(n_1+n_2)},
\end{equation}
where the negative $(-)$ branch gives the right solution in our problem. It is not guaranteed that every profile likelihood problem is differentiable and has a closed form solution, but this turned out to be the case here. The profile log-likelihood ratio is illustrated in Figure~\ref{fig:profile_IFR}, which has asymmetric 95\% confidence interval endpoints around the maximum likelihood value.

\subsection{3-dimensional Monte Carlo simulation}

\paragraph{Notation}~Q68, Q95 are used for pure density quantiles and CI68, CI95 for a parameter estimate confidence (frequentist) or credible (Bayesian) intervals.
\\

\noindent This elementary simulation approach starts with three Bernoulli random numbers: tests $\sim B_T$, infections $\sim B_I$ and deaths $\sim B_F$, which are together per person modelled as a 3-dimensional Bernoulli distribution. To remind, the Bernoulli distribution is the underlying distribution behind the binomial distribution, which turns into a Poisson distribution when $p$ is small and $n$ is large. Thus, this approach is ab initio in this hierarchy of distributions. Now in general, a $D$-dimensional Bernoulli requires $2^D-1$ free parameters. However, we do not have enough measurements here to constrain all the parameters. To simplify this problem, we \textit{factorize} $B_T$ to be independent of $B_I$ and $B_F$
\begin{equation}
B(T,I,F) \rightarrow B(T) \otimes B(I,F).
\end{equation}

That is, tests do not (hopefully) affect infections or fatalities. This leaves us with one and two dimensional sub-problems which require together 1 + 3 parameters. The two-dimensional problem can be parametrized directly with four probabilities of $(B_I,B_F)$-binary combinations, which sum to one. Another parametrization uses the expectation values $\mathbb{E}[B_I]$, $\mathbb{E}[B_F]$ and the correlation coefficient $\lambda[B_I,B_F] \in [-1,1]$. We use both in order to be able to sample correlated Bernoulli variables in the direct (multinomial) basis, which satisfy by definition the conservation of probability, and to give an interpretation of the problem in the correlation basis. The formulas are given in Appendix~\ref{appendix:sec:IFR_definitions} and \ref{appendix:sec:bernoulli_representations}.

The four `dynamic' parameters of the simulation are fixed in the correlation basis according to their maximum likelihood values
\begin{align}
\langle B_T \rangle &\leftarrow \ntest / \npop \approx 0.07295 \\
\label{eq.MC_I_extrapolation}
\langle B_I \rangle &\leftarrow \ninf / \ntest \approx 0.15016 \\
\langle B_F \rangle &\leftarrow \nfatal / \npop \approx 0.00055 \\
\label{eq.MC_rho_max}
\rho(B_I, B_F) &\leftarrow \text{`maximum coupling'} \approx 0.0559,
\end{align}
where the count variables and their associations to the underlying sets of tested $\mathbf{T}$, all infected $\mathbf{I}$ in the city and all fatal $\mathbf{F}$ are as described in Section~\ref{sec:Heinsberg}. Here, one must pay attention to Eq.~\ref{eq.MC_I_extrapolation}, where the study sample infection rate is assumed to be representative in the simulation for the whole population by assuming homogeneity between samples. A systematic uncertainty could be associated here. By choosing in Eq.~\ref{eq.MC_rho_max} the maximum possible positive correlation coupling (see Appendix~\ref{appendix:sec:bernoulli_representations}), the simulation output in Table~\ref{table:MC_Bernoulli} reproduces the event count observables, which enter as the input variables. That is, its value is fixed by data. Also, a boundary condition is used
\begin{equation}
\mathbb{P}(B_I = 0 \wedge B_F = 1) \equiv 0
\end{equation}
which states that no fatalities happen without getting infected. This forbids the combinations 1 and 5 in Table~\ref{table:MC_Bernoulli} from appearing. The total number of people $n_P=|\mathbf{P}|$ is kept fixed for each MC run. We have also fixed the test sample size $n_T=|\mathbf{T}|$ to be constant in these simulations, to follow more closely the Gangelt setup, but turning on the Bernoulli fluctuations is implemented in the code as an option. However, with the given event counts the difference is not significant for the IFR. Type I and II errors of tests are not simulated here. Once calibrated, including their effect as a post processing step is trivial with two free parameters and an additional coin flipping per tested person. For more details, see Appendix~\ref{appendix:sec:test_inversion}.
\begin{figure}[t!]
\centering
\includegraphics[width=0.55\textwidth]{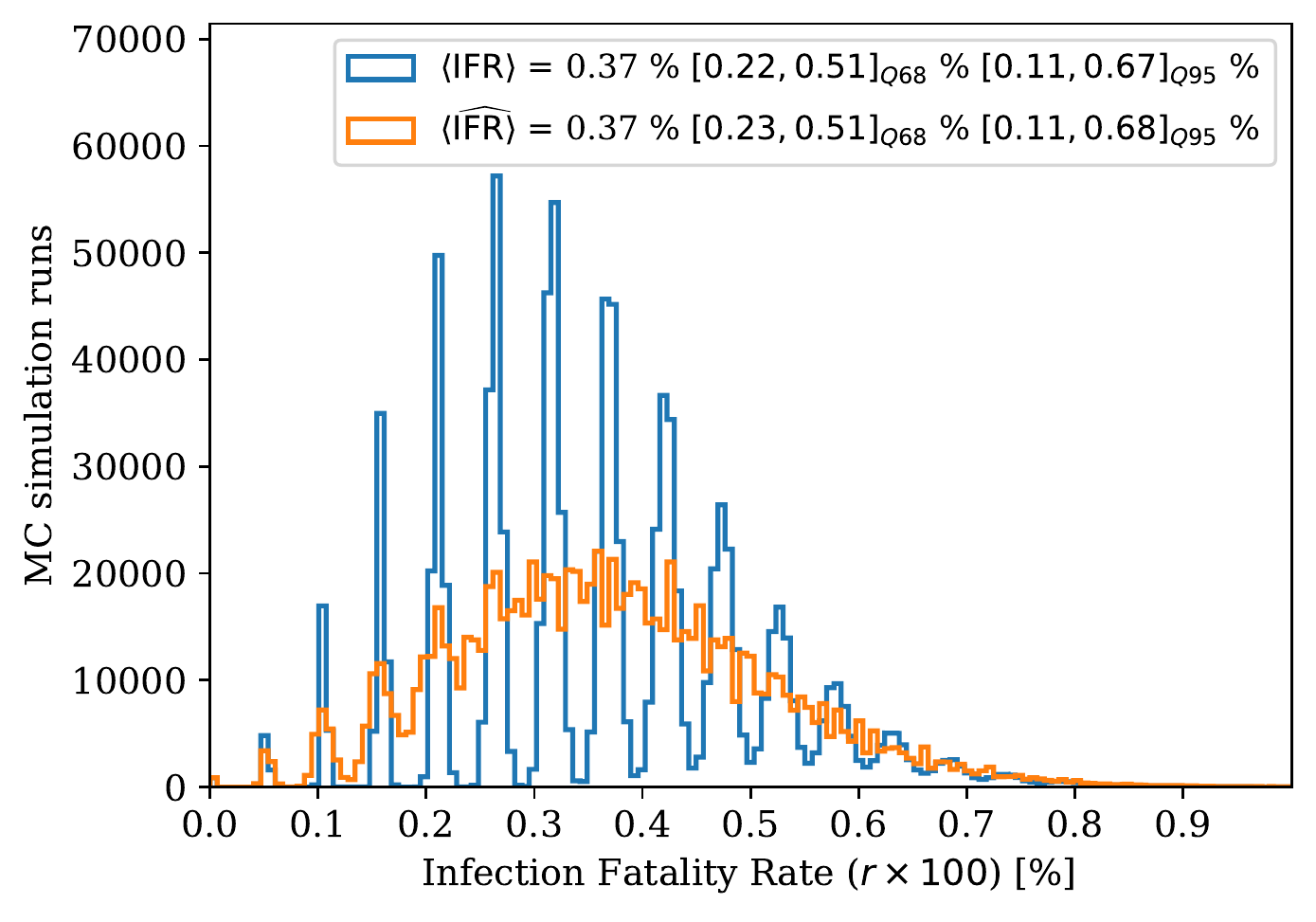}
\includegraphics[width=0.41\textwidth]{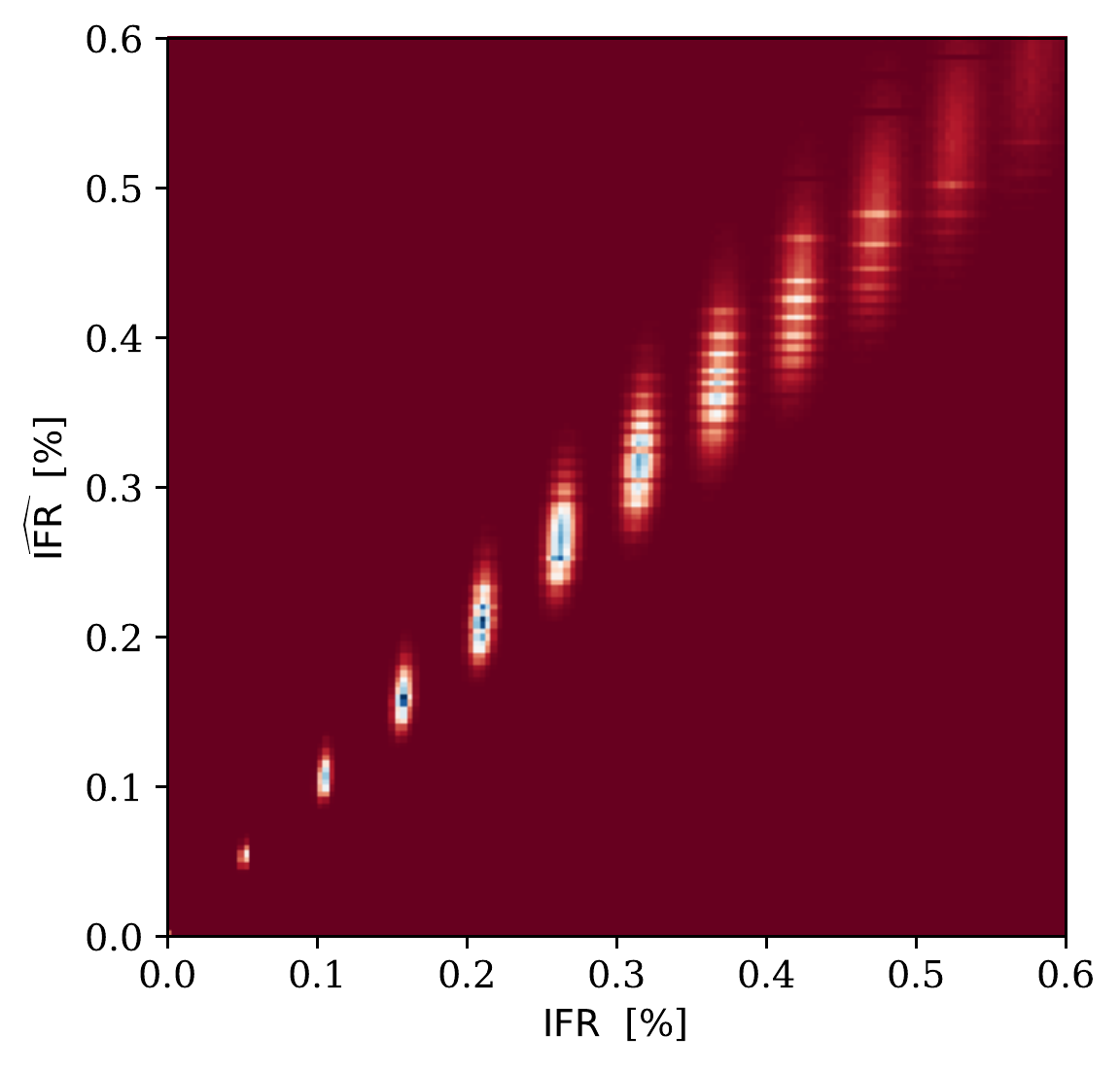}
\caption{(Gangelt setup) The IFR distributions obtained from Bernoulli Monte Carlo simulations on the left. The blue distribution simulates the full population with a complete test coverage and the orange distribution is obtained using the finite test sample, and its extrapolation to the full population using Eq.~\ref{eq:main_formula}. The same folding effect is visualized on the right in 2D, where the full population realized IFR value is on the horizontal axis and the test sample driven extrapolated estimate $\widehat{\text{IFR}}$ on the vertical axis.}
\label{fig:MC_Bernoulli_IFR}
\end{figure}
\begin{table}[htb!]
\center
\begin{tabular}{|cc|ccc|}
\hline
ID  & $TIF$ & $\langle \text{Counts} \rangle$ & Q68 & Q95 \\
\hline
0 & 000 & 9923.5  & [9878.0,  9969.0] & [9833.0, 10013.0] \\
1 & 001 & -     & - & - \\
2 & 010 & 1748.1  & [1710.0,  1787.0] & [1673.0,  1825.0] \\
3 & 011 & 6.5     & [4.0,     9.0]    & [2.0,    12.0] \\
4 & 100 & 780.8   & [754.0,   808.0]  & [728.0,   834.0] \\
5 & 101 & -     & - & - \\
6 & 110 & 137.6   & [126.0,   149.0] & [115.0,   161.0] \\
7 & 111 & 0.5     & [0.0,     1.0]   & [0.0,     2.0] \\
\hline
  &  $\Sigma$   & 12597   &                    & \\
\hline
\end{tabular}
\caption{(Gangelt setup) 3-dimensional Monte Carlo simulation summary results of event counts for eight different mutually exclusive $(B_T,B_I,B_F)$-categories. Combinations [0-3] do not belong to the test sample, whereas combinations [4-7] do belong, by definition.}
\label{table:MC_Bernoulli}
\end{table}

Using the generated Monte Carlo samples, arbitrary observables such as the IFR are computed by simply counting numbers from an $(8 \times N_{MC})$-dim matrix, here $N_{MC}=10^6$, and accumulating numerically the relevant point estimates such as mean values and percentiles. Note that this aggregated matrix is the fully \textit{sufficient statistic} and contains all simulation information, due to binary random variables. These results are given Table~\ref{table:MC_Bernoulli}. The simulated distributions for the infection fatality rates are shown in Fig.~\ref{fig:MC_Bernoulli_IFR}, which illustrates the non-trivial Dirac's comb discrete characteristics of the problem, but also the finite sample smearing effect on the extrapolation estimate. The smeared IFR-distribution is calculated from the test samples and the reference IFR-distribution from the inaccessible (full) population statistics, which are both obtained simultaneously in the simulation. See Appendix~\ref{appendix:sec:IFR_definitions} for the exact definitions.

To this end, we may summarize that the power of this simulation is the `full phase space' modelling of partially overlapping sets of tested, infected and fatal, which is not possible with independent binomial ratios. The simulation is based on describing the most elementary classical stochastic process involved, namely correlated Bernoulli coins. The optimal confidence interval estimator can be based on the simulations as described in Sec.~\ref{sec:toymc} where each simulation with fixed input parameters simply generates a sample for a single null hypothesis $H_0$. Alternatively, faster bootstrap approximations can be used.

\subsection{Bayesian inference}
\label{sec:bayesian}

In the exact Bayesian inference within two independent binomial distributions, we keep the number of observed events $k_1,k_2$ as fixed numbers, also $n_1,n_2$, and calculate the joint posteriori density for the binomial parameters $p_1$ and $p_2$. Their joint posteriori density is described with a product of two beta distributions
\begin{equation}
\label{eq:bayesian_joint_2D}
P(p_1,p_2|\{k,n,\alpha,\beta\}_{i=1,2}) = \prod_{i=1,2} \text{Beta}(p_i|k_i+\alpha_i,n_i-k_i+\beta_i),
\end{equation}
given generic beta priors Beta$(\alpha_i,\beta_i)$ and binomial likelihoods. The derivation of this and priors are given in Appendix~\ref{appendix:sec:bayesian}. Given this joint posterior, the ratio $r \equiv p_1/p_2$ density is obtained via change of variables such that $p_1=ry$ and $p_2=y$. Writing down the Jacobian determinant gives $|\partial (p_1,p_2)/\partial (r,y)| = |y|$. Then we substitute these new variables in Eq.~\ref{eq:bayesian_joint_2D}, include the determinant and integrate over $y$
\begin{equation}
\label{eq:bayesian_posterior_ratio}
P(r| \{k,n,\alpha,\beta\}_{i=1,2}) = \int_0^1 dy \, |y| \, P(r y, y|\{k,n,\alpha,\beta\}_{i=1,2}).
\end{equation}
\begin{figure}[t!]
\centering
\includegraphics[width=0.485\textwidth]{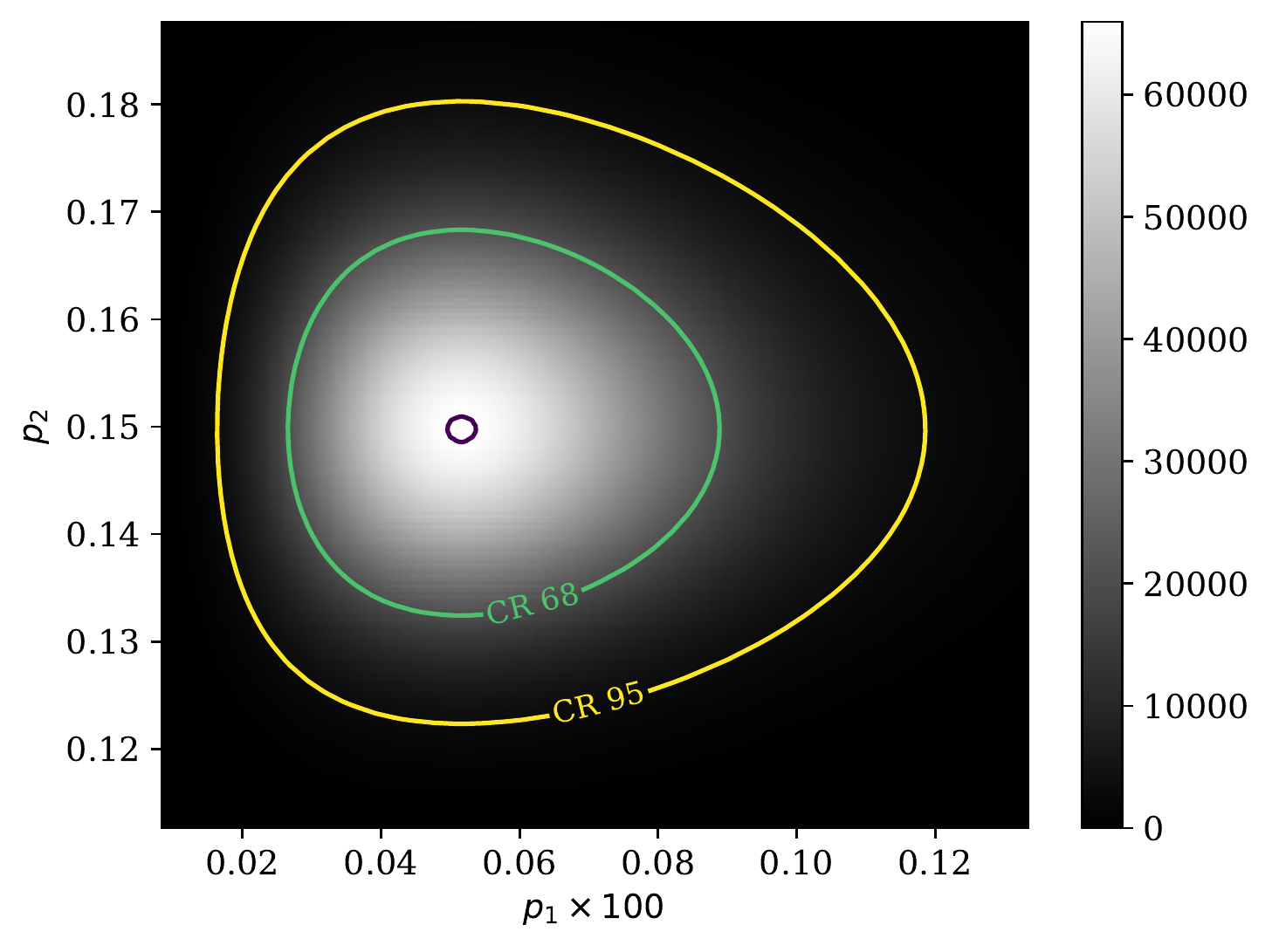}
\includegraphics[width=0.50\textwidth]{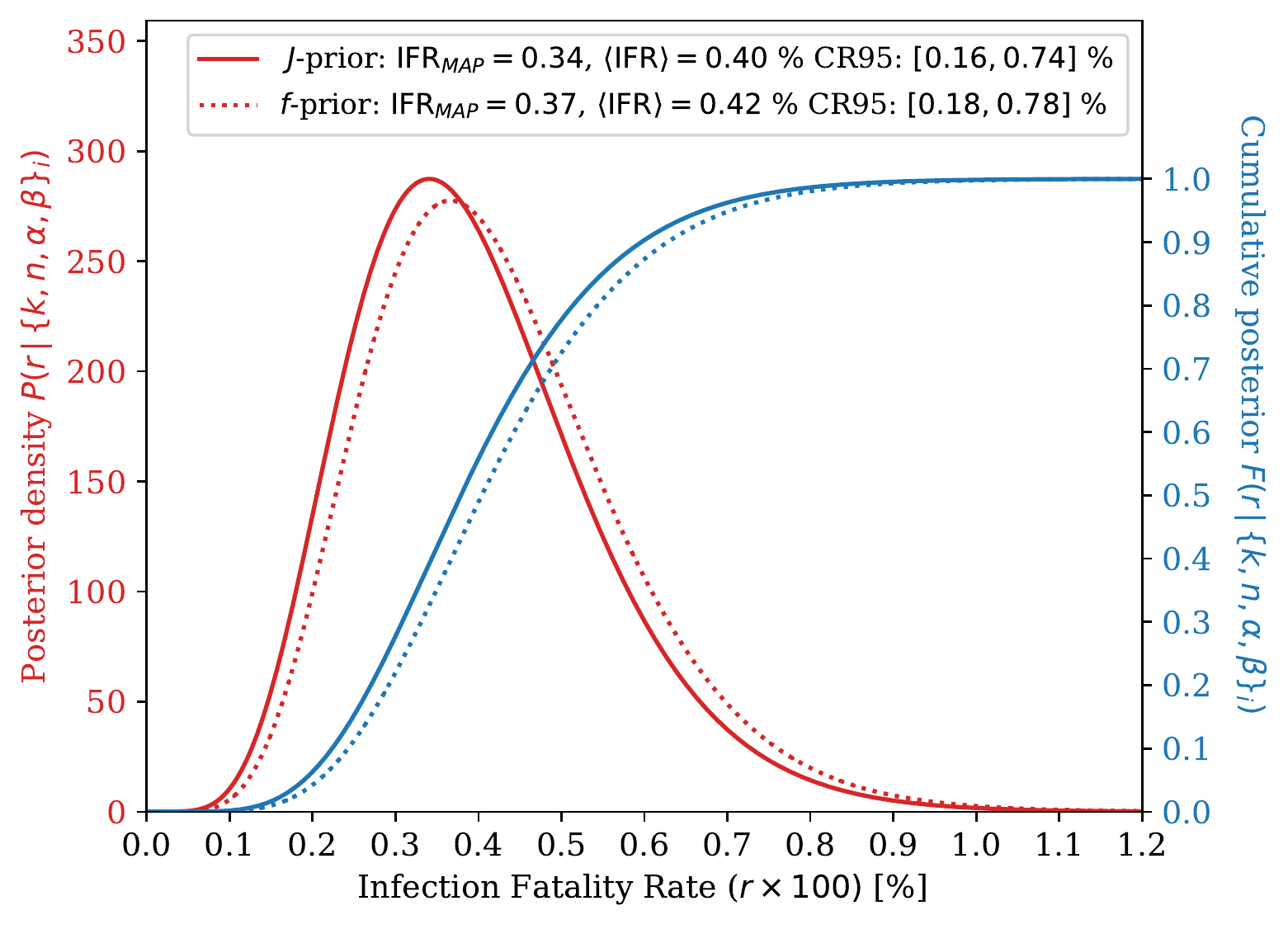}
\caption{(Gangelt setup) On the left, the Bayesian joint posteriori distribution using Jeffreys prior. On the right, the ratio posteriori density in red and the cumulative distribution in blue for Jeffreys and flat priors.}
\label{fig:bayesian}
\end{figure}
Using Mathematica, we obtain for this integral a representation
\begin{align}
\label{eq:posterior_ratio_density}
\nonumber
P(r| \{k,n,\alpha,\beta\}_{i=1,2}) = &r^{\alpha_1+k_1-1} \frac{\Gamma(\alpha_1+\alpha_2+k_1+k_2) \Gamma(\beta_2-k_2+n_2)}{B(\alpha_1+k_1,\beta_1-k_1+n_1) B(\alpha_2+k_2,\beta_2-k_2+n_2)} \times \\
&_2\widetilde{F}_1(\alpha_1+\alpha_2+k_1+k_2,1-\beta_1+k_1-n_1,\alpha_1+\alpha_2+\beta_2+k_1+n_2,r),
\end{align}
where $_2\widetilde{F}_1$ is the regularized Gauss hypergeometric $_2F_1$ function and $B$ is the Euler beta function. The regularized version is $_2\widetilde{F}(a,b,c,r) \equiv \, _2F_1(a,b,c,r) / \Gamma(c)$. Equation~\ref{eq:posterior_ratio_density} represents the master formula, which can be evaluated numerically with high precision special function libraries and credible intervals can be obtained with standard numerical integration techniques. However, we found that numerically it is easier to use directly Eq.~ \ref{eq:bayesian_posterior_ratio}.

The results are given Figure~\ref{fig:bayesian}, where the joint posteriori distribution includes the credible regions (CR), which encapsulate 68 and 95 percent of the probability mass. The shape is constructed according to the natural contour lines. The right figure shows the ratio posteriori density and the corresponding cumulative distribution by using two different prior distributions. The solid line is obtained using the non-informative Jeffreys prior $\text{Beta}(1/2,1/2)$, which is invariant under coordinate transformations. It is proportional to the square root of the Fisher's information determinant $p(\theta) \propto \sqrt{\det I(\theta)}$, where the determinant represents abstract information volume (here in one dimension the determinant is trivial). The results with dashed lines are obtained using a unit flat prior, which is not completely non-informative and results in slightly larger values. Its maximum (mode) gives numerically the same estimate for the IFR as the simple ML estimate.

\paragraph{Nuisance parameters and systematic uncertainties}~Bayesian framework allows one to add nuisance parameters and systematic uncertainties into the formulation. For example: the death counts $k_1$ may be need to be scaled with a parameter $\gamma$ due to time delays. Note that scaling the parameter $p_1$ instead is ambiguous, which is seen using the binomial pdf and by computing the Fisher information matrix, which will turn out to be singular. Which means that the corresponding parameter estimation problem would be rank deficient. Similarly the positive test counts $k_2$ may be multiplied with another scale $\lambda$. Using auxiliary measurements or prior judgement, the uncertainty information on the nuisance parameter is often modelled using a Gaussian prior $\pi_{\gamma}(\gamma ; \mu_{\gamma}, \sigma_{\gamma})$ with fixed $\mu_{\gamma},\sigma_{\gamma}$. However, with a positive definite scale, using the gamma prior could be more suitable, although practical difference can be small. This is applied on the Bayesian inference master formula as an additional prior constraint term and by replacing $k_1 \rightarrow \gamma k_1$ everywhere. Computationally, each nuisance parameter requires typically an additional integral when marginalizing the posterior and in the normalization (Bayes denominator). For more details, see Appendix~\ref{sec:systematic_bayesian_priors}.

\section{Estimator comparisons}
\label{sec:comparisons}

\begin{figure}[t!]
    \centering
    \includegraphics[width=0.63\textwidth]{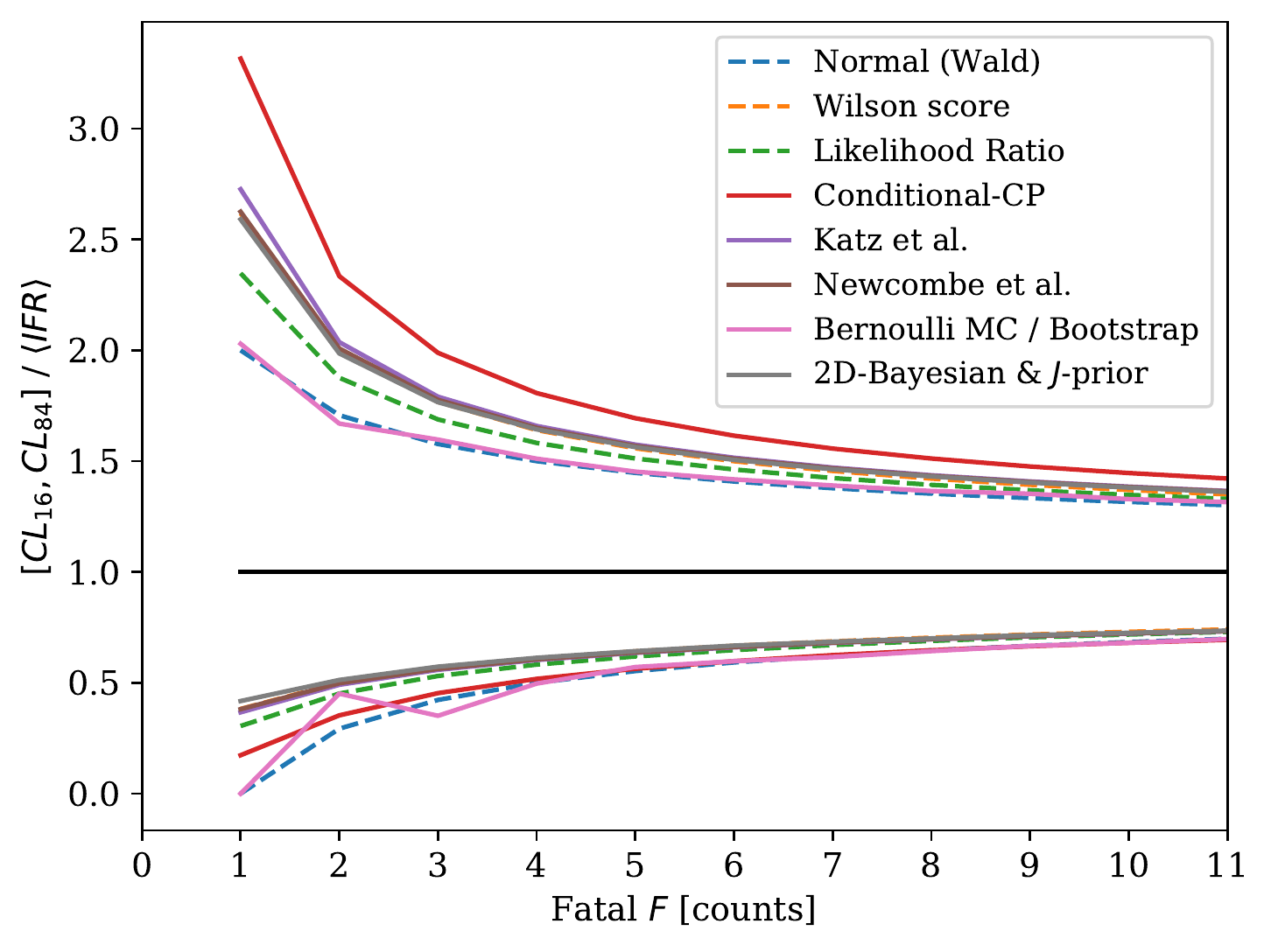}
    \includegraphics[width=0.63\textwidth]{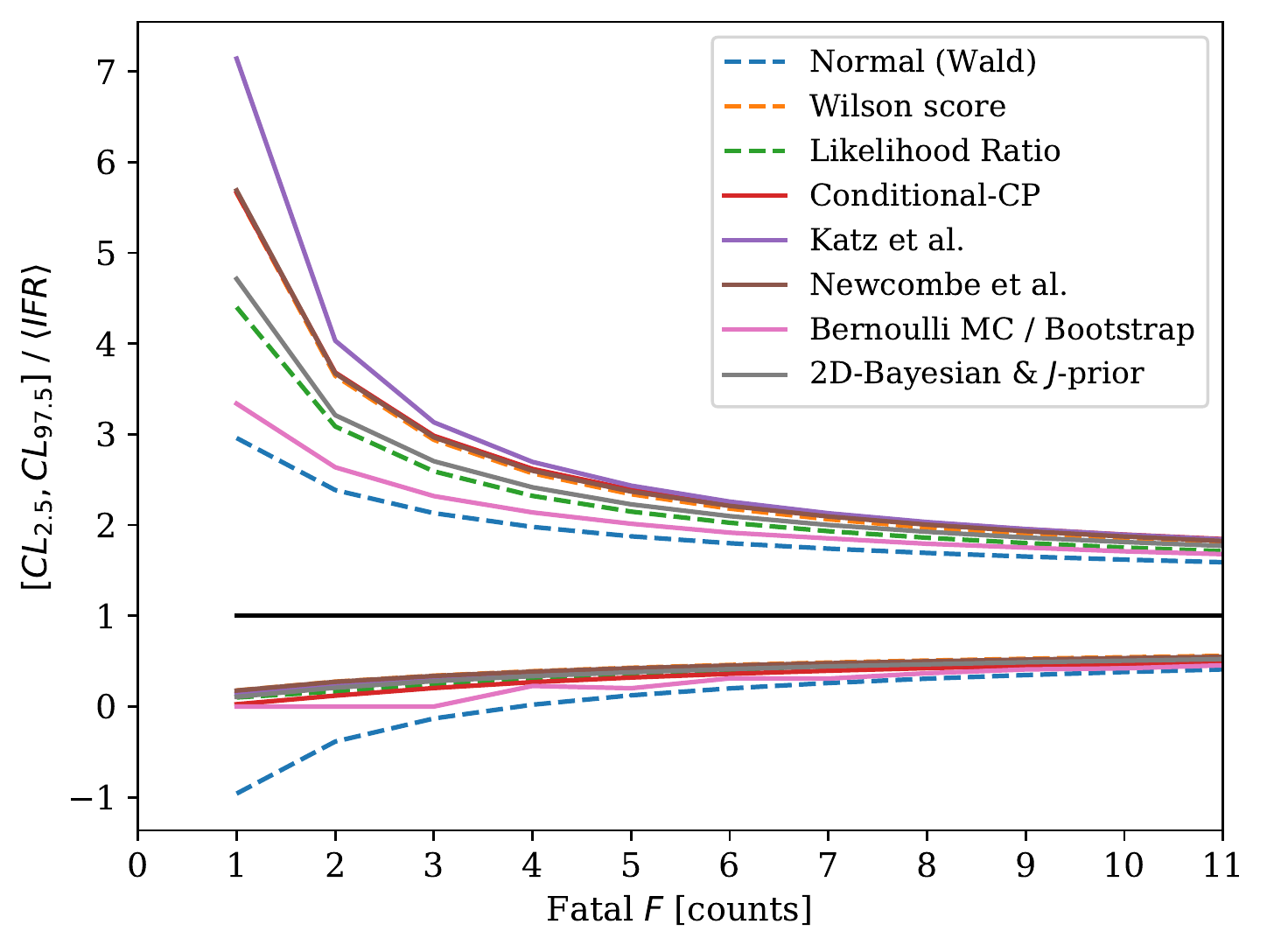}
    \caption{(Gangelt setup) Relative interval widths compared for a subset of the estimators with running death counts, other numbers kept fixed. At the top, CI68 endpoints and at the bottom, CI95 endpoints. The likelihood ratio is based on $\chi^2$-approximation and the bootstrap is constructed using the percentile approximation.}
    \label{fig:running_F}
\end{figure}

Table~\ref{tab:single_and_double_estimators} shows the numerical results for different confidence interval estimators, using count data of the Gangelt study and Figure~\ref{fig:running_F} similarly, but as a function of death counts (rather than for just the observed $F=7$). Clear outliers in the group of these estimators are the normal (Wald) test based and the bootstrap percentile estimator. Their interval is shifted toward smaller IFR values, which is especially visible with CI95 intervals. The Wald test will also give negative (unphysical) values at small $F$. This can be expected from their mathematical construction. The impact of the bias correction and acceleration for the bootstrap is clear. The rest of the estimators yield numerically similar values for the Gangelt input data and small differences are more easily seen from Figure~\ref{fig:running_F}. Compared with the Monte Carlo simulations of Figure~\ref{fig:MC_Bernoulli_IFR}, the Bayesian distributions of Figure~\ref{fig:bayesian} are completely smooth because the observed event counts are considered fixed and the continuous binomial parameters $p_1,p_2$ are considered random. In contrast, the simulations are closer to a frequentist inference, because the model parameters are fixed and the discrete event counts are random. Though technically speaking, the underlying simulation can be used within both philosophies.

\begin{table}[bt!]
\centering
\begin{tabular}{|l|c|c|l|}
\hline
$\widehat{\text{IFR}}$ interval estimator & CI68 [\%] & CI95 [\%]  & Type\\
\hline
Normal (Wald)    & [0.23, 0.51] &	 [0.10, 0.64] & Single binomial \\
Wilson score     & [0.25, 0.54] &	 [0.18, 0.76] & Single binomial \\
Likelihood Ratio ($\chi^2$) & [0.25, 0.53] &    [0.16, 0.72] & Single binomial (asymptotic approx.) \\
Likelihood Ratio (MC) & [0.23, 0.54] &    [0.14, 0.73] & Single binomial (exact Monte Carlo) \\
Clopper-Pearson  & [0.23, 0.57] &    [0.15, 0.76] & Single binomial \\
\hline
\hline
Conditional-mid-$P$ &  [0.25, 0.54]  & [0.16, 0.75] & Conditional ratio with mid-$P$ \\
Conditional-CP & [0.23, 0.58] &	 [0.15, 0.78] & Conditional ratio with Clopper-Pearson \\
Katz log       & [0.25, 0.54] &	 [0.17, 0.79] & Transform ratio \\
Newcombe sinh$^{-1}$ & [0.25, 0.54] & [0.18, 0.78]    & Transform ratio \\
Profile LR ($\chi^2$)       & [0.25, 0.53] & [0.16, 0.73]    & Profiled log-likelihood ratio \\
Bootstrap (prc)     & [0.23, 0.51] & [0.11, 0.68] & MC ratio percentiles \\
Bootstrap (bc)  & [0.25, 0.53] & [0.14, 0.71] & MC ratio perc. \& bias corrected \\
Bootstrap (bca) & [0.25, 0.55] & [0.16, 0.76] & MC ratio perc. \& bias cor.  \& accelerated \\
2D-Bayesian \& $J$-prior & [0.25, 0.54] & [0.16, 0.74] & Full posterior ratio \\
\hline
\end{tabular}
\caption{(Gangelt setup) Infection Fatality Rate (IFR) confidence interval estimation results. Methods above the break line treat uncertainty only in the numerator of the double ratio. No systematic uncertainties included in these estimates.}
\label{tab:single_and_double_estimators}
\end{table}

Coverage probability simulations and interval widths are given in Appendix~\ref{appendix:sec:coverage} for the single binomial based estimators. These illustrate significant undercoverage of the Wald test based estimator, the conservative coverage of the Clopper-Pearson based estimator and the `bracketing' coverage behavior of the Wilson score and Bayesian estimators. The likelihood ratio with an asymptotic $\chi^2$ approximation has behavior similar to the Wilson score, but significantly undercovers at very small values of $p$, similarly to the the Wald test.

\clearpage

\section{Time evolution}
\label{sec:time_evolution}

The previous discussion in Sections~\ref{sec:Heinsberg} and \ref{sec:methods} is only fully applicable under the asymptotic time $t \rightarrow \infty$ limit or instantaneous action, i.e., no time delays. To be concrete, by time asymptotic we mean the tail of a single epidemic outbreak and neglect additional possible complications (immunology evolution, different viral strains) which result from overlapping epidemic `waves'. However, purely mathematical overlap is automatically handled by our description, that is, we do not assume any specific epidemic shape for the time-series input. In this section, we briefly outline how the IFR estimation and its uncertainty is implemented during an evolving epidemic with finite time delays. For the interested reader, further details of our time evolution study can made available upon request.

A combined \textit{double delay effect} can be summarized with one ratio function
\begin{equation}
\label{eq:delay_scale}
\psi(t,\Delta t) = \frac{(K_F \ast \hat{I})(t+\Delta t)}{(K_S \ast \hat{I})(t)} \equiv \frac{\int_0^{\infty} d\tau \, K_F(t+\Delta t-\tau) \hat{I}(\tau)}{\int_0^{\infty} d\tau \, K_S(t-\tau) \hat{I}(\tau)},
\end{equation}
where $K_F$ is the delay kernel (pdf) from infections to deaths, $K_S$ is the delay kernel from infections to seroprevalence (antibodies) and the symbol $\ast$ denotes a linear time-lag convolution integral. These kernels are extracted from data by fitting them typically with Weibull or log-normal distributions, see e.g. supplementary material of the Geneva study in~\cite{Perez-Saez2020.06.10.20127423} and references there, where a similar convolution calculus was used. The kernels are given in Appendix~\ref{appendix:sec:deconvolution}, which shows explicitly the expected delays. The relative differences between $K_S(t)$ and $K_F(t)$ drive Eq.~\ref{eq:delay_scale}. The calculus here is general, and one may replace the antibody type tests with PCR type tests by replacing the delay kernel $K_S$. In that case an additional multiplicative effect to include is the `viral shedding' (loading) period probability, i.e., how long an infected person gives a positive test result. That evaporation factor can be neglected with antibody based serology (seroreversion effect), if the half-life involved is large enough on the scale of epidemic. For some specific antibodies, this may not be the case. The denominator of Eq.~\ref{eq:delay_scale} can be extended to incorporate it, see Appendix~\ref{sec:seroreversion} for that construction.

Basic numerical integration is used here for the convolutions. Time $t$ denotes the time of the seroprevalence determination and $\Delta t \geq 0$ denotes how many days later the population cumulative death count is taken. Because our problem here is essentially an inverse problem, the underlying cumulative infection count $\hat{I}(t)$ can be estimated computationally by regularized deconvolution of the reported positive cases $dC(t)/dt$ of PCR viral tests. Daily counts need to be used in the inversion instead of cumulative counts, to conserve all information. Although even if one assumes a constant reporting rate, $\hat{I}(t)$ can be estimated only up to an unknown scale (probability), which however fortunately cancels in Eq.~\ref{eq:delay_scale}. In principle, one may also use the reported daily death counts $dF(t)/dt$ to obtain the deconvolution inverse estimate of $\hat{I}(t)$, albeit the statistics might be too limited. For technical details about the deconvolution algorithm, see Appendix~\ref{appendix:sec:deconvolution}. The algorithm is based on non-negative linear least squares with Tikhonov smoothness regularization. Regularization is needed, because basically all naive inversion procedures always amplify the counting fluctuations (noise). No fine structure recovery is needed, thus smoothness is a good functional prior in this problem.

By using Eq.~\ref{eq:delay_scale}, the delay corrected non-equal time IFR estimate is now
\begin{equation}
\label{eq:IFR_estimate_non_equal_time}
\widehat{\text{IFR}}(t,\Delta t) = \frac{1}{\psi(t, \Delta t)} \frac{F(t+\Delta t)}{\hat{I}_S(t)},
\end{equation}
where $F(t)$ is the population cumulative death count and $\hat{I}_S(t)$ is the population level seroprevalence (extrapolated) estimate $\hat{I}_S(t) = \npop \times \ninf / \ntest$. Here $\npop$ is the population size, $\ninf$ is the number of infection positive in the demographically randomized test sample and $\ntest$ is the test sample size. By non-equal time we refer here to the shift by $\Delta t$, which can be optimized after the seroprevalence test.

No delay correction is needed, if $t$ or $\Delta t$ are chosen (or happen to be) with certain lucky values. This depends on interplay between three factors: 1. the delay kernel $K_F(t)$ of deaths, 2. the delay kernel $K_S(t)$ of antibodies (seroprevalence) and 3. the cumulative epidemic curve $\hat{I}(t)$. In our SARS-CoV-2 case, using kernels from~\cite{Perez-Saez2020.06.10.20127423}, the kernels give a functional shape for $\psi(t,\Delta t)$ which peaks above one for small $t$, then decreases below one, and asymptotically approaches one when $t \rightarrow \infty$. However, eventually the antibodies will vanish from the body (seroreversion), so realistic times scales must be used, also for other obvious reasons.

\begin{figure}[h!]
\centering
\includegraphics[width=0.8\textwidth]{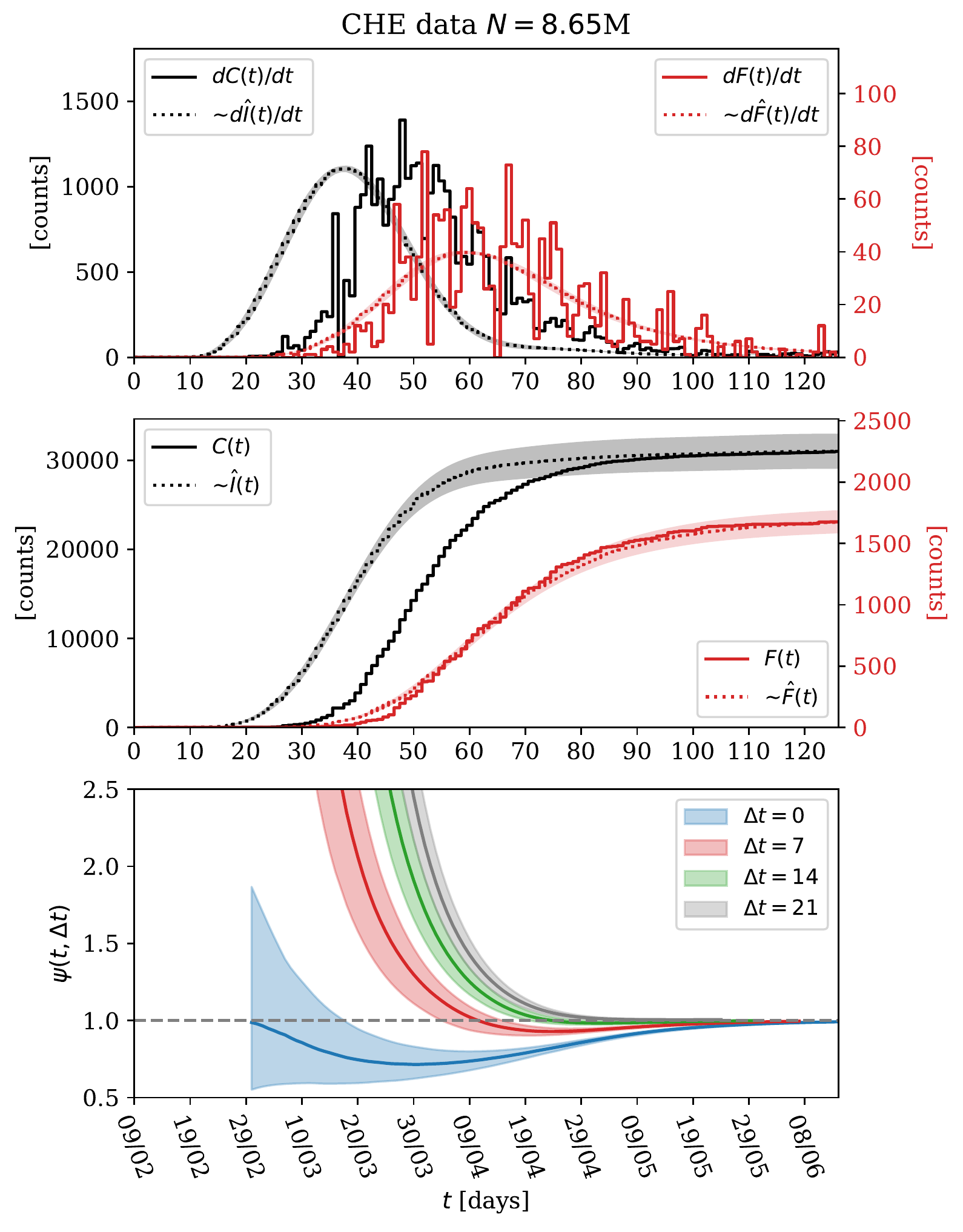}
\caption{Switzerland data deconvolution results (gray dotted), forward re-projection (red dotted) and in the bottom, the double delay scale function $\psi(t,\Delta t)$ contour estimates (CI95). The overall scale normalization of estimates (dashed) is matched with the measured (solid) functions of cases $C(t)$ and deaths $F(t)$, for visualization purposes. The top figure illustrates that our regularized deconvolution estimate of the infection rate $d\hat{I}(t)/dt$ is stable and that the delay kernels given in Appendix~\ref{appendix:sec:deconvolution} are realistic -- a good match is obtained between the measured cumulative death counts $F(t)$ and the forward projected shape $\hat{F}(t) = (K_F \ast \hat{I})(t)$ in the middle figure. Based on these observations, the obtained delay scale function $\psi(t,\Delta t)$ values in the bottom figure can be considered realistic. Raw data is from~\cite{owidcoronavirus}. }
\label{fig:deconvolution_CHE}
\end{figure}

The systematic uncertainty estimates should include perturbation of the kernels and the estimated $\hat{I}(t)$ function, most easily studied via toy Monte Carlo, propagated through the deconvolution algorithm and Eqs.~\ref{eq:delay_scale} and \ref{eq:IFR_estimate_non_equal_time}. The full procedure to estimate $\psi(t,\Delta t)$ is illustrated in Figure~\ref{fig:deconvolution_CHE}, based on kernel data from~\cite{Perez-Saez2020.06.10.20127423} and Switzerland data from~\cite{owidcoronavirus}. For the uncertainties, we used approximately 20 \% Gaussian equivalent uncertainties in the Weibull kernel parameters and fluctuated the input data with Poisson uncertainties, propagated via toy Monte Carlo. A good re-projection of the deconvolved $\hat{I}(t)$ to deaths $\hat{F}(t)$ is observed shape wise, as a `closure test'. We emphasize that this closure test would be trivial, if the daily death count $dF(t)/dt$ would have been used as the algorithm input. But we used the reported daily PCR cases $dC(t)/dt$ as the input, so the result is non-trivial. The absolute normalization for $\hat{I}(t)$ and $\hat{F}(t)$ is matched to data for the visualization, because it is not obtained as a part of the procedure. The scale function $\psi(t,\Delta t)$ has larger uncertainty and larger values earlier in the epidemic, which is natural.

\paragraph{Optimal and practical procedures} ~ Now in principle, after observing the epidemic time series, we can determine the optimal delay argument $\Delta t$ of the $\psi$-function for each fixed prevalence determination time point $t$ by setting $\psi(t,\Delta t) = 1$, and inverting the best $\Delta t$ value numerically. This gives us the time point $t + \Delta t$ to read out the death counts. Alternatively, we use Equation~\ref{eq:IFR_estimate_non_equal_time} directly with some chosen $\Delta t$ value and obtain the correction factor given by $\psi$, which is a more flexible option. This is because numerically equivalent $\Delta t$ value can be in the asymptotic future, if the epidemic evolution has already saturated (no more counts). Also in practise the right hand tail truncation (causality) of data must be treated explicitly e.g. in kernel estimation in very early phase.

The strategy of using the inversion machinery discussed here can be somewhat non-conservative. As a more conservative strategy, Figure~\ref{fig:deconvolution_CHE} shows that using a fixed read-out delay $\Delta t = 7$ days gives already a quite good choice as long as $t$ is after the peak of the daily deaths. Before that point, it yields upward biased IFR values. Similar behavior was observed also with other public datasets, which basically follows from the underlying functional shape of the epidemic curve. However, these conclusions are not without uncertainties and depend ultimately on data, as formulated in Eq.~\ref{eq:delay_scale}. The uncertainties related to kernels and their parametrizations are never very rigorous with a novel virus. Thus, from a prevalence test design point of view, an optimal choice for precision IFR estimates is to use a prevalence determination which has not been done too early in the local epidemic evolution.

In what follows, we explicitly evaluate the IFR values with different fixed $\Delta t$ values as a transparent and practical procedure. In addition we show results with an optimal delay solved from $\psi(t,\Delta t)$ function using the same kernels globally for each region, as an approximation. Both of these procedures have their pros and cons, as discussed here. The fixed delay case is essentially a special case of the latter, and even the complete procedure with fully known (oracle) kernels relies on a specific assumption of delay kernels being invariant (constant) over time. This time invariance is the defining property of the convolution integral. Also, whenever the daily reported positive PCR cases are used in the inversion, it may be necessary to \textit{normalize} the counts by non-constant test rates e.g. due to active policy changes of public test campaigns.

\clearpage

\section{Combination analysis}
\label{sec:combination}

In general, the value of IFR for a given individual is dependent on a number of factors such as age, sex, viral load, diet, genetics etc. For example, estimates of IFR in SARS-CoV-2 were found to have a strong age dependence in~\cite{Perez-Saez2020.06.10.20127423} and~\cite{stockholmstudy}, varying over two to three orders of magnitude as a function of age. Mathematically, let there be a multivariate IFR function
\begin{equation}
\text{IFR}\left(\mathbf{X} = (\text{age, sex, diet, genes, \dots}) \right): \mathbb{R}^d \rightarrow [0,1],
\end{equation}
which takes as an input a random human feature vector $\mathbf{X}$ and returns the expected probability to die $Y$ if infected.

In Bayesian modelling, the random feature vector is often composed into measured $\mathbf{X}$, and identified to be important but not necessarily measured variables $\mathbf{Z}$, such that,
\begin{equation}
\label{eq:bayesian_latent}
P(Y|\mathbf{X}) = \int d\mathbf{Z} \, P(Y|\mathbf{X},\mathbf{Z}) P(\mathbf{Z}|\mathbf{X}).
\end{equation}
Given enough, well sampled data, the function $\text{IFR}(\mathbf{X}) \equiv P(Y|\mathbf{X})$ can be modelled using simple histograms, (conditional) logistic regression, deep learning learning techniques or other methods. However, given the naturally limited data available early on in a pandemic, the IFR function can be integrated over its dependents to obtain a local expected IFR for a particular population,
\begin{equation}
\mathbb{E}[\text{IFR}]_j = \int d\mathbf{X} \, \text{IFR}(\mathbf{X}) \, f_j(\textbf{X}|\text{city}),
\end{equation}
where $f_j(\mathbf{X}|\text{city})$ is the normalized sampling density of the population. A city is chosen here to represent a realistic system size in terms of sample statistics, which can be considered as independent from other systems.

Any significant difference in the population densities $f(\mathbf{X}|\text{city})$, between cities, will yield different empirical $\langle \text{IFR}\rangle_j$ values. If the IFR function has a strong dependence over a particular feature, a bias with respect to the other city will be present. When comparing studies implemented in different cities, this intrinsic sampling bias can be compensated in two ways:
\begin{enumerate}
\item physically, \textit{a priori}, using carefully designed sampling (selection) of the population
\item mathematically, \textit{a posteriori}, using an inverse weight or sampling function, which can be modelled using detailed demographic statistics of the test sample and the city.
\end{enumerate}

Using either the strategy 1 or 2, one must also choose a reference population density or a `standard template' to represent a typical demography. This provides the golden reference for the sampling procedures. Once this sampling or stratification bias is accounted for, a truthful comparison of IFR values can be obtained. The expected raw global value without any re-sampling schemes can be modelled with a mixture density model as
\begin{equation}
\label{eq:mixture_density_model}
\mathbb{E}[ \text{IFR} ]_G = \int d\mathbf{X} \, \text{IFR}(\mathbf{X}) \, \sum_{j=1}^{K} w_j f_j(\mathbf{X}| \text{city}) = \sum_{j=1}^K w_j \mathbb{E}[\text{IFR}]_j,
\end{equation}
where the weights are $w_j = N_j/\sum_i N_i$ and $N_j$ is the total number of people in the city.

\subsection{Data}

Table~\ref{tab:prevalence_data} shows a number of studies performed to determine the prevalence of SARS-CoV-2. The datasets are chosen to represent Western cities with varying population sizes and reasonably similar demographics, and all studies, except for Iceland and Stockholm, are based on an antibody type blood tests. Count data for each dataset are given in Table~\ref{tab:count_data}. The daily reported cases and death counts for different cities and regions are collected from public databases~\cite{owidcoronavirus, newyorktimes, stockholmdata, swissdata} with full time series information, except Gangelt, for which we use only the death counts as given in their report. All data we used is available within our online analysis code. We do not do any further adjustments to count data, such as try re-correct or verify type I (false positive) or type II (false negative) errors of the antibody tests but we do evaluate their estimated effect on the uncertainties (see Appendix~\ref{appendix:sec:test_inversion}), compensate for underlying health conditions of the individuals (cause of death ambiguity) or try to adjust for demographic sampling differences. The effective population counts $\npop$ are from Wikipedia, which may be biased for some regions. It should be noted that the true prevalence can be determined only by full population (antibody) testing with unit sensitivity and specificity, which is not attainable with current technology. For more information about possible hidden sources of systematic uncertainties, perhaps correlated between different studies, see Appendix~\ref{appendix:sec:systematics}. 

\begin{table}[b!]
\centering
\begin{tabular}{|l||c|c|c|c|}
\hline
\textsc{Dataset} & Prevalence test period $T$ & Type & Age \\
\hline
Finland (FIN)~\cite{finnish_thl} & [2020-06-01, 2020-06-14] & IgG & 0-69 \\
Los Angeles (LAC)~\cite{10.1001/jama.2020.8279} & [2020-04-10, 2020-04-11] & IgG/IgM & all \\
Santa Clara (SCC)~\cite{Bendavid2020.04.14.20062463} & [2020-04-03, 2020-04-04] & IgG/IgM & all \\
San Francisco (SFR)~\cite{10.1001/jamainternmed.2020.4130} & [2020-04-23, 2020-04-27] & IgG & all \\
Iceland (ISL)~\cite{gudbjartsson2020spread} & [2020-04-04, 2020-04-04] & PCR & all \\
Gangelt (GAN)~\cite{Streeck2020.05.04.20090076} & [2020-03-31, 2020-04-06] & IgG/IgA & all \\
Geneva (GVA)~\cite{Stringhini2020.05.02.20088898} & [2020-05-04, 2020-05-09] & IgG & all \\
New York City (NYC)~\cite{10.1001/jamainternmed.2020.4130} & [2020-03-23, 2020-04-01] & IgG & all \\
Miami-Dade (MIA)~\cite{10.1001/jamainternmed.2020.4130} & [2020-04-06, 2020-04-10] & IgG & all \\
Region Stockholm (STK)~\cite{stockholmstudy} & [2020-03-26, 2020-04-02] & PCR & all \\
Philadelphia (PHI)~\cite{10.1001/jamainternmed.2020.4130} & [2020-04-13, 2020-04-25] & IgG & all \\
\hline
\end{tabular}
\caption{Studies for the determination of SARS-CoV-2 prevalence together with the type of the test used. The references for each study are provided in the table. Test type acronyms: Ig(G,M,A) is immunoglobulin type-(G,M,A) antibodies and PCR is polymerase chain reaction based amplification of viral RNA. We use a global fixed values for the test sensitivity $v=(0.892 \pm 0.02)$ and the specificity $s=(1 - 6\cdot 10^{-3} \pm 1.4 \cdot 10^{-3})$, obtained by averaging and taking standard error on the mean with input from~\cite{imperial_report_34}. Alternatively, each dataset could be associated local central values and their estimated uncertainties, which is supported by our code. Studies have corrected their count data for sensitivity and specificity with methods similar to derived in Appendix~\ref{appendix:sec:test_inversion}.}
\label{tab:prevalence_data}
\end{table}

The data are collected over different time periods, $T=[t_0,t_1]$, are of different sample sizes and cover different points during the developing epidemic. We collect the death counts at times $t+\Delta t$, and take a moving average of deaths over the prevalence determination period $T$
\begin{equation}
\nfatal(\Delta t) \equiv \frac{1}{|T|} \sum_{t=t_0}^{t_1} F(t+\Delta t),
\end{equation}
to take into account the finite span of the test period. These averaged and rounded death counts are shown in Table~\ref{tab:count_data}. We study the dependence of the results for different $\Delta t$ values to be able to explicitly show the IFR estimate sensitivity on the epidemic time-evolution. The datasets are with different prevalence rates, some of them quite low. This means that especially then the test specificity can be a problem, i.e. test errors cannot be anymore reliable corrected for. See Appendix~\ref{appendix:sec:test_inversion} for more information on this important aspect. We estimate this uncertainty using the methods described in the appendix, with the test sensitivity and specificity obtained by taking global averages from~\cite{imperial_report_34}.

\begin{table}[t!]
\centering
\begin{tabular}{|l||r|r|c||r|r|r|r|r|r|}
\hline
\textsc{Data} & \small{Pos.} & \small{Tests} $\ntest$ & \small{Prevalence} & \small{Population} $\npop$ & $\nfatal (\Delta t=0)$ & $\Delta t=7$ & $\Delta t=14$ & $\Delta t=21$ \\
\hline
FIN & 13 & 388 & 3.4E-02 & 5528737 & 323 & 325 & 327 & 328 \\ 
LAC & 35 & 863 & 4.1E-02 & 10039107 & 368 & 660 & 962 & 1260 \\ 
SCC & 50 & 3330 & 1.5E-02 & 1928000 & 40 & 51 & 74 & 100 \\ 
SFR & 12 & 1224 & 9.8E-03 & 883305 & 22 & 28 & 33 & 36 \\ 
ISL & 13 & 2283 & 5.7E-03 & 364134 & 4 & 7 & 8 & 10 \\ 
GAN & 138 & 919 & 1.5E-01 & 12597 & 7 & 7 & 8 & 8 \\ 
GVA & 84 & 775 & 1.1E-01 & 499480 & 278 & 286 & 292 & 294 \\ 
NYC & 171 & 2482 & 6.9E-02 & 19979477 & 805 & 3312 & 8286 & 12650 \\ 
MIA & 33 & 1742 & 1.9E-02 & 2716940 & 58 & 157 & 251 & 335 \\ 
STK & 18 & 707 & 2.5E-02 & 2370000 & 94 & 306 & 588 & 949 \\ 
PHI & 26 & 824 & 3.2E-02 & 1584000 & 339 & 501 & 699 & 896 \\ 
\hline
\end{tabular}
\caption{Count data for each dataset, with different read-out delay $\Delta t$ [days] choices for the death counts $\nfatal$ in the population. The observed fatalities are counted for the whole population as indicated and the number of infection positive are counted within the test sample. Columns are: positive counts (pre-corrected), number of tests, prevalence, population count and death counts for different read-out delays. Datasets with low prevalence are relatively more unstable under the test error inversion corrections. }
\label{tab:count_data}
\end{table}

Certain regions such as New York, Los Angeles or Stockholm were in a fast evolving stage in the epidemic evolution, when the prevalence studies were performed. This is seen in the Table~\ref{tab:count_data} fatality counts, as the counts change significantly for different values of $\Delta t$. In contrast, Geneva or Finland were already in a stable stage, thus time delays will have a minimal impact for estimating the IFR. 

In the following, we will determine the IFR for each dataset, and study the impact of fixed time delay to account for time evolution. Based on Figure~\ref{fig:deconvolution_CHE}, the choice of $\Delta t \simeq 7$ days can be considered a reasonable conservative approximation, which most likely overestimates the IFR very early on the epidemic curve, then being close to optimal choice at larger $t$ values. The optimal $\Delta t$ for each dataset is solved as outlined in Section~\ref{sec:time_evolution}. Future precision estimates in terms of time delays require careful kernel extraction for each dataset (region) individually.

\subsection{Combination strategies}\label{sec:combination-strategies}

The datasets from the studies in Table~\ref{tab:prevalence_data} can be considered simultaneously to study the global IFR. In this section, we describe several strategies to do so.  In general, we assume that the IFR values in each dataset are integrated values over all physical properties, and that the sample used in the study is chosen such that the single IFR value obtained will be representative of the whole population.
Furthermore, we assume statistically independent infection rates in each city, because the systems are physically long distance isolated and do not contain common infection sources. An exact decomposition of the sources of variance in the IFR, e.g. due to different demographies both within (local) and across (global) populations, is clearly not uniquely solvable. However, meaningful statistical estimates can be obtained. For comparisons, see~\cite{kenward1997small}.  

Fundamentally, we can write down an illustrative global-local-random-sampling additive model to study the datasets as a whole,
\begin{equation}
\label{eq:combination_model}
\boxed{
\underbrace{r}_{\text{true global IFR}} \rightarrow r + \overbrace{\delta_j}^{\text{local shift}}  = \underbrace{\theta_j}_{\text{true local IFR}} \rightarrow \theta_j + \overbrace{e_j}^{\text{sample noise}} = \underbrace{r_j}_{\text{observed IFR}}.
}
\end{equation}
\\
The following is a brief overview of the methods used in Section~\ref{sec:results}. In the following, $r_{j}$ represents the observed IFR values for $j=1...K$ independent studies. For some early work on the comparison analysis of similar experiments, we refer to Cochran \cite{cochran1937problems}.

\paragraph{Method of Moments} ~ This non-parametric estimator for the meta-analysis was in its simplest form proposed by DerSimonian and Lard~\cite{dersimonian1986meta}. This method aims to determine the mean and variance of the parent distribution of $r$. The mean is determined as, 

\begin{align}
\hat{r} &= \sum_{j=1}^K w_j r_j / \sum_{j=1}^K w_j,
\end{align}
with the global variance or `heterogeneity' given by~\cite{langan2019comparison}
\begin{align}
\hat{\Delta}^2 &= \max \left \{0, \frac{Q - \sum_{j=1}^K w_j s_j^2 + \sum_{j=1}^K w_j^2 s_j^2 / \sum_{j=1}^K w_j }{\sum_{j=1}^K w_j - \sum_{j=1}^K w_j^2 / \sum_{j=1}^K w_j} \right\}, 
\end{align}
where the test statistic is $Q = \sum_{j=1}^K w_j \left( r_j - \hat{r} \right)^2$ and $s_j^2$ represents the estimated variance within each study. 
We use a two step, iterative procedure in which $\hat{r}$ and $\hat{\Delta}^2$ are first determined using,
\begin{equation}
w_{j} = 1/s_j^2, 
\end{equation}
and then both $\hat{r}$ and $\Delta^2$ are updated by setting 
\begin{equation}
w_{j} = 1/(s_j^2 + \hat{\Delta}^2).
\end{equation}
Local convergence is obtained typically after a few iterations. The asymptotic standard error on $\hat{r}$ is given by
\begin{equation}
\hat{\text{se}}(\hat{r}) = \left( \sum_{j=1}^K w_j \right)^{-1/2},
\end{equation}
which provide Wald test-like confidence intervals. This method does not yield uncertainty on $\hat{\Delta}^2$. In general, a finite sample error is to be expected directly based on the sampling error in the study-specific variance estimates $s_j^2$. Several weighting scheme variants of the DL estimator have been proposed. See Ref.~\cite{langan2019comparison} for a recent comparison study, where the two-step DL estimator was found to be among the best, but the original DL performed weakly in some scenarios.


\paragraph{Normal Likelihood model} ~ Hardy and Thomson~\cite{hardy1996likelihood} used parametric normal-normal hierarchy with sampling densities
\begin{align}
r_j &\sim N(\theta_j, s_j^2) \\
\theta_j &\sim N(r, \Delta^2).
\end{align}
After integrating out the latent $\theta_j$, this gives a normal marginal distribution $r_j \sim N(r, s_j^2 + \Delta^2)$. The total joint log-likelihood with $K$ contributing studies is
\begin{equation}
\ln L(r,\Delta^2) = \ln \prod_{j=1}^K L_j(r, \Delta^2; r_j, s_j^2) = - \sum_{j=1}^K \frac{1}{2} \ln 2\pi(s_j^2 + \Delta^2) - \sum_{j=1}^K \frac{(r_j - r)^2}{2(s_j^2 + \Delta^2)}.
\end{equation}
It is easy to change the underlying sampling densities e.g. to a log-normal which takes effectively into account the physical boundary $r > 0$. The maximum likelihood solution can be obtained via standard optimization techniques or by iterating the following equations
\begin{align}
\hat{r} &= \sum_{j=1}^K \frac{r_j}{s_j^2 + \hat{\Delta}^2} / \sum_{j=1}^K (s_j^2 + \hat{\Delta}^2)^{-1} \\
\hat{\Delta}^2 &= \sum_{j=1}^K \frac{(r_j - \hat{r})^2 - s_j^2}{(s_j^2 + \hat{\Delta}^2)^2}  / \sum_{j=1}^K (s_j^2 + \hat{\Delta}^2)^{-2}.
\end{align}
The two-dimensional confidence region on these parameters is obtained with the log-likelihood ratio
\begin{equation}
2\ln L(r,\Delta^2) > 2\ln L(\hat{r},\hat{\Delta}^2) - \chi_{2,1-\alpha}^2.
\end{equation}
By profiling, we obtain the individual confidence intervals for $r$ and $\Delta^2$. It is also straight-forward to extend this normal-normal model to fully Bayesian hierarchies as described in~\cite{smith1995bayesian}. In that case Markov Chain MC sampling is typically used for obtaining the posterior density, which requires special technical care and is most easily dealt with specialized libraries.

As an alternative to the simple normal likelihood based, the so-called Restricted Maximum Likelihood (REML) method was introduced by Patterson and Thomson~\cite{patterson1971recovery} for unbiased estimates of variance components in linear mixed models. See Ref.~\cite{harville1977maximum} for a detailed derivation within the correlated and full multivariate formulations.

\paragraph{Wasserstein-Fr\'echet mean} ~ The Wasserstein metric barycenter or the Fr\'echet mean~\cite{agueh2011barycenters, dowson1982frechet} is an optimal transport (OT) based approach, which solves the optimization problem
\begin{equation}
\widetilde{P}(r) = \argmin_{P} \sum_{j=1}^K w_j W_p^p[P(r),P_j(r)],
\end{equation}
where $\widetilde{P}$ is the optimally combined new density under the $p$-Wasserstein metric $W_p$, which is a geodesic transport metric in the space of densities.  See Appendix~\ref{appendix:sec:optimal_transport} for more details and Ref.~\cite{panaretos2019statistical} for statistical properties. The weights can be taken as $w_j \propto 1/s_j^2$, if the solution is taken to be (inversely) proportional to the sample variances, for example. The solution is found by discretizing the 1D-posterior densities for each dataset and constructing the barycenter as an average in the inverse CDF space of quantile functions. We tried also a Sinkhorn iteration based algorithm using an entropy regularized transport cost formulation known as Bregman projections~\cite{benamou2015iterative}, with minimum regularization set such that a numerically stable output was obtained. However, this approach resulted seemingly in an over-smoothed output which is mathematically expected due to the entropic approximation.

\paragraph{Arithmetic mean of posteriors} ~ The arithmetic mean of posterior densities is
\begin{equation}
\label{eq:arithmetic_mean}
\widetilde{P}(r) = \sum_{j=1}^K w_j P_j(r|\{k_i,n_i,\alpha_i,\beta_i\}_j),
\end{equation}
which has a mixture model interpretation and the weights can be taken as in the optimal transport case. The density interpolation properties can be more limited compared to the optimal transport case, which can be crucial if the idea behind combining the posterior densities is to find one common data generating distribution. For multimodal densities, the mean estimator is an inclusive, probability mass covering estimator.
\vspace{1em}

\paragraph{Product of posteriors} ~ The normalized product (geometric mean) of posterior densities is
\begin{equation}
\widetilde{P}(r) = \frac{1}{Z} \prod_{j=1}^K P_j^{w_j}(r|\{k_i,n_i,\alpha_i,\beta_i\}_j),
\end{equation}
where $Z = \int_0^\infty dr \, \prod_j P_j^{w_j}(r|\{k_i,n_i,\alpha_i,\beta_i\})$ provides the re-normalization. This approach is known in machine learning as the product of experts model by Hinton~\cite{hinton2002training}, where several simple models are combined to `vote' together. It is also called logarithmic pooling. Thus for multimodal densities, the product estimator is an exclusive, single mode seeking estimator. By using information theory, this approach can be derived using e.g. the so-called $\alpha$-divergence of Chernoff, which is a generalization of the standard Kullback-Leibler (KL) divergence (relative entropy). The work by Amari~\cite{amari2007integration} shows how the product is an optimal solution to a divergence risk minimization problem with $\alpha=1$ (reverse KL) and that the arithmetic mean of Eq.~\ref{eq:arithmetic_mean} is also a minimal risk solution, but under its dual $\alpha = -1$ (forward KL).

\paragraph{Joint likelihood ratio} ~ This approach uses a product over the likelihood for each independent dataset, 
\begin{equation}
\label{eq:joint_likelihood}
-2 \left[ \sup_{ \{ p_1^{(j)} \} } \ln \prod_{j=1}^K L(r_0,p_1^{(j)}; X_j) - \sup_{ \{ r^{(j)}, p_{1}^{(j)} \} } \ln \prod_{j=1}^K L(r^{(j)},p_1^{(j)}; X_j) \right],
\end{equation}
where $X_j = \{k_1,n_1,k_2,n_2 \}_j$ is the $j$-th dataset. We compute the profile likelihood ratio test statistic of Eq.~\ref{eq:profile_likelihood} independently for each dataset. The combined IFR maximum likelihood value and confidence intervals are then obtained easily by comparing the total product (sum) of Eq.~\ref{eq:joint_likelihood} against the $\chi_1^2$-distribution quantiles as described in Section~\ref{sec:profile_likelihood}.
\\

\noindent We can summarize that the optimal choice of the combination method depends on the underlying assumptions, which are encoded by the implicit or explicit algebraic, information theoretic or probabilistic aspects of the method. Here the first class of combiners is more inclusive, the second class more exclusive, in their output decision. See Appendix~\ref{sec:risk_functions} for some illustrative properties of the probabilistic risk functions, which may be used for formal motivations.

As already mentioned, the methods we considered can be roughly classified into two scenarios for combining the data from the different studies. The first scenario aims to determine a parent distribution, from which each IFR observed for a particular city is assumed to be sampled from.  In the second scenario, we assume $\delta_j \equiv 0$ for all $j$ in Equation~\ref{eq:mixture_density_model} which represents a model in which the the true global and local IFR values are the same. Under this assumption, the individual measurements of IFR can be combined to provide a more precise but possibly overoptimistic measurement of IFR. In the following section, we use the \textbf{Method of moments}, the \textbf{Normal likelihood}, the \textbf{Wasserstein-Fr\'echet mean}, and the \textbf{Arithmetic mean of posteriors} methods for the first scenario, and the \textbf{Product of posteriors} and \textbf{Joint likelihood ratio} for the second scenario. 

\clearpage

\subsection{Results}\label{sec:results}

The individual observed IFR result for each dataset is given in Table~\ref{tab:IFR_results}, where the 95\% confidence (credible) intervals are obtained using the Bayesian estimator described in Section~\ref{sec:bayesian}. The results are given using four different choices of the fixed time delay $\Delta t$, and with an adaptive delay determined with the inverse machinery described in Section~\ref{sec:time_evolution}. In general, some datasets are strongly dependent on the chosen read-out delay. The reason for this is the underlying local epidemic evolution and its time derivatives. The adaptive delays are given in Table~\ref{tab:optimal_delays} where the confidence intervals are based on propagating Poisson fluctuations and kernel uncertainties through the whole deconvolution chain with Monte Carlo sampling as described in Section~\ref{sec:time_evolution}. This estimated uncertainty $\delta \gamma$ on the death count scale is included as a Gaussian prior in the Bayesian estimator as described in Section~\ref{sec:bayesian} and Appendix~\ref{sec:systematic_bayesian_priors}, whereas the fixed read-out delay estimates here are `bare' and do not include time scale uncertainties. This difference is manifest in the width of the individual densities. Measurements done in the very end of the local epidemic, generate larger optimal read-out delay values and correspondingly smaller delay uncertainties on the death counts because the daily increase in deaths has reached the slow asymptotic regime. The test inversion systematic scale uncertainty $\delta \lambda$ is included as a Gaussian prior affecting the IFR denominator, numerically larger for low prevalence datasets (see Table~\ref{tab:optimal_delays}).

\begin{figure}[t!]
    \centering
    \includegraphics[width=0.49\textwidth]{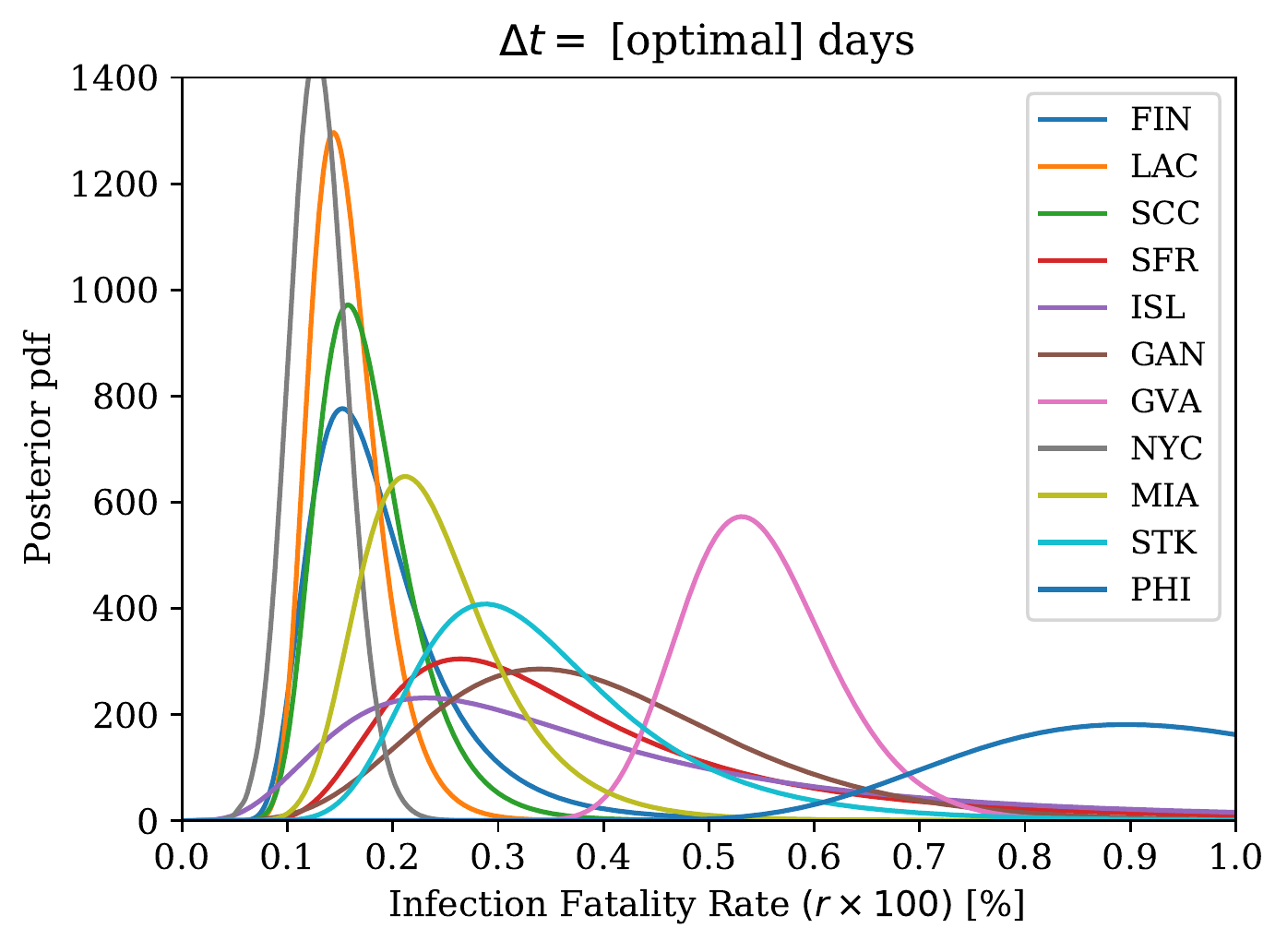}
    \includegraphics[width=0.49\textwidth]{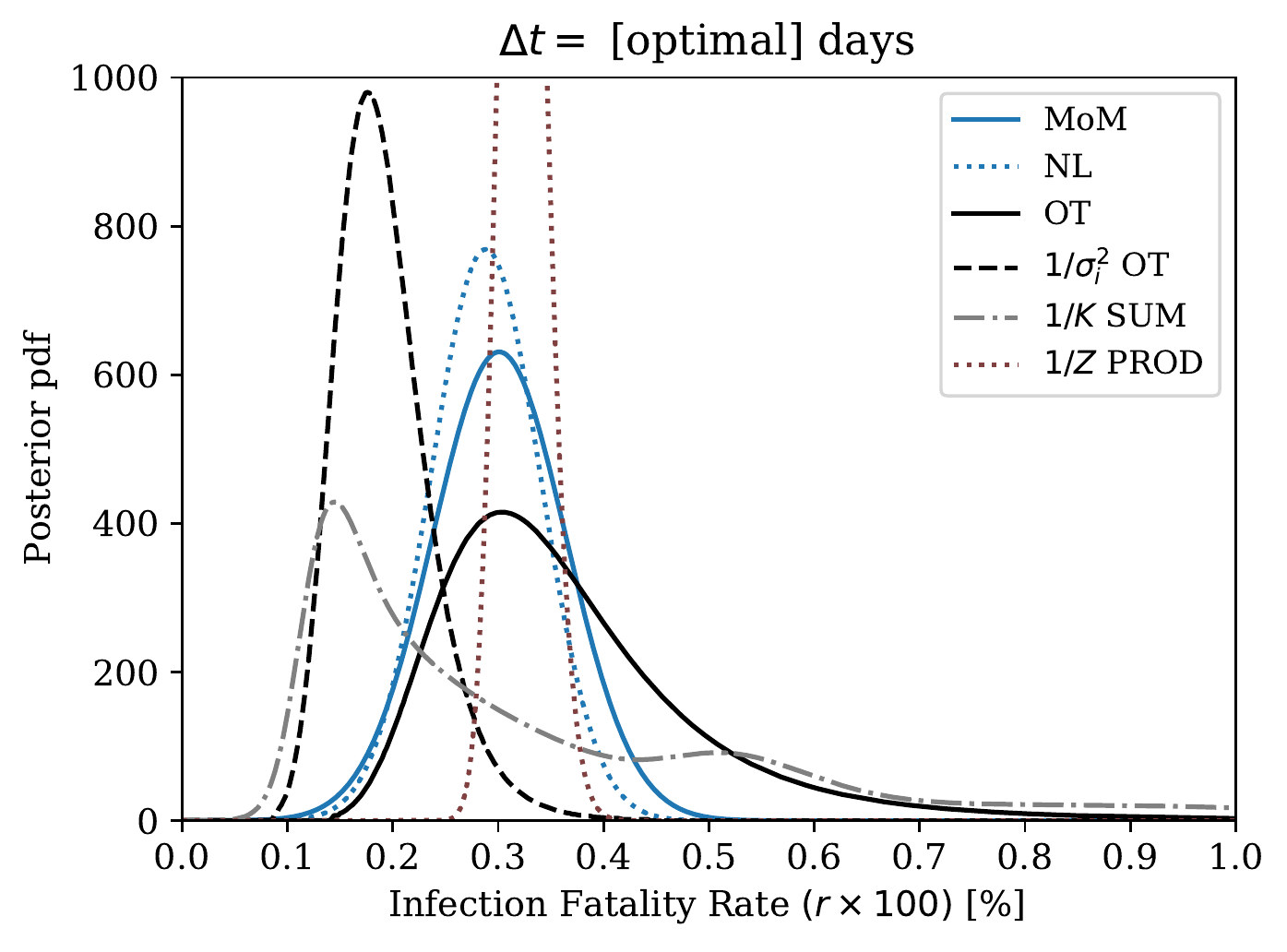}
    \caption{On the left, Bayesian posteriori densities for different datasets under non-informative Jeffreys prior. On the right, the method of moments (MoM) and the normal likelihood (NL) model based point estimates of $\hat{r}$ visualized with Gaussian uncertainties, the combined densities using unweighted and variance weighted optimal transport (OT), the mean of posteriors (SUM) and the normalized product (PROD). With the optimal $\Delta t \leftarrow \psi(t,\Delta t)$ read-out delay solution from deconvolution.}
    \label{fig:combination_adaptive_days}
\end{figure}

\begin{figure}[t!]
    \centering
    \includegraphics[width=0.49\textwidth]{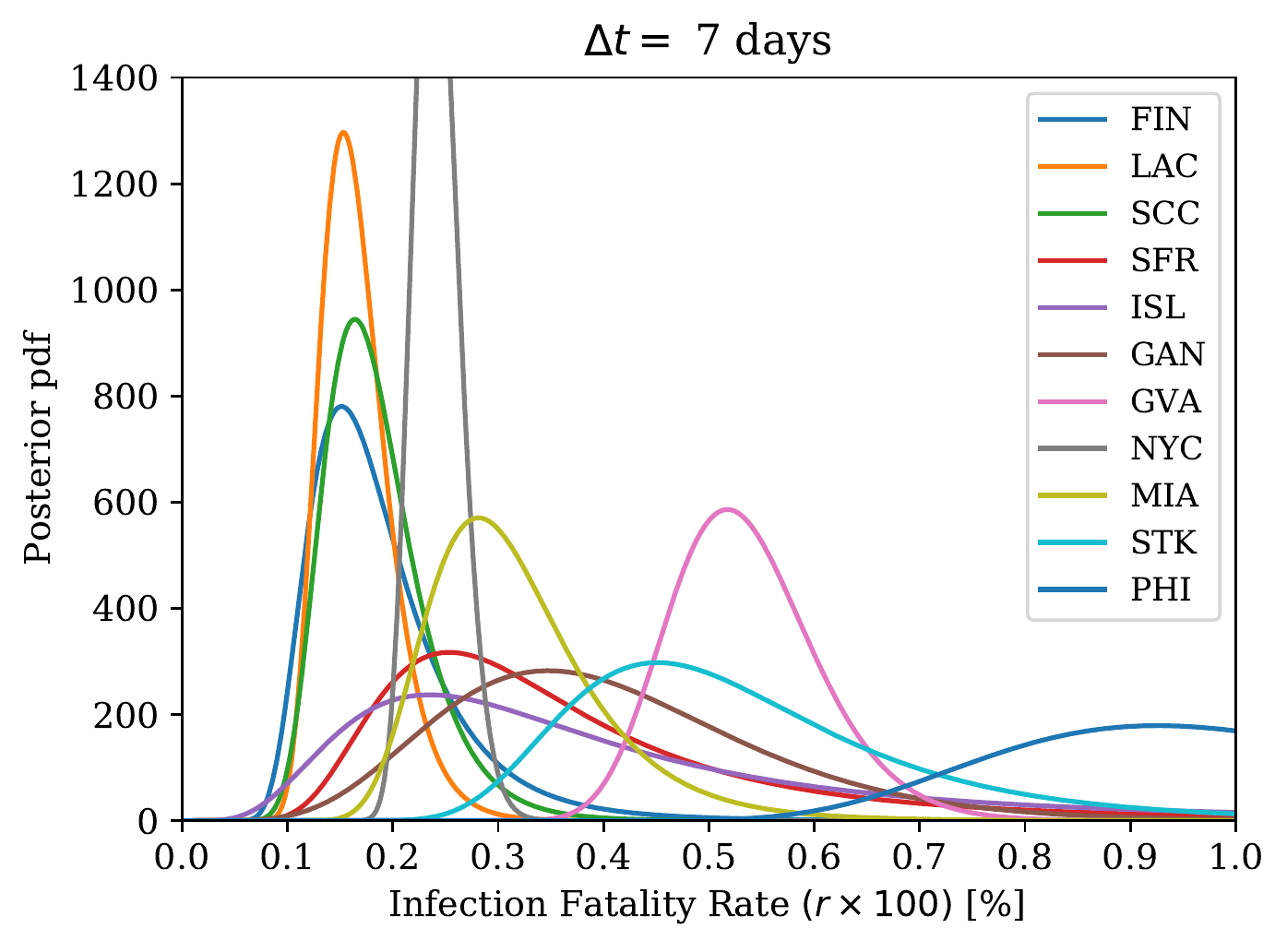}
    \includegraphics[width=0.49\textwidth]{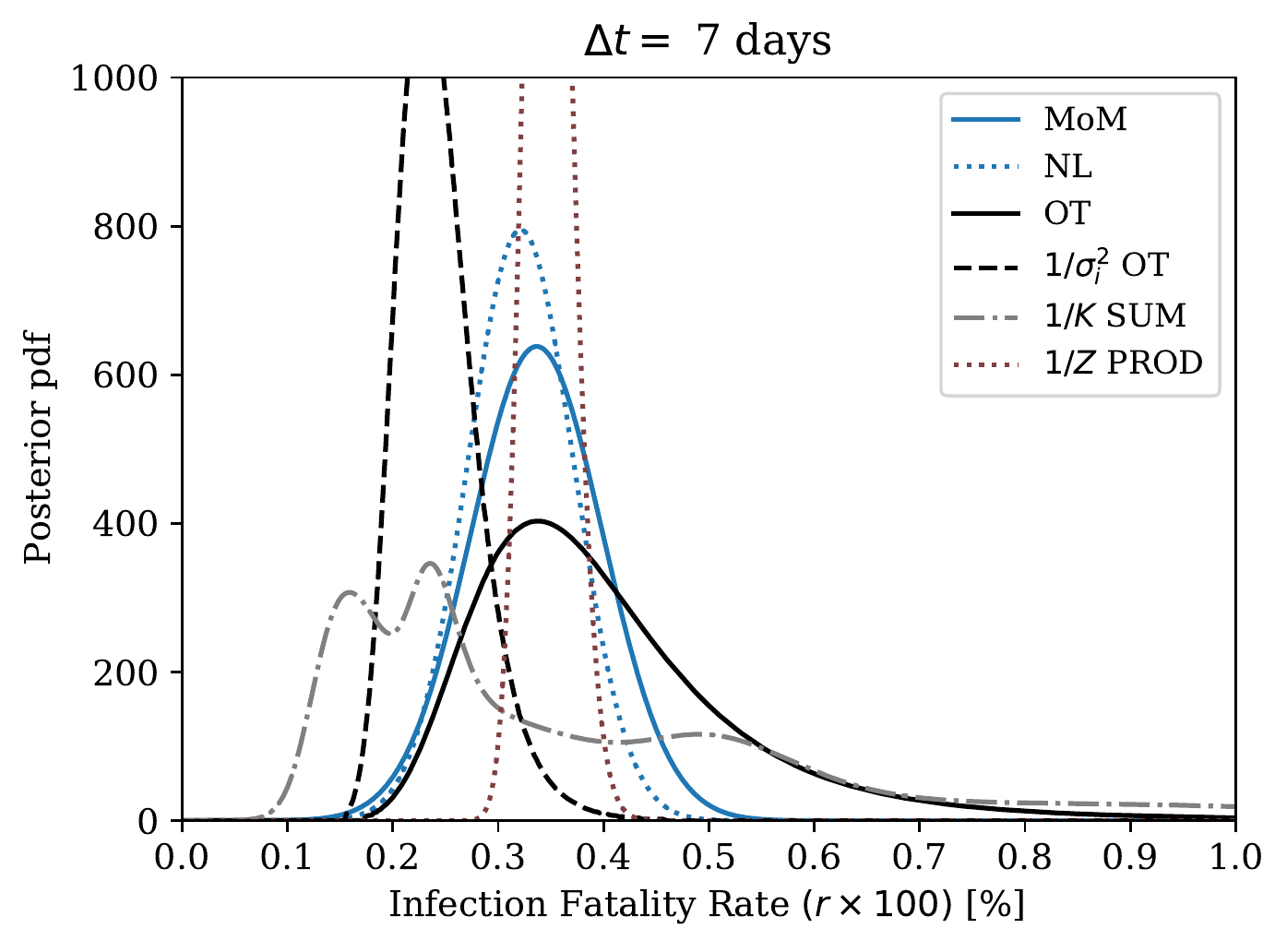}
\caption{Same as Fig. \ref{fig:combination_adaptive_days}, but using the read-out delay $\Delta t = 7$ days. }
    \label{fig:combination_7_days}
\end{figure}

\begin{figure}[t!]
    \centering
    \includegraphics[width=0.49\textwidth]{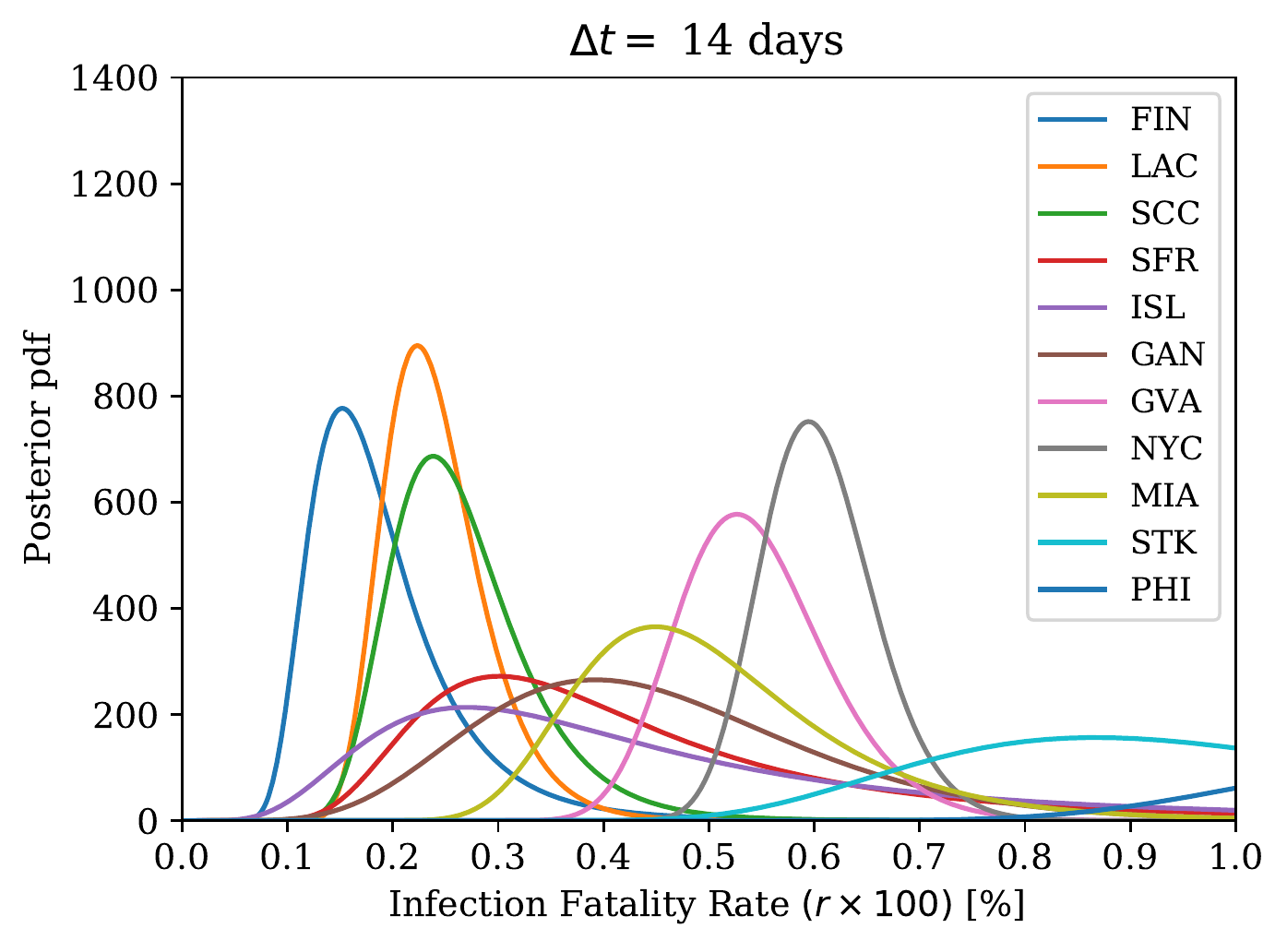}
    \includegraphics[width=0.49\textwidth]{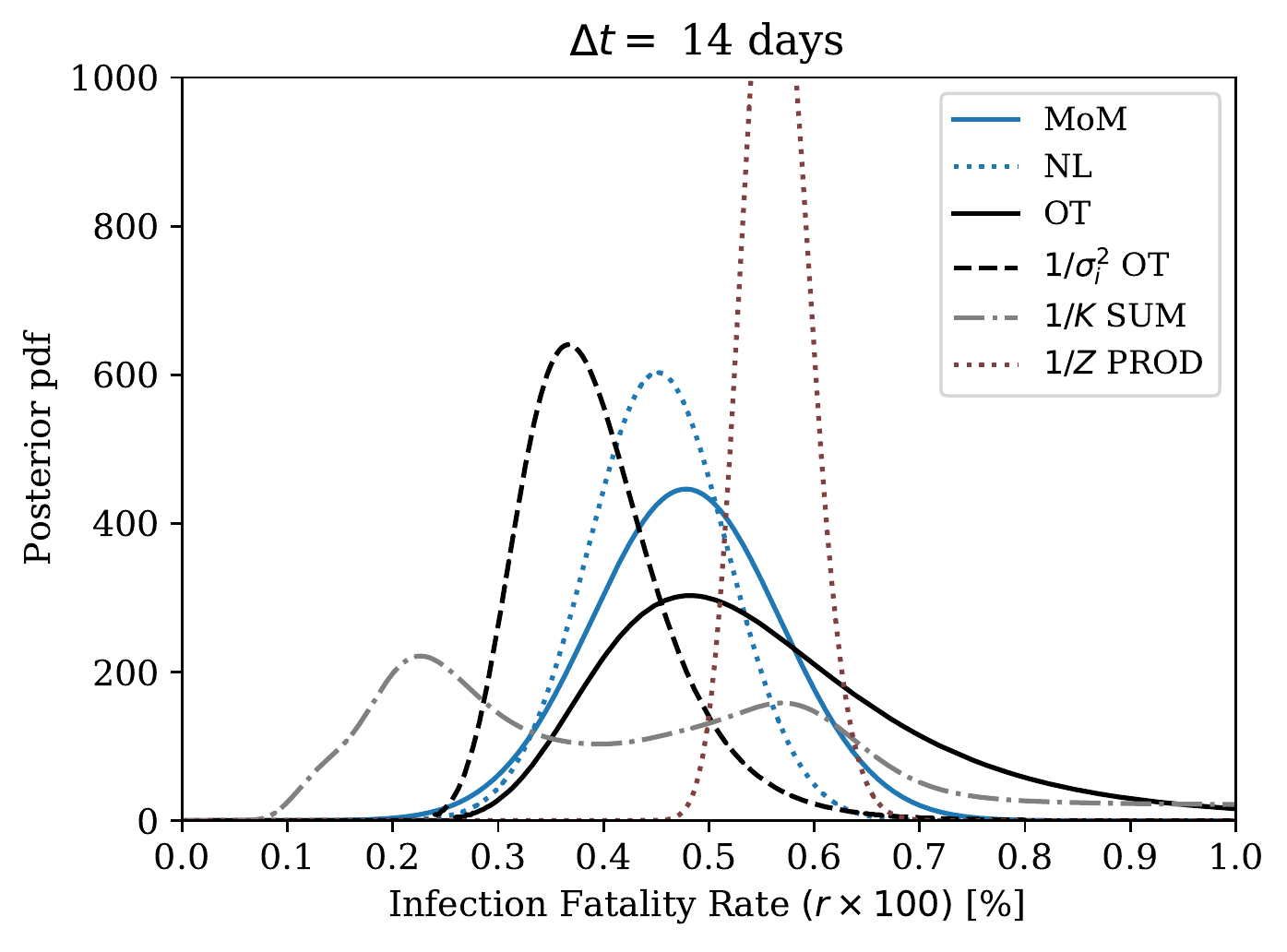}
\caption{Same as Fig. \ref{fig:combination_7_days}, but using the read-out delay $\Delta t = 14$ days.}
    \label{fig:combination_14_days}
\end{figure}

The posterior distributions from each city are presented in Figures~\ref{fig:combination_adaptive_days}, \ref{fig:combination_7_days} and \ref{fig:combination_14_days} for the choices of adaptive, $\Delta t=7$ and $\Delta t=14$ days, respectively. The sensitivity to the time delay effects highlights the importance of including these effects into any complete study of the IFR. Philadelphia yields significantly larger IFR peak values than the rest, while Los Angeles, Santa Clara and Finland all yield much smaller IFR peaks. This could point to a relatively strong IFR  dependence on the local population, and further highlights the importance of sampling the population within an individual study. The Gangelt study is approximately in the middle and fairly constant with choice of $\Delta t$. It is worth noting that the kernels used in adaptive delay inversion are the same (global) for each dataset, which makes it locally biased for PCR test based prevalence data (Stockholm, Iceland) and due to local reporting delays.

\begin{table}[b!]
\centering
\begin{tabular}{|l||c|c|c|c|c|}
\hline
\textsc{Data} & $\widehat{\text{IFR}}_{\text{MEAN}}$ $\Delta t=0$ & $\Delta t=7$ & $\Delta t=14$ & $\Delta t=21$ & $\Delta t \leftarrow \psi(t,\Delta t)$ \\
\hline
FIN & 0.19 [0.10, 0.37] & 0.19 [0.10, 0.37] & 0.19 [0.10, 0.37] & 0.19 [0.10, 0.37] & 0.19 [0.10, 0.37] \\
LAC & 0.09 [0.06, 0.14] & 0.17 [0.11, 0.25] & 0.24 [0.17, 0.36] & 0.32 [0.22, 0.47] & 0.16 [0.10, 0.24] \\
SCC & 0.14 [0.08, 0.24] & 0.18 [0.11, 0.30] & 0.27 [0.17, 0.43] & 0.36 [0.23, 0.57] & 0.18 [0.11, 0.29] \\
SFR & 0.31 [0.11, 0.87] & 0.40 [0.15, 1.08] & 0.47 [0.18, 1.27] & 0.50 [0.19, 1.35] & 0.41 [0.15, 1.13] \\
ISL & 0.29 [0.05, 1.08] & 0.47 [0.11, 1.66] & 0.52 [0.13, 1.81] & 0.63 [0.17, 2.08] & 0.47 [0.10, 1.67] \\
GAN & 0.40 [0.16, 0.75] & 0.41 [0.17, 0.76] & 0.45 [0.20, 0.82] & 0.45 [0.20, 0.82] & 0.40 [0.16, 0.75] \\
GVA & 0.52 [0.40, 0.67] & 0.53 [0.41, 0.69] & 0.54 [0.42, 0.70] & 0.55 [0.42, 0.71] & 0.55 [0.42, 0.71] \\
NYC & 0.06 [0.05, 0.07] & 0.24 [0.20, 0.29] & 0.61 [0.51, 0.72] & 0.92 [0.78, 1.10] & 0.13 [0.08, 0.19] \\
MIA & 0.12 [0.07, 0.20] & 0.32 [0.20, 0.52] & 0.51 [0.32, 0.83] & 0.68 [0.43, 1.10] & 0.24 [0.14, 0.42] \\
STK & 0.17 [0.09, 0.31] & 0.54 [0.30, 0.97] & 1.03 [0.59, 1.86] & 1.66 [0.95, 2.92] & 0.35 [0.18, 0.66] \\
PHI & 0.70 [0.44, 1.14] & 1.04 [0.66, 1.68] & 1.45 [0.92, 2.34] & 1.86 [1.18, 2.96] & 1.01 [0.63, 1.64] \\
\hline
\end{tabular}
\caption{Infection Fatality Rate (IFR) [\%] estimates (CR95) for each dataset, where columns denote different fixed read-out delays $\Delta t$ [days] and a data adaptive (inverse solved) read-out $\Delta t \leftarrow \psi(t,\Delta t)$ using global kernels. The credible interval takes into account the statistical counting uncertainty in the double ratio via Bayesian posterior estimator under non-informative Jeffreys prior, the test inversion related uncertainty $\delta \lambda$ and also the delay uncertainty $\delta \gamma$ in the case of adaptive delay estimation in the rightmost column.}
\label{tab:IFR_results}
\end{table}

\begin{table}[ht!]
\centering
\begin{tabular}{|l||c|c|c|c|}
\hline
\textsc{Strategy} & $\widehat{\text{IFR}}_{\text{MODE}}$ [\%] & $\widehat{\text{IFR}}_{\text{MEAN}}$ [\%] & Q68 [\%] & Q95 [\%] \\
\hline
 & \multicolumn{4}{c|}{$\Delta t = 7$ days} \\
\hline
MoM                                 & 0.34 & 0.34 & $[0.27, 0.40]$ & $[0.21, 0.46]$ \\
NL                                  & 0.32 & 0.32 & $[0.27, 0.37]$ & $[0.22, 0.42]$ \\
OT                                  & 0.34 & 0.41 & $[0.29, 0.52]$ & $[0.23, 0.78]$ \\
$1/\sigma_i^2$ OT                   & 0.23 & 0.24 & $[0.21, 0.28]$ & $[0.18, 0.34]$ \\
$1/K$ SUM                           & 0.24 & 0.41 & $[0.17, 0.62]$ & $[0.12, 1.23]$ \\
$1/Z$ PROD                          & 0.35 & 0.35 & $[0.33, 0.37]$ & $[0.31, 0.39]$ \\
Joint LLR                           & 0.34 & 0.34 & $[0.32, 0.35]$ & $[0.31, 0.37]$ \\
\hline
\hline
 & \multicolumn{4}{c|}{$\Delta t = 14$ days} \\
\hline
MoM                                 & 0.48 & 0.48 & $[0.39, 0.57]$ & $[0.30, 0.65]$ \\
NL                                  & 0.45 & 0.45 & $[0.39, 0.52]$ & $[0.32, 0.58]$ \\
OT                                  & 0.48 & 0.57 & $[0.42, 0.72]$ & $[0.34, 1.05]$ \\
$1/\sigma_i^2$ OT                   & 0.37 & 0.39 & $[0.33, 0.46]$ & $[0.28, 0.56]$ \\
$1/K$ SUM                           & 0.23 & 0.57 & $[0.22, 0.91]$ & $[0.14, 1.72]$ \\
$1/Z$ PROD                          & 0.56 & 0.56 & $[0.53, 0.60]$ & $[0.51, 0.63]$ \\
Joint LLR                           & 0.56 & 0.56 & $[0.54, 0.59]$ & $[0.51, 0.61]$ \\
\hline
\hline
 & \multicolumn{4}{c|}{$\Delta t \leftarrow \psi(t,\Delta t)$} \\
\hline
MoM                                 & 0.30 & 0.30 & $[0.24, 0.36]$ & $[0.18, 0.42]$ \\
NL                                  & 0.29 & 0.29 & $[0.24, 0.34]$ & $[0.19, 0.39]$ \\
OT                                  & 0.30 & 0.37 & $[0.26, 0.48]$ & $[0.20, 0.73]$ \\
$1/\sigma_i^2$ OT                   & 0.18 & 0.19 & $[0.15, 0.23]$ & $[0.12, 0.30]$ \\
$1/K$ SUM                           & 0.14 & 0.37 & $[0.14, 0.59]$ & $[0.10, 1.19]$ \\
$1/Z$ PROD                          & 0.32 & 0.32 & $[0.30, 0.34]$ & $[0.28, 0.37]$ \\
Joint LLR                           & 0.28 & 0.28 & $[0.26, 0.29]$ & $[0.25, 0.30]$ \\
\hline
\end{tabular}
\caption{Infection Fatality Rate (IFR) [\%] combined results with different strategies. The columns are the distribution mode, mean and 68 and 95 quantile intervals. The intervals have different interpretations and formal definitions depending on the underlying method (see text for details). N.B. Joint LLR is without systematic scale uncertainties $\delta \gamma$ and $\delta \lambda$ of Table~\ref{tab:optimal_delays} and includes only statistical counting uncertainties. The fixed $\Delta t$ delay estimates are without including the systematic scale uncertainty $\delta \gamma$ on the death count.}
\label{tab:combined_results_table}
\end{table}
\clearpage

\begin{table}[b!]
\centering
\begin{tabular}{|l||c||c|c|}
\hline
\textsc{Data} & $\Delta t \leftarrow \psi(t,\Delta t)$ [days] (CI68) & $\delta \gamma$ [\%] (CI68) & $\delta \lambda$ [\%] (CI68) \\
\hline
FIN & 15 [14, 17] & 0.066 & 17 \\
LAC & 6 [5, 7] & 7 & 10 \\
SCC & 6 [5, 6.5] & 2.6 & 15 \\
SFR & 8 [7, 9] & 3.1 & 32 \\
ISL & 7 [6, 8] & 14 & 43 \\
GAN & 1 [1, 1] & 0 & 4.3 \\
GVA & 29 [27, 30] & 0 & 5.4 \\
NYC & 3.6 [2.6, 4.5] & 19 & 4.9 \\
MIA & 5 [4, 5.8] & 13 & 15 \\
STK & 4.1 [3.2, 5] & 14 & 16 \\
PHI & 6 [5.1, 6.9] & 4.7 & 13 \\
\hline
\end{tabular}
\caption{Left column: the optimal read-out delay in days for each dataset based on the deconvolution inversion and its uncertainty. The Gangelt data includes no detailed time-series for deconvolution, thus $\Delta t=1$ is used instead. Center column: deconvolution and Monte Carlo propagated relative systematic scale uncertainty $\delta \gamma \equiv \sigma [\nfatal(\Delta t) ]/ \nfatal(\Delta t)$ on the cumulative death counts due to causal time-delays at time $t+\Delta t$. Right column: Type I and II test error inversion relative systematic scale uncertainty $\delta \lambda$ constructed with an error propagated uncertainty and a pure binomial Wilson uncertainty on the corrected prevalence fraction $p_2$, according to Eq.~\ref{eq:renormalization}. Test sensitivity and specificity values are as given in Table~\ref{tab:prevalence_data}, which are used in the error propagation described in Appendix~\ref{appendix:sec:test_inversion}, to obtain individual $\delta \lambda$ values given here. These systematic uncertainties are finally applied with the Bayesian priors as described in Appendix~\ref{sec:systematic_bayesian_priors}.}
\label{tab:optimal_delays}
\end{table}


\begin{figure}[ht!]
    \centering
    \includegraphics[width=0.49\textwidth]{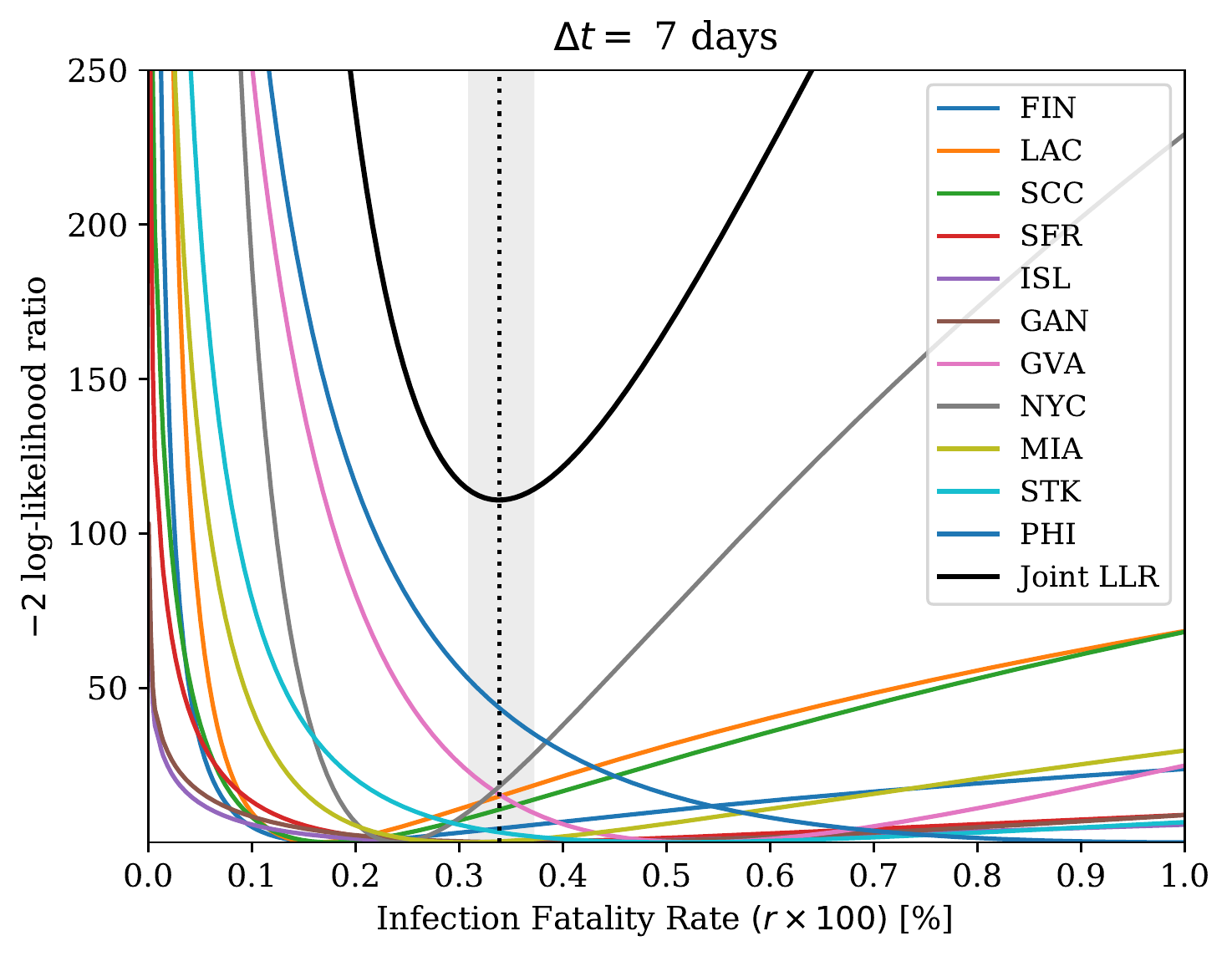}
    \includegraphics[width=0.49\textwidth]{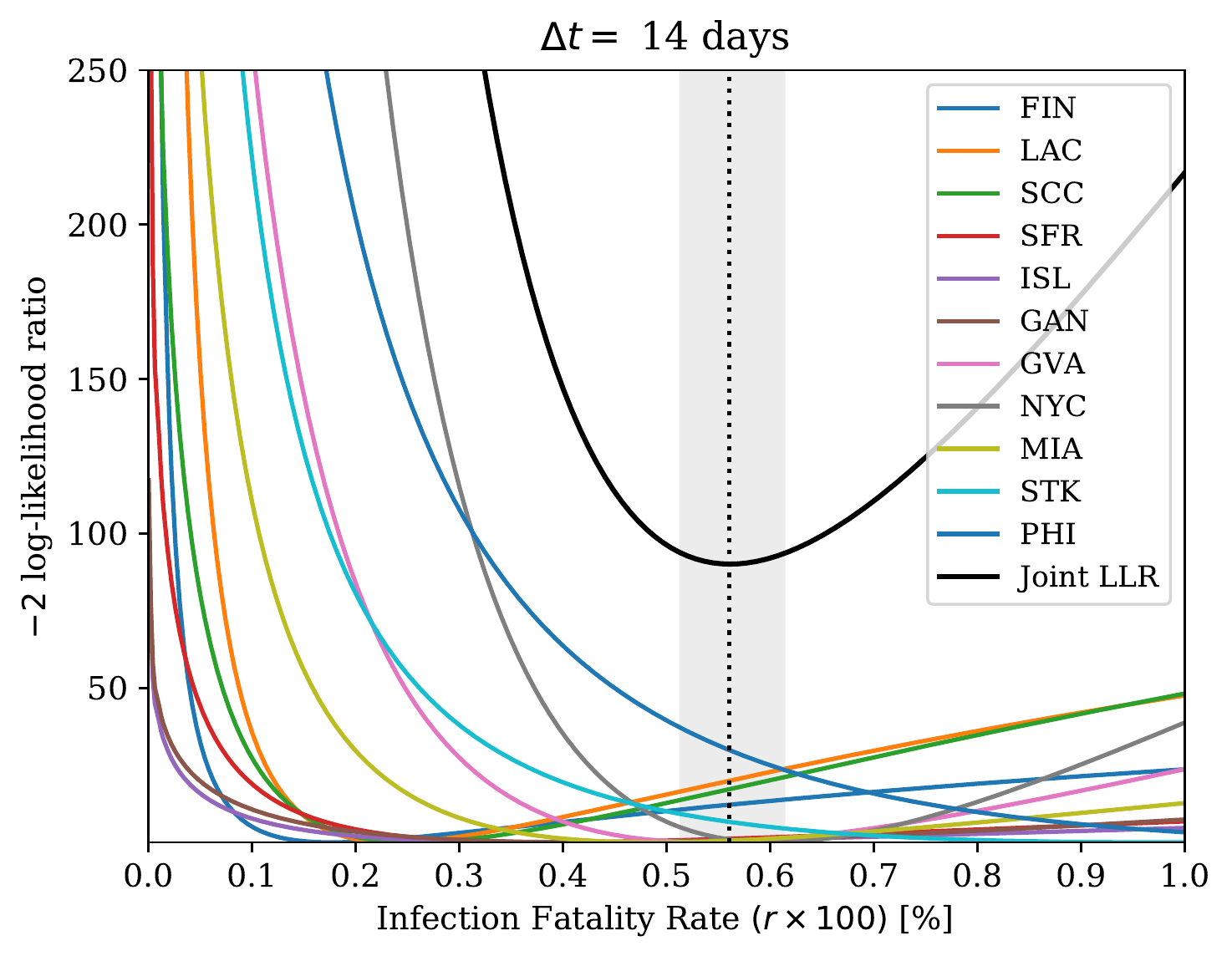}
\caption{Same as Figs. \ref{fig:combination_7_days} and \ref{fig:combination_14_days}, but using the joint LLR combination and without scale uncertainties $\delta \gamma$ and $\delta \lambda$. The vertical lines are the maximum likelihood estimate of the IFR and its CI95 confidence intervals.}
    \label{fig:combination_LLR}
\end{figure}

\begin{figure}[ht!]
    \centering
    \includegraphics[width=0.49\textwidth]{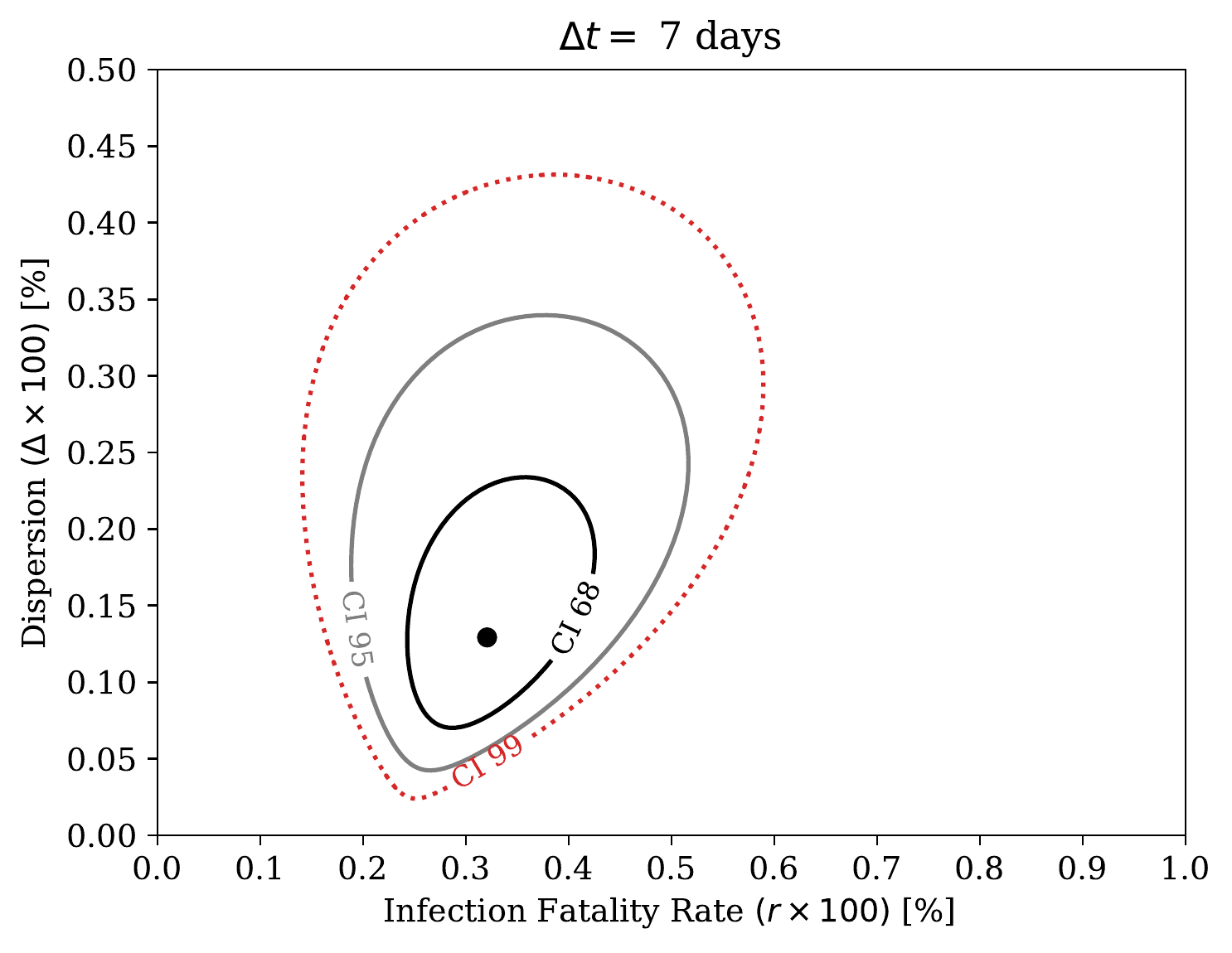}
    \includegraphics[width=0.49\textwidth]{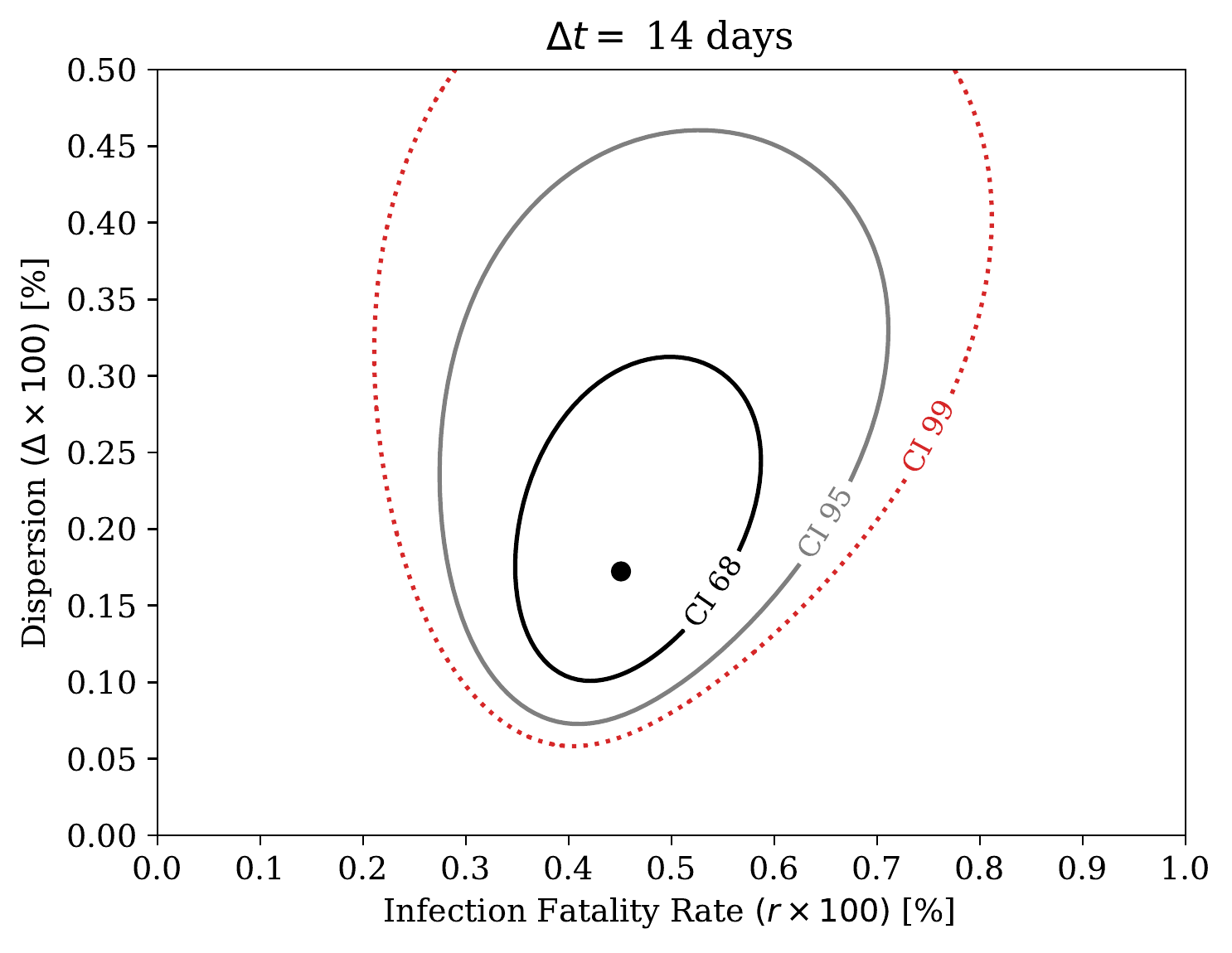}
\caption{The normal likelihood model parameters $r$ (IFR) and $\Delta$ (global dispersion) uncertainty regions (CI68, CI95 and CI99) based on contouring the corresponding log-likelihood 2D-function against $\chi_{2,1-\alpha}^2$ distribution quantiles. The parameter maximum likelihood values are shown with black dots.}
    \label{fig:combination_2D}
\end{figure}

The results of the different combination strategies are given in Table~\ref{tab:combined_results_table} and in Figures~\ref{fig:combination_7_days} and \ref{fig:combination_14_days}, obtained by combining the individual Bayesian posterior densities. The meta-analysis method of moments (MoM), and normal likelihood (NL) based strategies are presented assuming a Gaussian distribution, with the mean being $\hat{r}$ and the standard deviation representing the uncertainty on $\hat{r}$, obtained using the methods described in Section~\ref{sec:combination-strategies}. The global dispersion parameter $\Delta$ and its uncertainty are shown instead in Figure~\ref{fig:combination_2D} together with the IFR parameter $r$, using the NL model full likelihood information. The optimal transport is fully non-parametric, thus its output distribution does not explicitly try to disentangle the IFR and global dispersion like components and their individual uncertainties. Instead, it yields distributions which look in their functional form similar to individual distributions. The inverse variance weighted variant prefers smaller values of $r$, which is well expected with this set of data. However, often there can be other more suitable weighting strategies, based on e.g. some auxiliary risk minimization and the available domain knowledge.

The mean of posteriors method gives a very different distribution to the others. This method is sensitive to the datasets chosen as input and in general will not give a single representative (unimodal) distribution when the individual distributions have small overlap regions. In particular this is seen in the close by double peak structure around $r=0.2$ in $\Delta t=7$ days results, which disappears in the $\Delta t=14$ days result. This double peak structure is caused mainly by the New York distribution being narrow enough and moving significantly between these two choices. The simplest distribution peak structure is obtained with the adaptive delays, shown in Figure~\ref{fig:combination_adaptive_days}, which has only one strong peak between $r=0.1$ and $r=0.2$.

The product of posteriors and joint likelihood methods yield distributions which are much more narrow than the others. Figure~\ref{fig:combination_LLR} show the combination results using the joint likelihood method, where the delay $\Delta t = 14$ days gives slightly smaller joint log-likelihood ratio than the delay $\Delta t = 7$ days. However, this cannot stand on its own as a method for determining a good effective global $\Delta t$, because some datasets are invariant under the choice of $\Delta t$, thus their IFR values would have a pivotal role. Based on our time-delay calculus, individual datasets all have different delay scale functions $\psi(t,\Delta t)$ and are evaluated at different $t$ values, which give optimal local $\Delta t$ values shown in Table~\ref{tab:optimal_delays}. 

Finally, the high similarity between the two product methods is a natural consequence of the underlying assumption of these strategies that the IFR is a global quantity and that individual studies can be combined to improve the precision of the measurement. We caution however that this is a strong assumption and in particular highlight the fact that several of the individual distributions show tension with the product like combinations, indicating that the assumption may be unjustified.

\section{Conclusions}
\label{sec:summary}

The demographic averaged infection fatality rate (IFR) of SARS-CoV-2 in Western societies has been estimated here to be approximately $0.4 \; [0.2, 0.8] \, \%$ (Q95), when using a fixed $\Delta t$ = 7 day read-out delay between the prevalence determination and public death counts and the optimal transport based posterior combination of individual datasets. In general, our methods included the Bayesian double ratio based binomial counting uncertainty, deconvolution of the underlying time-series for the optimal read-out delay determination, systematic uncertainties in the death counts and systematic uncertainties in the corrected positive test counts. This estimate is a factor of $3-5$ larger than the IFR of a typical seasonal influenza, if one assumes its often referred value of $\sim0.1$ \%. Our result is numerically similar to analyses e.g. in~\cite{verity2020estimates} or~\cite{Meyerowitz-Katz2020.05.03.20089854}, but differs in the methodology. However, even if we constructed our methodology rigorously, one should not take our estimate as an ultimate measurement of the IFR. In addition of requiring better control of the underlying details of collected data, we identified also other factors which more extensive analysis would take into account such as demographic differences, population counts used in the normalization and regional time-delays. 

Our IFR estimation is based on several random sampled seroprevalence determination datasets combined with different statistical techniques, such as Bayesian estimation of counting uncertainties and modern algorithmic optimal transport driven probability density fusion. We recommend the Wilson estimator for fast but reasonable confidence interval estimation of binomial counting experiments and the Bayesian double ratio estimator for more extensive counting uncertainty estimates in the context of the IFR, because it provides access to a full posterior distribution. Also additional effects can be more easily incorporated with the Bayesian formulation. When analysing multiple datasets, the choice of data combination tools depends on the underlying scenarios. If it is reasonable to assume that there exists one global integrated IFR value, the product of posterior distributions or the joint likelihood method are perhaps the most natural approaches to use for improving the individual estimates. However, if that is not the case, then the Wasserstein-Fr\'echet mean (optimal transport) of posteriors, the classic meta-analysis models and the linear mean of posteriors can be more suitable, as we also reasoned with results from the information theory.

An accurate estimate also requires careful consideration of the significant time delays observed between infection and death. The required time delay convolutions necessitate the extraction (i.e. fitting) of delay kernel functions that may need to be locally adjusted because of differences in health care and administrative procedures. Poorly estimated kernels used within (de)convolution results in unknown bias, naturally. A more transparent solution is to always show in addition results with several fixed delays such as one or two weeks, as presented here. However, we provided and analyzed the necessary inverse problem methods for advanced, close to optimal delay corrections and demonstrated this machinery with data using fixed global kernels. The best solution experimentally is a cross-sectional seroprevalence trial that minimizes time-domain effects, namely, not done too early in the epidemic and which is tightly localized (not spread) over time, if possible.

Based on studies in Stockholm~\cite{stockholmstudy} and Geneva~\cite{Perez-Saez2020.06.10.20127423}, and global comparisons~\cite{imperial_report_34}, binning (stratifying) over age seems to be a crucial selection variable for the SARS-CoV-2 IFR, as expected. All statistical methodology developed here can be applied also in a stratified analysis. However, age stratification is not enough; although it gives strong ordering in IFR, it does not provide a proper explanation of the underlying dynamics. Possible body response differences and uncertainties in the antibody type tests are crucial factors, as well as the crucial extrapolation to the total population level. A recent study has found a new type of genetic defect risk in type I interferon (IFN) pathways, inducing a life-threatening COVID-19 pneumonia \cite{Bastardeabd4585}, but not being an active mechanism with influenza viruses.

From a larger perspective, a direct one-to-one comparison e.g. with seasonal flu is non-trivial, because for that the population has more natural immunity and there are seasonal vaccinations. However, an indirect comparison is possible and for understanding the total risk and harm on the society, one should understand the multiplicative reproduction differences between these viruses. A virus with a seemingly relatively small IFR can still be of high risk, if it is easily transmitted. A driving factor might not just be the mean $R_0$, but also independently the variance and tails of the transmission chain multiplicities. This possibly overdispersed case (compared to e.g. Poisson) is often modelled with a negative binomial distribution, which in physics describes at a phenomenological level the charged particle final state multiplicities of high energy proton-proton collisions. By using sharp analogies, tools from high energy physics can be useful on modelling and analyzing the epidemic production side of the problem.

Physically, the ultimate solution for the future is to increase massively the testing capabilities for real-time monitoring. This basically requires new type of non-invasive personal health care technology, perhaps based on promising new techniques such as CRISPR based diagnostics~\cite{myhrvold2018field}. More measurements is not just a requirement to obtain minimally extrapolated IFR estimates and reliable values for the epidemic parameters such as $R_0$ and understanding the critical role of multiplicative fluctuations, but more crucially, to obtain control of the epidemic with minimal lock-down measures. All the analyzed and developed tools here were constructed essentially from the first principles, and thus these methods should stay highly relevant also in the future.

\paragraph{Acknowledgements}~We thank Heather Battey and Yoshi Uchida for reading and comments on the manuscript, and Allen Caldwell for providing further information on their study~\cite{caldwell2020infections}.

\paragraph{Notes}~Open source Python code which reproduces all algorithms, figures and tables shown here, and beyond, is available at: \href{https://github.com/mieskolainen/covidgen}{github.com/mieskolainen/covidgen}.

\bibliographystyle{JHEP}
\bibliography{IFR-CL-paper}

\clearpage

\appendix

\section{Infection fatality rate observable}
\label{appendix:sec:IFR_definitions}

Definitions of the infection fatality rate estimators are
\begin{align}
\label{eq:IFR_true}
\text{Full statistics $\text{IFR}$}=& \; \frac{|\mathbf{I} \bigcap \mathbf{F}| }{|\mathbf{I}|} \;\;\;\;\;\;\;\;\;\; (\text{only via simulations or by complete testing}) & \\
\label{eq:IFR_estimate}
\text{Limited statistics $\widehat{\text{IFR}}$}=& \; \frac{|\mathbf{F}|}{|\mathbf{P}|} / \frac{|\mathbf{I} \bigcap \mathbf{T}|}{|\mathbf{T}|} \;\;\, (\text{a test sample based extrapolation estimate}),
\end{align}
where $\mathbf{F}, \mathbf{I}, \mathbf{T}, \mathbf{P}$ denote the sets of fatal, infected, tested and all people in the city, respectively. The number of elements of a set is denoted with $|\cdot|$ and the intersect of two sets with $\bigcap$. To show that the construction is consistent, consider the limit where all people in the city are tested by substituting $\mathbf{T} \rightarrow \mathbf{P}$ in Eq. \ref{eq:IFR_estimate}. Then the limited statistics IFR coincides with the full statistics IFR by construction, because also always holds that $\mathbf{I} \bigcap \mathbf{F} \equiv \mathbf{F}$ and $\mathbf{I} \bigcap \mathbf{P} \equiv \mathbf{I}$. These definitions are purely formal and assume perfect test sensitivity \& specificity and no time-delays. These necessary corrections are discussed in other sections of this paper.

The implicit assumption made in the extrapolation is that the infections observed in the test sample represent truthfully the stochastic infection process in the full sample. In essence some ergodicity (time-average equals ensemble average) and sample homogeneity must be assumed.

\section{Sampling two dimensional Bernoulli random numbers}
\label{appendix:sec:bernoulli_representations}

Using the expectation values $\mathbb{E}[X],\mathbb{E}[Y]$ and the correlation coefficient $\rho[X,Y]$ between two correlated Bernoulli random variables $X$ and $Y$, the direct (hypercube) basis parametrization is
\begin{align}
P_3 &= \rho[X,Y] \left( \mathbb{E}[X] \mathbb{E}[Y](\mathbb{E}[X] - 1)(\mathbb{E}[Y] - 1) \right)^{-1/2} + \mathbb{E}[X]\mathbb{E}[Y] \\
P_2 &= \mathbb{E}[X] - P_3 \\
P_1 &= \mathbb{E}[Y] - P_3 \\
P_0 &= 1 - (P_1 + P_2 + P_3) \\
0 &\leq P_0,P_1,P_2,P_3 \leq 1.
\end{align}
The sampling of vectors $(X,Y)$ is now multinomial, such that $[(0,0),(0,1),(1,0),(1,1)] \sim [P_0,P_1,P_2,P_3]$ are four corners of the hypercube. Any multinomial distribution sampling algorithm can be used.

\section{Details of the Bayesian estimator}
\label{appendix:sec:bayesian}

\subsection*{Binomial posterior density}
We start with the generic Bayesian inference formula for the posterior density
\begin{equation}
\label{eq:Bayes}
P(\theta | X_1,\dots,X_n,\gamma) = \frac{f(X_1,\dots,X_n|\theta,\gamma) \pi(\theta|\gamma)}{p(X_1,\dots,X_n)} = \frac{f(X_1,\dots,X_n|\theta,\gamma) \pi(\theta|\gamma)}{\int d\theta \,  f(X_1,\dots,X_n | \theta, \gamma) \pi(\theta|\gamma)},
\end{equation}
where $X_i$ contains the observed data, $\theta$ the parameters of true interest and $\gamma$ the hyperparameters or nuisance parameters. The likelihood function $L(\theta,\gamma) = f(X_1,\dots,X_n|\theta,\gamma) = \prod_i f(X_i|\theta,\gamma)$ is the sampling density evaluated as a function of $\theta,\gamma$ for $n$ iid observations and the prior density is $\pi(\theta|\gamma)$. In what follows, we will set
\begin{align}
\label{eq:likelihood}
\text{sampling density:}& \;\; f(X = k | \theta = p,n) = \begin{pmatrix} n \\ k \end{pmatrix} p^k (1-p)^{n-k} \\
\text{prior density:}& \;\; \pi(\theta=p) = \text{Beta}(p|\alpha,\beta).
\end{align}
A computationally easy conjugate prior pair for the binomial is the beta distribution yielding a beta posterior density, which we show below. In a same way, the gamma and Poisson distributions are conjugate pairs. The flat prior case is Beta$(p|1,1)=1$. The Jeffreys coordinate equivariant prior corresponds to the case Beta($p|1/2,1/2$) = $[\pi \sqrt{p(1-p)}]^{-1}$, which is an important one when considering non-informativeness under coordinate transforms.

To obtain the denominator (evidence) of Eq. \ref{eq:Bayes}, we marginalize over the parameter $p$-space
\begin{align}
&\int_0^1 \, dp \, \begin{pmatrix} n \\ k \end{pmatrix} p^k (1-p)^{n-k} \text{Beta}(p|\alpha,\beta) \\
\label{eq:evidence}
&= \begin{pmatrix} n \\ k \end{pmatrix} \frac{1}{\text{B}(\alpha,\beta)} \frac{\Gamma(\alpha+k) \Gamma(\beta+n-k) }{\Gamma(\alpha+\beta+n)}.
\end{align}
The posterior distribution is obtained by substituting Eq. \ref{eq:likelihood} and Eq. \ref{eq:evidence} into Eq. \ref{eq:Bayes}, giving $P(p|k,n,\alpha,\beta) = \text{Beta}(p|k+\alpha,n-k+\beta)$ distribution.

\subsection*{Beta-Binomial posterior mean values}

Using a generic Beta$(\alpha,\beta)$ prior and the binomial likelihood will give us the posterior density Beta$(k+\alpha, n-k+\beta)$ with the mean value
\begin{equation}
\hat{p}_{\text{Bayes}|\text{Prior Beta}(\alpha,\beta)} = \frac{k+\alpha}{k+\alpha + n-k+\beta} = \frac{k+\alpha}{n+\alpha+\beta}.
\end{equation}
Different priors will give
\begin{align}
\label{eq:flat_prior}
\text{Beta}(1,1):& \;\;\; \hat{p} = \frac{k+1}{n+2} \;\;\;\;\;\; (\text{Flat prior}) \\
\text{Beta}(1/2,1/2):& \;\;\; \hat{p} = \frac{k+1/2}{n+1} \;\;\; (\text{Jeffreys prior})\\
\text{Beta}(0,0):& \;\;\; \hat{p} = \frac{k}{n} \;\;\;\;\;\;\;\;\;\;\;\;\, (\text{Haldene's prior}).
\end{align}
The result in Eq. \ref{eq:flat_prior} was presumably first found by Laplace in his `law of succession' and inverse probabilities, which was considered somewhat controversial at that time because it does not coincide with the intuitive maximum likelihood answer $k/n$. 

\subsection*{Prior and posterior predictive distributions}

For arbitrary new data $x_{\text{new}}$, the prior predictive distribution is
\begin{equation}
p(x_{\text{new}}) = \int_\Theta d\theta \, \ell(x_{\text{new}}|\theta) \pi(\theta).
\end{equation}
Then, using a measured sample $\mathbf{X} \equiv (X_1,X_2,\dots,X_n)$, the posterior predictive distribution for new data is
\begin{equation}
p(x_{\text{new}}|\mathbf{X}) = \int_\Theta d\theta \, \ell(x_{\text{new}}|\theta) P(\theta|\mathbf{X}).
\end{equation}
The posteriori predictive distribution allows one to draw values $x$ from the sampling density with the parameter $\theta$ uncertainty described by the posteriori density $P(\theta|\mathbf{X})$. Thus, strictly speaking there exists no direct frequentist equivalent of this expression. 

\section{Systematic uncertainties via Bayesian priors}
\label{sec:systematic_bayesian_priors}

Additional systematic uncertainties on counts $k_1$ and $k_2$, are applied by multiplying and integrating over the Bayesian posteriori ratio IFR formula of Eq.~\ref{eq:bayesian_posterior_ratio} with
\begin{equation}
\label{eq:dressed_IFR}
P(r) \propto \int_{\epsilon}^\infty 
d\gamma \int_{\epsilon}^\infty d\lambda \int_0^1 dy |y| \, 
P\left( ry,y| \gamma k_1, n_1,  \lambda k_2, n_2, \{\alpha_i, \beta_i\} \right) G(\gamma; \mu_{\gamma}, \delta{\gamma})G(\lambda; \mu_{\lambda}, \delta{\lambda}),
\end{equation}
where $G(x,\mu,\sigma)$ is a normal density, $\epsilon$ a small positive scalar and the integrals are computed numerically. These additional Gaussian distributed parameters model multiplicative scale corrections $\gamma$ and $\lambda$ on the death counts $k_1$ and on the positive test counts $k_2$, respectively. The triple integral gives the posteriori ratio probability density up to the overall normalization, which is obtained numerically. The normal prior densities here can be replaced with gamma densities, for example.

The mean values are taken $\mu_{\gamma} = \mu_{\lambda} = 1$, typically, if the the counts are corrected prior this formula. The 1-sigma uncertainties on these corrections are described by $\delta {\gamma}$ and $\delta {\lambda}$, which come from auxiliary procedures or calibration measurements. In our case, $\delta {\gamma}$ is obtained via Monte Carlo error propagation of the deconvolution procedure and its kernel uncertainties in Section~\ref{appendix:sec:deconvolution}, and $\delta {\lambda}$ as described in Section~\ref{appendix:sec:test_inversion}. One needs to pay attention to possible double counting of statistical uncertainties in Equation~\ref{eq:dressed_IFR}, when estimating these parameters.

\section{Credible and confidence intervals}
\label{appendix:sec:interval_definition}

\paragraph{Bayesian}~A Bayesian credible interval (CR) at the level $1-\alpha$ is defined as an integral over the posterior density
\begin{equation}
\mathbb{P}(\theta \in \mathcal{C}|X) = \int_{\mathcal{C}} d\theta \, P(\theta | X, \gamma) = 1 - \alpha,
\end{equation}
where $\mathcal{C}$ defines the credible interval or multidimensional region, which contains the true parameter with $(1-\alpha) \times 100$ \% probability. There is usually an infinite number of such intervals, but often the tail masses are fixed to be equal $\alpha/2$. A given credible interval is not constructed to contain the parameter with the same probability if the experiment is repeated, which is what a frequentist confidence interval tries to construct. However, the Bayesian construction may have also strong frequentist coverage properties, as the well-known `Jeffreys interval' demonstrates~\cite{brown2001interval}.
\\

\paragraph{Frequentist}~A basic property of frequentist confidence intervals (CI) is their coverage. This is a property of statistical procedures for extracting intervals for parameters of interest $\theta$ at some confidence level $1 - \alpha$; it does not apply to a single confidence interval from a specific experiment. For a repeated set of measurements, each with its own fluctuations, the position of the intervals will vary. The coverage is defined as the fraction of intervals that contain the true value of $\theta$. Coverage can vary with the value of $\theta$, but for frequentist intervals from a Neyman construction~\cite{neyman1937x}, it will never be smaller than $1-\alpha$. i.e.
\begin{equation}
\lim_{n \rightarrow \infty} \inf_{\theta} \frac{1}{n} \sum_{i=1}^n I(\theta \in \mathcal{C}_i) = 1 - \alpha,
\end{equation}
where $I$ is the indicator function $I: \mathbb{R} \rightarrow \{0,1\}$ and $n$ is the number of repeated experiments (with differing intervals). Formally, for a given $\alpha$, the confidence interval or region $\mathcal{C}_i$ is the one which gives the infimum (the greatest lower bound) of the coverage probability. The interval and its lower and upper endpoints $L(X) \leq U(X)$ are random variables depending on the random data $X$, where as the true parameter $\theta$ itself is not a random variable in this picture. Finally, it is instructive to show that combining two one-sided bounds
\begin{equation}
\inf_\theta \mathbb{P}(L(X) \leq \theta)) = 1-\alpha/2 \;\; \text{and} \;
\inf_\theta \mathbb{P}(U(X) \geq \theta)) = 1-\alpha/2,
\end{equation}
gives the expected confidence interval
\begin{align}
\nonumber
&\mathbb{P}(L(X) \leq \theta \leq U(X)) \\
\nonumber
&=1 - \mathbb{P}(L(X) > \theta \cup U(X) > \theta) \\
\nonumber
&=1 - \left[\mathbb{P}(L(X) > \theta) + \mathbb{P}(U(X) < \theta) \right] \\
&= 1 - [\alpha/2 + \alpha/2] = 1-\alpha.
\end{align}
Unlike the Bayesian credible intervals, the frequentist confidence intervals do not explicitly estimate the probability for the parameter to be within some range.

\section{Acceptance set ordering principles}
\label{appendix:sec:ordering}

The optimal frequentist confidence interval acceptance set construction, used in the inverse construction of the Neyman confidence belts, can be derived briefly as follows~\cite{kendall1961advanced}.

\begin{enumerate}
\item There is a one-to-one mapping between tests and confidence intervals.
\item Uniformly most accurate (UMA) confidence region minimizes the probability of false coverage.
\item By using Property 1, UMA set is found by inverting the uniformly most powerful (UMP) test.
\item According to the Neyman-Pearson lemma~\cite{neyman1937x}, the likelihood ratio test is the UMP when \textit{both} the hypothesis $H_0$ and alternative $H_A$ are simple (not composite). The UMP also exists for a composite $H_A$, if the underlying distributional family has the so-called monotone likelihood ratio property. In the most general case, no UMP test is guaranteed to exist.
\end{enumerate}

Several other acceptance set constructions or ordering principles also exist, such as the shortest expected length and various pdf based orderings, perhaps optimal under some very specific condition such as certain interval topology. Also randomized intervals can be constructed, but which are mostly used only in theoretical analysis of (discrete) problems.

\subsection*{Explicit construction}
Let our parameter of interest be $\theta \in \Omega$, the random measurement be $X$, and let us use here the likelihood ratio based ordering. We can formalize the confidence interval as a set
\begin{equation}
S(X=x) = \{ \theta : LR(x,\theta) \geq c(\theta) \}
\end{equation}
having the corresponding coverage probability
\begin{equation}
\mathbb{P}_\theta(\theta \in S(X)) = \mathbb{P}_\theta \left( LR(X, \theta) \geq c(\theta) \right) \geq 1 - \alpha \;\; \forall \theta \in \Omega.
\end{equation}
To construct the set, the likelihood ratio is considered at each value of $\theta$, for each value of $X$
\begin{equation}
LR(x,\theta) = \frac{f(x,\theta)}{f(x,\hat{\theta})},
\end{equation}
where $\hat{\theta}$ is the maximum likelihood estimate. The crucial piece above is the confidence level $1-\alpha$ constructing local threshold
\begin{equation}
c(\theta) = \sup_r \mathbb{P}_\theta \left( LR(X,\theta) \geq r \right) \geq 1 - \alpha,
\end{equation}
which is explicitly dependent on $\theta$. This value can be constructed with asymptotic approximations or with Monte Carlo. For more information, see e.g.~\cite{spjotvoll1972unbiasedness, feldman1998unified}.

\section{Type I and type II test error inversion}
\label{appendix:sec:test_inversion}

Let $p=P(V_+)$ be the true viral prevalence of the population, let $q=P(T_+)$ be the fraction of positive tests in the test sample. Let \textit{specificity} be $s \equiv P(T_-|V_-) = 1-\alpha = 1 - \mathbb{P}(\text{type I error})$ and let \textit{sensitivity} be $v \equiv P(T_+|V_+) = 1-\beta = 1 - \mathbb{P}(\text{type II error})$. Using alternative terminology, $\alpha$ is known as False Positive Rate and $1-\beta$ as True Positive Rate. These symbols should not be mixed with the parameters of the Beta priors, to be clear. The following derivation uses pure probability calculus, without specifying the underlying density or mass functions.

The four different conditional probabilities can be combined under the Bayes' rule
\begin{equation}
P(V_i|T_j) = \frac{P(T_j|V_i)P(V_i)}{P(V_j)} = \frac{P(T_j|V_i)P(V_i)}{\sum_{k \in \{-,+\}} P(T_j|V_k)P(V_k)}, \;\; \text{for} \; i,j \in \{-,+\},
\end{equation}
with the law of total probability expanded for the true prevalence
\begin{equation}
P(V_{+}) = P(V_{+}|T_{-})P(T_{-}) + P(V_{+}|T_{+})P(T_{+}).
\end{equation}

Using these, a well-known inverse estimator (see e.g.~\cite{rogan1978estimating}) for the true prevalence is
\begin{equation}
\label{eq:test_inversion}
\hat{p} = \frac{q+s-1}{v+s-1},
\end{equation}
which has a physical solution $0 \leq \hat{p} \leq 1$, if and only if
\begin{align}
&1-s \leq q \leq v, \;\; \text{i.e.} \\
\nonumber \text{False Positive Rate $\alpha$} \leq \; &\text{Positive Test Fraction $q$} \leq \text{True Positive Rate $(1-\beta)$}.
\end{align}
Otherwise the problem is physically ill-posed. Especially the FPR lower bound is problematic when the viral prevalence is low. As a simple estimate of the related uncertainty, we can use the first order Taylor expansion (error propagation) with $q,s,v$ taken independent. We get
\begin{equation}
\label{eq:error_propagation}
\widetilde{\sigma}_{p}^2 = \frac{(v+s-1)^2 \sigma_q^2 + (q-v)^2 \sigma_s^2 + (q+s-1)^2 \sigma _v^2}{(v+s-1)^4},
\end{equation}
where $\sigma_q^2, \sigma_s^2, \sigma_v^2$ are the individual 1-sigma uncertainties squared. The first one is driven by the binomial counting, the two other by the uncertainty in the laboratory calibration of the test error rates. Instead of using dichotomic (binary) test output decisions and Eq.~\ref{eq:test_inversion}, alternative inversion strategies can be based on a test-by-test weighted inversion, according to the conditional probabilities of Bayes' rule and Expectation-Maximization (maximum likelihood) iteration of the prevalence fraction. This requires that the test provides a probabilistic output (e.g. multivariate analysis). Different strategies should be simulated with Monte Carlo sampling.

\paragraph{Renormalization procedure}~Given already inverted prevalence rate $\hat{p}$ (or counts $k = n\hat{p}$) together with known (or assumed) sensitivity and specificity and their uncertainties, we can estimate the relative systematic multiplicative scale uncertainty $\delta {\lambda}$ due to type I and II errors, by first computing the corresponding raw rate $q$ by (re)inverting Eq.~\ref{eq:test_inversion}, compute its binomial uncertainty $\sigma_q$, then apply Eq.~\ref{eq:error_propagation} and finally find out the additional (orthogonal) relative uncertainty
\begin{equation}
\label{eq:renormalization}
\delta {\lambda} \equiv \left[ \left( \frac{\widetilde{\sigma}_p}{\hat{p}}\right)^2 - \left( \frac{\sigma_p}{\hat{p}} \right)^2 \right]^{1/2},
\end{equation}
by remembering that in the multiplicative case relative uncertainties add in quadrature. The pure binomial reference uncertainty $\sigma_p$ can be computed e.g. with the Wilson estimator. The re-inversion step is needed only if no raw data is available. The idea behind this renormalization procedure is to protect against double counting the statistical uncertainty component, when multiplicatively `dressing' the Bayesian IFR estimates (with the corrected counts as input) as explained in Section~\ref{sec:systematic_bayesian_priors}.

\section{Regularized non-negative deconvolution}
\label{appendix:sec:deconvolution}

The deconvolution here is implemented as a non-negative least squares with Tikhonov regularization. We found this classic approach to be by far the most stable of standard methods in this problem, including regularized Fourier space methods and early stopping regulated maximum likelihood EM-iteration (Richardson-Lucy). The EM-iteration driven formulation assumes Poisson noise, which in principle should be more optimal, however, the explicit regularization properties of the method shown here seemed to play a bigger role.

The regularized least squares solution for the discretized infection rate $\mathbf{x} \sim dI(t)/dt$ is obtained by inverting the linear convolution equation $A\mathbf{x} = \mathbf{y}$, by minimizing
\begin{equation}
\label{eq:nonneg_least_squares}
\hat{\mathbf{x}} = \argmin_{\mathbf{x}} \, || A \mathbf{x} - \mathbf{y}||^2 + \lambda_R^2 || L (\mathbf{x} - \mathbf{x}_0)||^2 \;\; \text{subject to} \;\; \mathbf{x} \geq \mathbf{0},
\end{equation}
where $\mathbf{y}$ is the measurement vector and $\lambda_R$ controls the regularization strength. The measurement vector is constructed from the daily PCR infection counts $\mathbf{y} \sim dC(t)/dt$, where the vector domain is extended (padded) with zeros before the first counts, in order to be able to describe the `pull-back' of deconvolution without a limiting boundary. The matrix $A$ is the convolution operation Toeplitz matrix constructed from the corresponding discretized kernel function $K(t)$. The auxiliary vector is $\mathbf{x}_0 = \mathbf{0}$ in our problem formulation. The regulate the solution smoothness (curvature), we use a finite difference second order derivative matrix
\begin{equation}
L = \begin{pmatrix}
1 & -2 & 1 & 0 & \dots \\
0 & 1 & -2 & 1 & \dots \\
 & & \ddots & \\
\dots & 0 &1 & -2 & 1
\end{pmatrix}.
\end{equation}
Other typical options for $L$ are the identity matrix and first order derivatives. The minimization is done through an active set method~\cite{lawson1995solving} which enforces the necessary Karush-Kuhn-Tucker (KKT) constrained optimization conditions. To be able to use standard optimization algorithms with the regularization term included, we use an augmented matrix formulation
\begin{align}
\hat{\mathbf{x}} = \argmin_{\mathbf{x}} \, &||\widetilde{A}\mathbf{x} - \widetilde{\mathbf{y}}||^2, \;\; \text{subject to} \;\; \mathbf{x} \geq \mathbf{0}, \;\; \text{where} \\
& \widetilde{A} \equiv \begin{pmatrix}
A \\
\lambda_R L
\end{pmatrix}, \;\;
\widetilde{\mathbf{y}} \equiv \begin{pmatrix}
\mathbf{y} \\
\lambda_R L \mathbf{x}_0
\end{pmatrix}.
\end{align}
Thus the regularization is fully explicit here. We use minimal parameter values for $\lambda_R$ yielding smooth inversion results without oscillatory behavior and remark that the statistical uncertainties of the inverse estimate are affected by the regularization procedure, due to the bias-variance trade-off. The regularization adds a small bias term into the solution, and correspondingly suppresses the statistical fluctuations. This makes the statistical uncertainty properties of inverse estimates non-trivial.

\begin{figure}[tb!]
    \centering
    \includegraphics[width=0.95\textwidth]{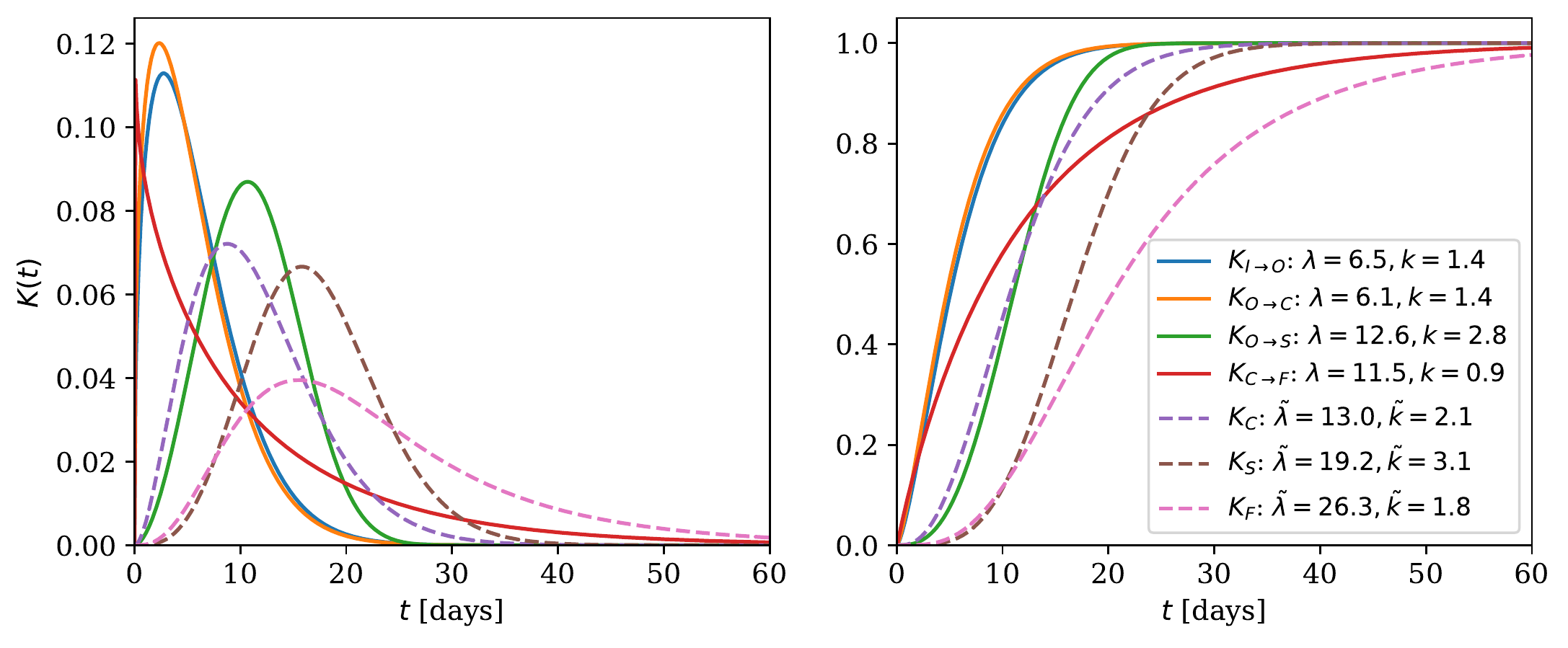}
\caption{Delay convolution Weibull kernels $K(t)$ fitted using the mean and standard deviation values given in \cite{Perez-Saez2020.06.10.20127423}, where $\lambda$ (scale) and $k$ (shape) denote the corresponding Weibull pdf parameters. The individual delays are: $K_{I \rightarrow O}$ is from the infection to the symptom onset (incubation period), $K_{O \rightarrow C}$ is from the symptom onset to the case report, $K_{O \rightarrow S}$ is from the symptom onset to seroconversion (antibodies) and $K_{C \rightarrow F}$ is from the case report to death. The combined delays are: $K_C = K_{I \rightarrow O} \ast K_{O \rightarrow C}$ is from the infection to the case report, $K_S = K_{I \rightarrow O} \ast K_{O \rightarrow S}$ is from the infection to seroconversion and $K_F = K_{I \rightarrow O} \ast K_{O \rightarrow C} \ast K_{C \rightarrow F}$ is from the infection to death. The combined kernels are solved by numerical convolution of the individual kernels, with tilded variables denoting the resulting Weibull parameters. }
    \label{fig:delay_kernels}
\end{figure}

The kernel extraction from data itself is its own problem, typically approached using e.g. Kaplan-Meier type non-parametric estimators~\cite{kaplan1958nonparametric} and functions generalizing the basic exponential (memoryless process) delay kernel, such as the Weibull pdf. Sequential delays are easy to model via cascaded convolutions, but the factorization and identifiability of the component kernels in terms of the underlying physically independent delay sources is not necessarily possible. The fitted kernels are shown in Figure~\ref{fig:delay_kernels}, which are also causal such that they are defined only for $t > 0$.

\label{sec:seroreversion}
\paragraph{Seroreversion}~An additional effect beyond the causal delays discussed earlier, is the finite half-life of antibodies. Taking this evaporation effect into account, the total measurable seroprevalence $\widetilde{I}_S(t)$ can be modelled with the following convolutions
\begin{align}
\label{eq:seroreverse}
\nonumber
\widetilde{I}_S(t) &= I_S(t) - I_{RS}(t) \\
&= (K_S \ast \hat{I})(t) - (K_R \ast (K_S \ast \hat{I}))(t),
\end{align}
where $\hat{I}(t)$ comes from the deconvolution procedure and $K_R$ is the new kernel function, modelling the finite lifetime of measurable antibodies in the body (e.g. exponential decay). The extraction of this requires time-dependent control studies where a group of test positive are monitored and continuously re-tested over a period of months. Convolution integrals are involved in the solution, because we deal with time-evolving input distributions, assume linearity of the system and time-invariance of the kernels. In Eq.~\ref{eq:seroreverse}, the first term $I_S(t)$ is the time-delayed seroprevalence without antibody decays and the second term $I_{RS}(t)$ is the delayed and decayed distribution part. The difference between these gives the actual measured seroprevalence $\widetilde{I}_S(t)$, and asymptotically $\widetilde{I}_S(t) \rightarrow 0$ when $t \rightarrow \infty$, due to the finite half-life. Naturally, when the half-life of antibodies $t_{1/2} \rightarrow \infty$, then we recover the case $\widetilde{I}_S(t) \rightarrow I_S(t)$, that is, the case without seroreversion.

Because the antibody decay kernel can have a very long tail, all computational convolution procedures with arrays should domain extend (pad) the daily counts with zeros, to handle properly the convolutional (un)winding of these tails.

\section{Wasserstein optimal transport}
\label{appendix:sec:optimal_transport}

Using standard notation, the $p$-Wasserstein metric~\cite{dobrushin1970prescribing} for $p\geq 1$ is given by
\begin{equation}
W_p(\mu,\nu) = \left(\inf_{\gamma \in \Gamma(\mu,\nu)} \int_{\chi \times \chi} d(x,y)^p \gamma(dx,dy) \right)^{1/p},
\end{equation}
where $d(x,y)$ is the basic cost between two points $x$ and $y$, for example $d(x,y) = \|x-y\|$. The so-called transport map $T:\mu \mapsto \nu$, maps a measure density $\mu$ to another measure density $\nu$ over the space $\chi$. The set of all possible couplings is $\Gamma(\mu,\nu)$, which has marginals $\mu$ and $\nu$, with a realization $\gamma(dx,\chi) = \mu(dx)$ and $\gamma(dy,\chi) = \nu(dy)$. The case $p=2$ and $\chi=\mathbb{R}^D$ has a unique minimum solution. In one dimension $D=1$, the metric can be written as
\begin{equation}
\label{eq:quantile_OT}
W_p(\mu,\nu) = \left(\int_0^1 dz \, |U^{-1}(z) - V^{-1}(z)|^{p} \right)^{1/p},
\end{equation}
where $U(z)$ and $V(z)$ are the cumulative distribution functions (CDF) of $\mu$ and $\nu$, and the comparison in Eq.~\ref{eq:quantile_OT} is between inverse CDFs (quantile functions).

\section{Optimality under risk functions}
\label{sec:risk_functions}

It may be tempting to choose only one of the estimator results. However, optimality of this decision depends on the risk function definition. To ease out with possible interpretations, here we list shortly some typical risk functions. Let our parameter of interest be $\theta$, the random variable of data be $X$, the decision function (estimator) be $\delta(X)$ and the loss function be $\xi(\theta, \delta(X))$, which encodes our cost definition. Beyond these probabilistic risks, there are related principles in information theory, such as the minimum description length (MDL) \cite{rissanen1978modeling} and other formulations e.g. in economics.
\\

\paragraph{Frequentist risk}~The frequentist risk is the loss integrated over the sampling density
\begin{equation}
R(\theta,\delta) = \mathbb{E}_{\theta}[\xi(\theta, \delta(X))] = \int dx \, \xi(x, \delta(X)) f(x|\theta).
\end{equation}
If the loss function is
\begin{equation}
\xi(\theta, \delta (X)) = (\theta - \delta(X))^2,
\end{equation}
then the optimal decision $\delta^*(X)$ minimizes the sum of (squared) bias and variance, which is trivial to show.

\paragraph{Posterior risk}~The posterior risk is the loss integrated over the posterior density
\begin{align}
B(\theta,\delta) &= \int d\theta \, \xi(\theta, \delta) P(\theta|X),
\end{align}
which has two typical solutions
\begin{align}
\xi(\theta,\delta(X)) &= (\theta - \delta(X))^2 \rightarrow \delta^*(x) = \int d\theta \, \theta P(\theta|x) \sim \text{posterior mean} \\
\xi(\theta,\delta(X)) &= |\theta - \delta(X)| \rightarrow \delta^*(x) \sim \text{posterior median}.
\end{align}
The optimal decisions $\delta^*(X)$ for these losses are obtained by the posterior mean and median.

\paragraph{Bayes rule risk}~The hybrid risk is the frequentist risk integrated over the prior density
\begin{equation}
H(\theta,\delta) = \int d\theta \, R(\theta,\delta(X)) \pi(\theta).
\end{equation}
Using certain specific priors $\pi(\theta)$, one can turn this into a minimax risk.

\paragraph{Minimax risk}
\begin{equation}
\sup_{\Theta} R(\theta,\delta),
\end{equation}
is the worst case (maximum) frequentist risk. As its name states, the minimax-optimal decision is the one which minimizes the maximum expected risk.

\section{Overview of systematic uncertainties}
\label{appendix:sec:systematics}

This section is a general summary of possible unknowns.
\\

\noindent \textbf{Sampling model and demographic variations}

\begin{itemize}

\item The number of tested people is a well-known quantity, but the total (effective) population size of the system is not fully known. This is the problem of open versus closed systems, or their idealization. In reality, not all citizens are in contact but there are locally isolated systems, which are not `thermalized' together. One may argue that to be able to define the IFR in a way as is typically done, by using a test sample and extrapolating to the full city population scale, the so-called ergodicity hypothesis of Boltzmann is assumed to hold implicitly. Another sampling issue is the local household clusterization effect, which can in principle induce both positive and negative correlations such as the average infection rate first increasing and then decreases as a function of the household size, due to children. Monte Carlo simulations can be used to study these issues, but we may expect other sources of uncertainties to be typically much larger, at least while comparing studies implemented in relatively similar sized and dense systems.

\item The demographic heterogeneity uncertainties and their regression modelling are discussed already in some detail in Section \ref{sec:combination}. It makes sense to compare the average IFR one-to-one between countries which have similar demographics. The population median age in the world spans approximately 32 years, between Niger $\sim 15$ years to Japan $\sim 47$ years, which is expected to have a large impact. Similarly, the provided health care are very different. The combination analysis, if implemented using studies done under similar demographic conditions, probes then the underlying and always partially unknown systematics in an empirical and effective way.

\end{itemize}

\noindent \textbf{Initial viral dose}

\begin{itemize}
\item It is currently an open question how large is the effect of the initial viral dose on the outcome of the disease development. It has been hypothesized \cite{gandhi2020facial} that using face masks effectively reduces, not just the number of infections, but in a more non-linear way also the infection fatality rate due to smaller doses transmitted and received. In this case, a person receiving a small dose, would allow their body to develop mild symptoms and even immunity. The serious condition would happen instead more likely with a large initial dose of the virus. A positive correlation can be expected with large viral load during the disease and the severity, but the transmission dose dependence instead is hard to analyze without dedicated studies.
\end{itemize}

\noindent \textbf{PCR and antibody tests}

\begin{itemize}
\item Sensitivity (true positive rate) and specificity (true negative rate), or the ROC-curve `receiver operating characteristics' working point of PCR or antibody tests, should be carefully calibrated and corrected for. Person by person, there are irreducible type I (false positive) and type II (false negative) classification errors to be made which cannot be avoided, however, for large samples it is possible to compensate these errors by inversion analysis. The corrections can be calculated as explained in Section~\ref{appendix:sec:test_inversion} or even perhaps more optimally, using weighted corrections test-by-test. Re-weighting or other corrections can be executed only if the test manufacturer has produced well calibrated tests and algorithms with a probabilistic output. In a review of five studies, SARS-CoV-2 PCR tests have been estimated to have a false negative rate up to 29 \%~\cite{Arevalo-Rodriguez2020.04.16.20066787}, however, this depends on the chosen working point of false positive rate.

\item The degree of personal variation on the antibody response is not yet well understood. As an alternative strategy to antibodies, the T-cell response for SARS-CoV-2 seems currently promising to combine with the antibody response~\cite{le2020sars}.

\end{itemize}

\noindent \textbf{Temporally induced biases}

\begin{itemize}
    \item Cumulative death counts [IFR bias $\downarrow \uparrow$] \\
    Relative undercounting of death counts happens simply due to the chosen analysis time interval endpoint and finite time delays according to Eq.~\ref{eq:delay_scale}, driven by biological and communication delays. Similarly, it is possible to do relative overcounting. This `efficiency' or `overcounting' type of counting error can be estimated and multiplicatively corrected, but its accuracy is limited by the quality of delay kernels extracted from data.
    
    \item Infection decoupling [IFR bias $\uparrow$] \\
    If a PCR type test is made too late, it can miss possibly (earlier) positive person. This effect is prominent in the tails, when the infection vanishes from the population. In this case, fatalities and infection counts will stay the same, but when the test count grows as a function of time, it leads to a growing IFR estimate. The associated time period is called also as the `duration of viral shedding'. To mitigate this problem, typically seroprevalence tests should be used primarily to determine the IFR. Alternative is continuous (daily) PCR testing, which is typically feasible only for high risk group individuals.
    
    \item Antibody development and half-life [IFR bias $\uparrow$] \\
    When an antibody (IgG, IgM, $\dots$) type seroprevalence determination is implemented, it is necessary to take into account the body response delays of developing the necessary amount of antibodies to pass the test thresholds but also the fact that the antibodies do also decay, i.e., their half-life is not necessarily insignificant on the time scale of the epidemic. Delays in development or vanishing of antibodies can bias the IFR estimate upward by downward biased prevalence count. In Ref.~\cite{Liu2020.06.13.20130252} it was concluded that after SARS-CoV-2 infection, long-lasting protective antibodies are not likely produced. This is currently an open question in precision terms. In Ref.~\cite{iyer2020persistence} was found that SARS-CoV-2 IgG responses decreased only 4\% within 90 days. However, IgA and IgM were short-lived with median decay times of 70.5 [58.5, 87.5] and 48.9 [43.8, 55.6] days (CI95). Neutralizing antibody titers had little decrease, being also highly correlated with IgG.
    
    \item Non-uniform sampling rate [IFR bias $\downarrow \uparrow$] \\
    Any precision procedure relying computing e.g. (de)convolution between time-series data and delay kernels, may need to take into account the non-uniform testing and reporting rates. However, the reported daily death count time series can be considered more reliable, assuming that deaths are correctly reported and placed in the time series. Inspecting public data, this evidently is not always the case, with anomalously large discontinuities seen in time-series.
\end{itemize}

\noindent \textbf{Cause of death ambiguity}

\begin{itemize}
    \item The conditional classification of the death itself, to be caused by COVID-19, is not fully unique. A person may develop simultaneous serious bacterial (Streptococcus etc.) pneumonia increasing the fatality risk, which is typical with seasonal influenza viruses and one of the most common causes of death~\cite{morris2017secondary}. The unique cause of death will be ambiguous or degenerate in this case. Similarly, any underlying chronic conditions can significantly affect the outcome, such as the metabolic syndrome. One future solution to this could be a more advanced bookkeeping scheme, which assigns data-driven probabilities with one or more international cause of death (ICD) codes, to tag simultaneous underlying conditions. The conditional probability $P(Y|X)$ is by definition the joint probability $P(Y,X)$ divided by the probability of the condition $P(X)$. As an approximation, there could be also just two categories of COVID-19 deaths, with and without existing chronic conditions. In addition, there can be also other systematic country or study level differences in basic bookkeeping of death counts. This issue is particularly relevant, when excess fatality rate comparisons due to COVID-19 are made against seasonal flu fluctuations.
\end{itemize}
\clearpage

\section{Coverage simulations}
\label{appendix:sec:coverage}
\begin{figure}[htb!]
    \centering
    \includegraphics[width=0.49\textwidth]{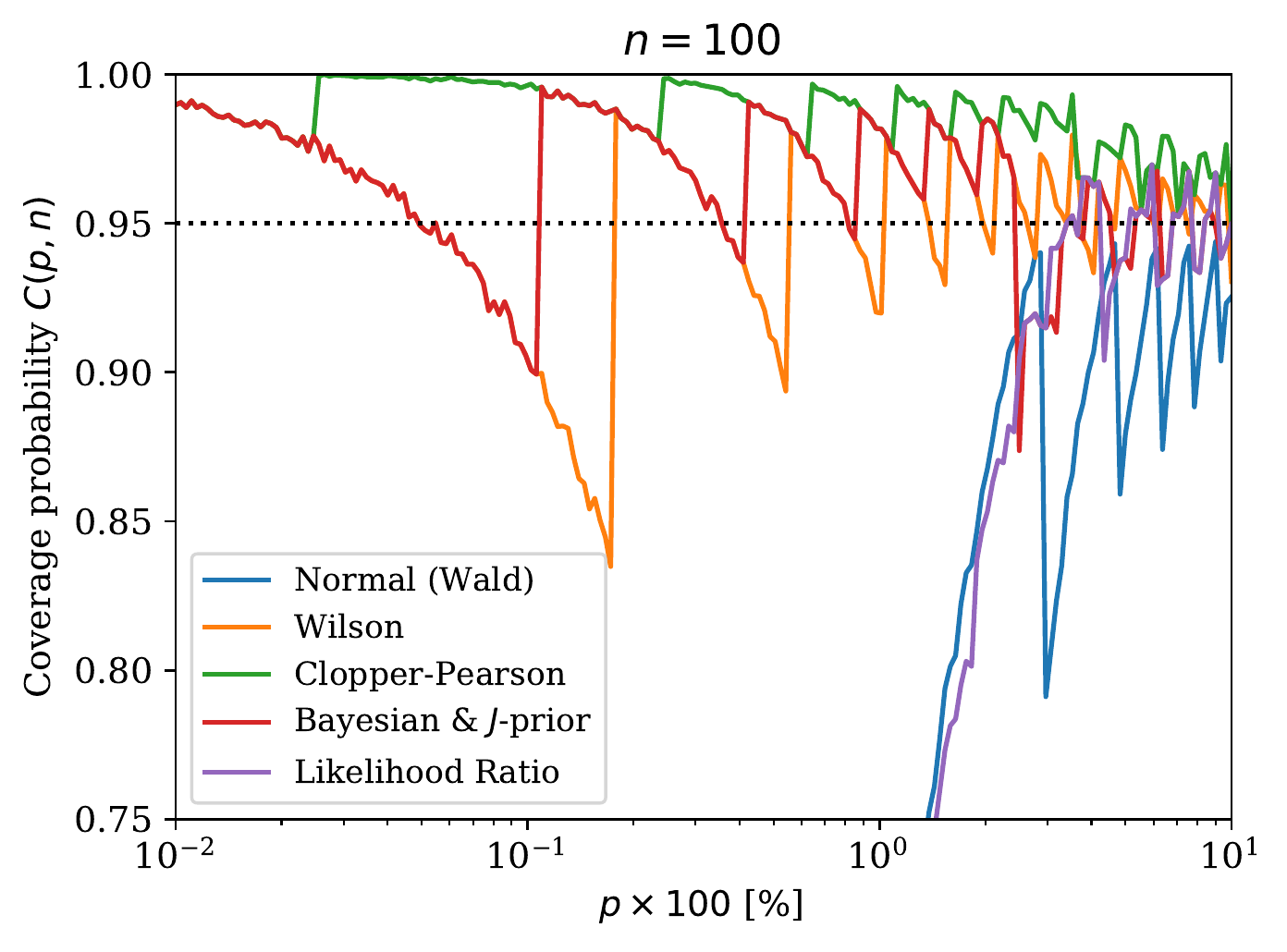}
    \includegraphics[width=0.485\textwidth]{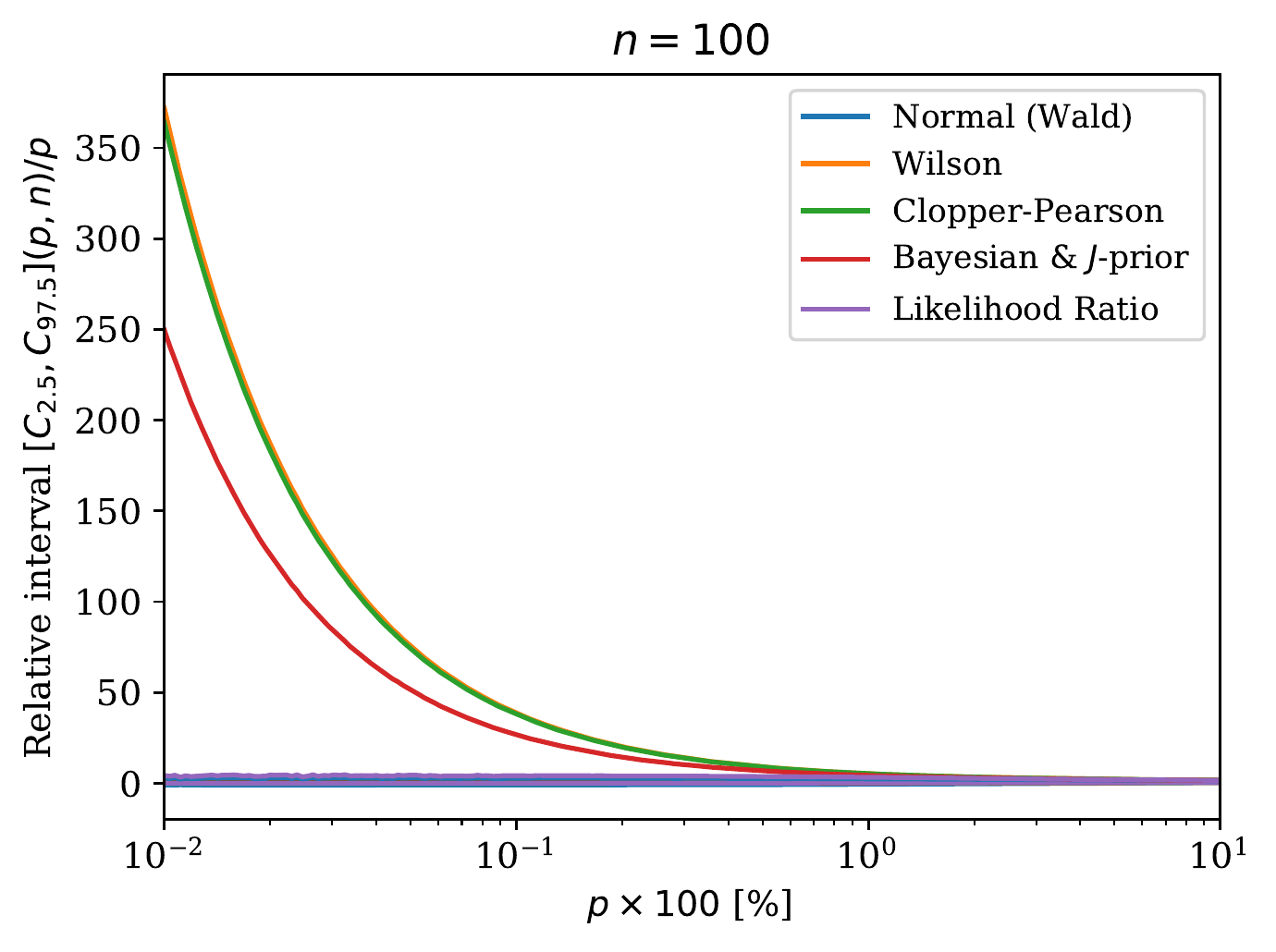} \\
    \vspace{1.5em}
    \includegraphics[width=0.49\textwidth]{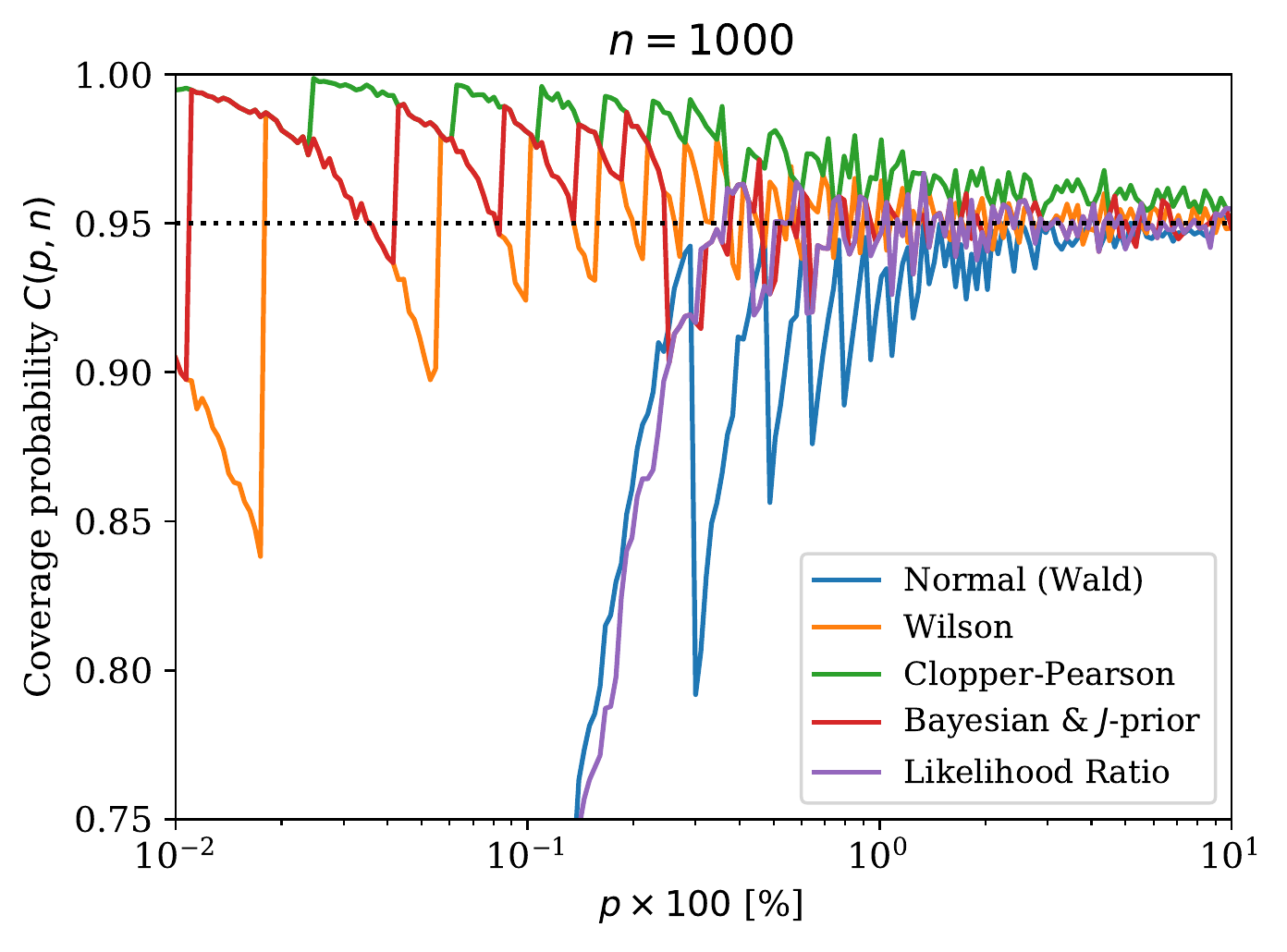}
    \includegraphics[width=0.475\textwidth]{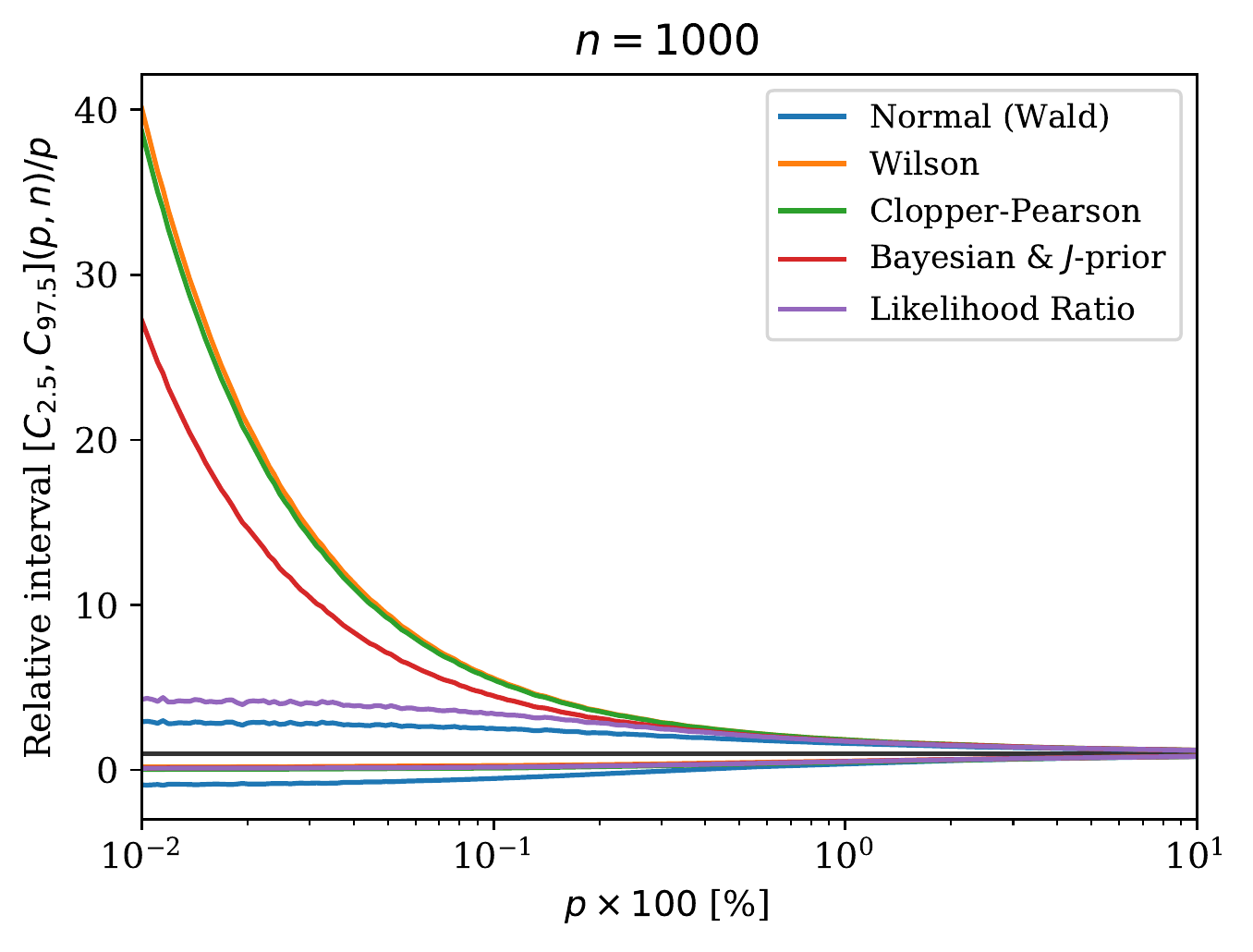} \\
    \vspace{1.5em}
    \includegraphics[width=0.49\textwidth]{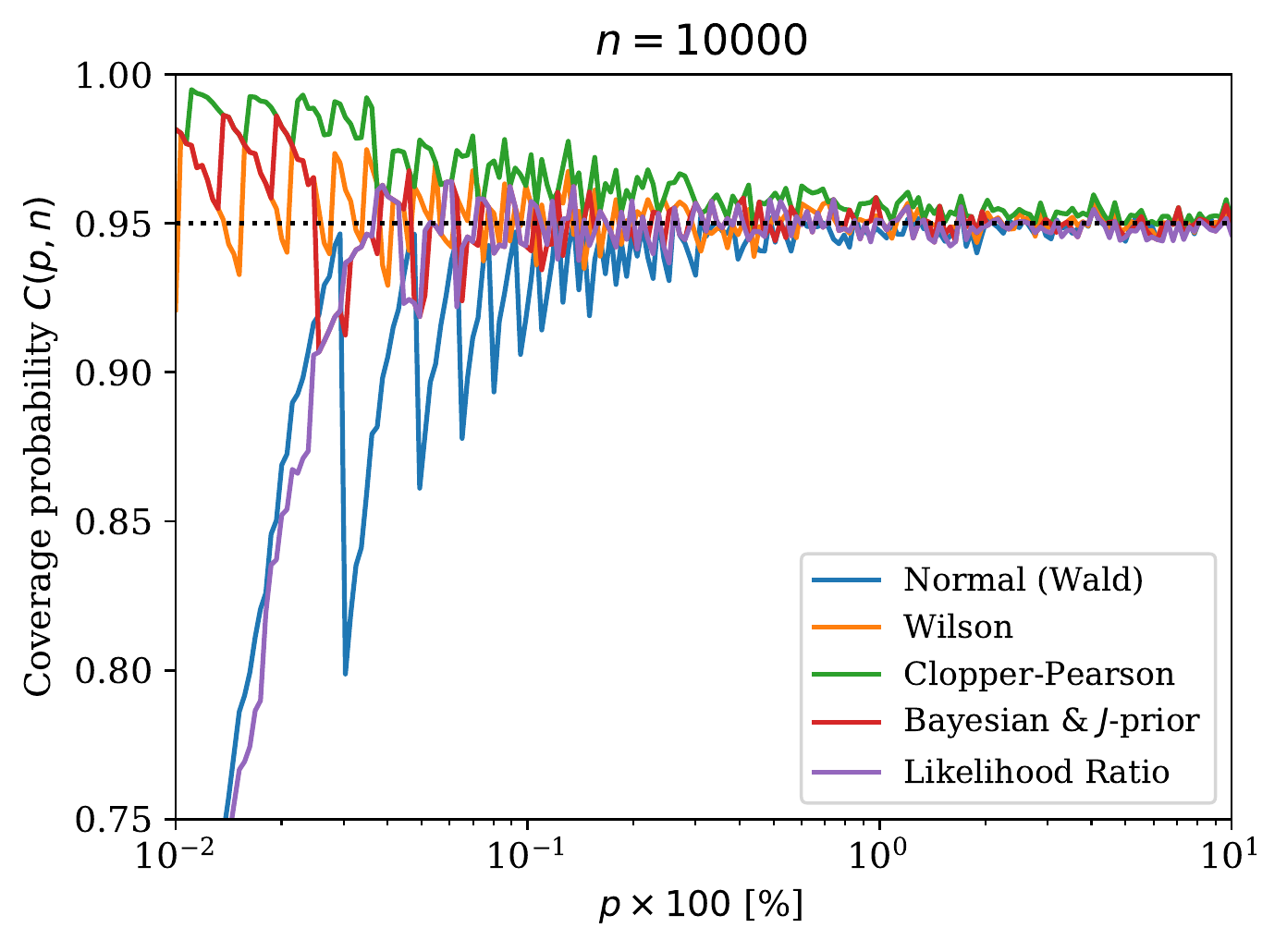}
    \includegraphics[width=0.468\textwidth]{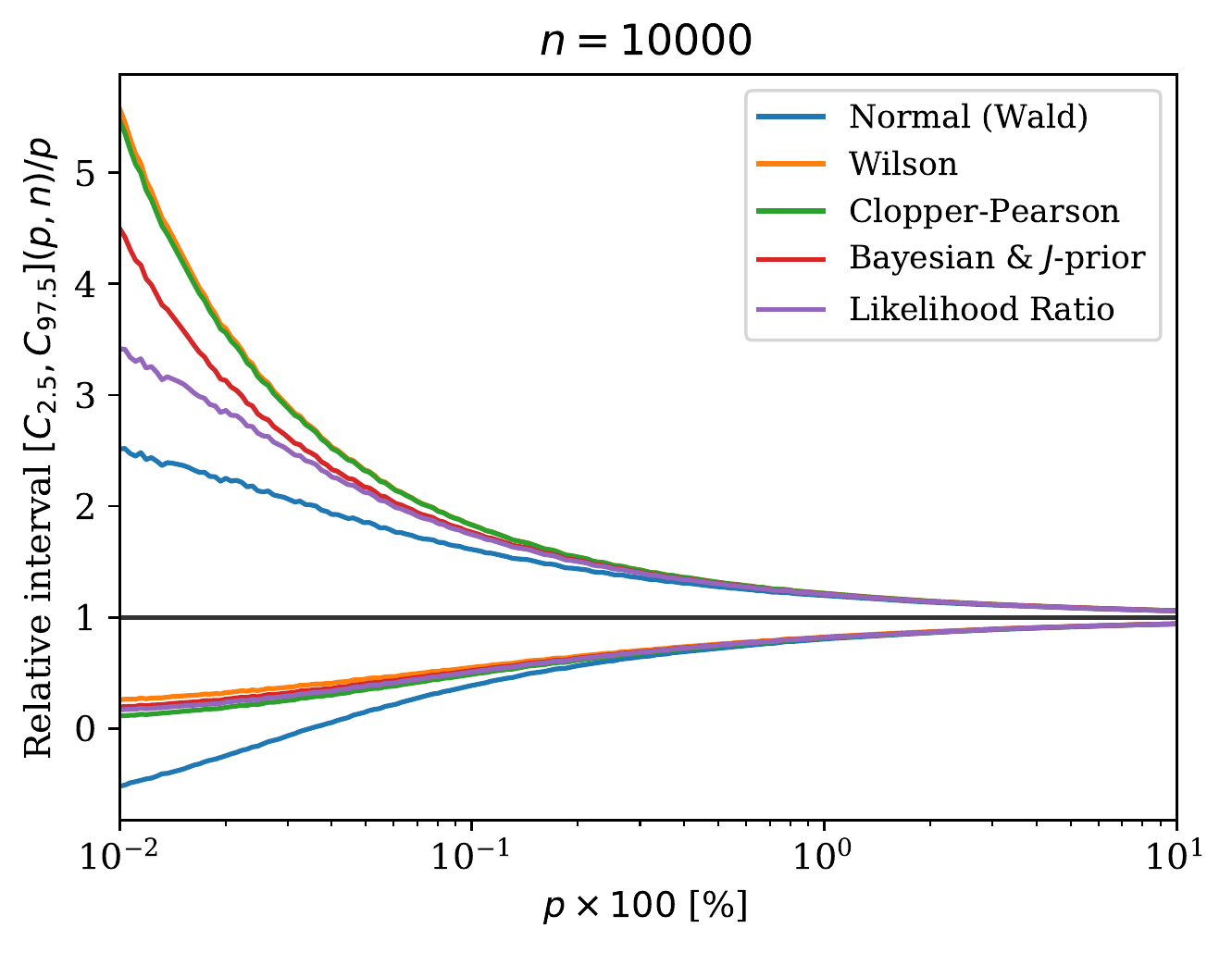}
    \caption{On the left, simulations of a single binomial proportion confidence interval CI95 coverage as a function of the binomial probability $p$. On the right, the corresponding average confidence interval CI95 relative widths (endpoints). Each row for $n$ number of binomial trials. The likelihood ratio is with $\chi^2$-approximation.}
    \label{fig:coverage_sim}
\end{figure}

\end{document}